\documentclass[a4paper]{article}
\pdfoutput=1

\usepackage{jheppub} 

\usepackage[T1]{fontenc} 
\usepackage[utf8]{inputenc}
\usepackage{pythonhighlight} 

\usepackage{autobreak}

\usepackage{autobreak}
\usepackage{physics}
\usepackage{feynmp-auto}
\usepackage{slashed}

\usepackage{ulem}
\usepackage{amsmath}
\usepackage{amssymb}
\usepackage{bm}
\usepackage{bbold}
\usepackage{multirow}
\usepackage{makecell}

\newcommand\Tstrut{\rule{0pt}{2.6ex}}         

\usepackage{graphicx}
\usepackage{hyperref}
\usepackage[american]{babel}
\usepackage{enumerate}
\usepackage{mathtools}
\usepackage{support-caption}
\usepackage{subcaption}
\usepackage{soul}
\setstcolor{red}
\usepackage{makebox}
\usepackage{comment}
\usepackage{booktabs}

\newcommand{\U}[1]{\mathrm{U}(1)_{\mathrm{#1}}}			
\newcommand{\SU}[2]{\mathrm{SU}(#1)_{\mathrm{#2}}}		

\renewcommand{\[}{\left[}

\newcommand{\del}{\partial}
\newcommand{\mean}[1]{\left \langle #1 \right \rangle }

\usepackage{pifont}

\newcommand\varpm{\mathbin{\vcenter{\hbox{%
  \oalign{\hfil$\scriptstyle\hspace{-0.2ex}+\hspace{-0.2ex}$\hfil\cr
          \noalign{\kern-.5ex}
          $\scriptscriptstyle({-})$\cr}%
}}}}

\newcommand\varmp{\mathbin{\vcenter{\hbox{%
  \oalign{\hfil$\scriptstyle\hspace{-0.2ex}-\hspace{-0.2ex}$\hfil\cr
          \noalign{\kern-.5ex}
          $\scriptscriptstyle({+})$\cr}%
}}}}

\DeclareMathSymbol{\comma}{\mathpunct}{letters}{"3B} 

\definecolor{ForestGreen}{rgb}{0.13, 0.55, 0.13}

\definecolor{amber}{rgb}{1.0, 0.45, 0.0}

\definecolor{plum}{rgb}{0.56, 0.27, 0.52}

\usepackage{xspace}
\newcommand*{\GW}{\ensuremath{\mathrm{GW}}\xspace}

\usepackage[capitalise]{cleveref}

\crefname{section}{Section}{Sections}
\crefname{table}{Table}{Tables}
\crefname{figure}{Fig.}{Figs.}
\crefname{equation}{Eq.}{Eqs.}
\crefname{appendix}{Appendix}{Appendices}

\graphicspath{{figures/}}

\allowdisplaybreaks

\title{Gravitational waves from supercooled phase transitions in conformal Majoron models of neutrino mass}

\author[a,b,e]{Jo\~ao~Gon\c{c}alves,}
\author[c]{Danny Marfatia,}
\author[a,d,e]{Ant\'onio~P.~Morais}
\author[b]{and Roman Pasechnik}

\affiliation[a]{Departamento de F\'{i}sica da Universidade de Aveiro and Centre  for  Research  and  Development  in  Mathematics  and  Applications  (CIDMA),
	Campus de Santiago, 3810-183 Aveiro, Portugal}
 \affiliation[b]{Department of Physics, Lund University, 221 00 Lund, Sweden}
\affiliation[c]{Department of Physics and Astronomy, University of Hawaii at Manoa, Honolulu, HI 96822, USA}
\affiliation[d]{Departamento de Física, Escola de Ciências, Universidade do Minho, 4710-057 Braga, Portugal}
\affiliation[e]{Laboratório de Instrumentação e Física Experimental de Partículas (LIP), Universidade do Minho, 4710-057 Braga, Portugal}
\emailAdd{jpedropino@ua.pt}
\emailAdd{dmarf8@hawaii.edu}
\emailAdd{amorais@fisica.uminho.pt}
\emailAdd{roman.pasechnik@fysik.lu.se}

\abstract{We study supercooled first-order phase transitions above the QCD scale in a wide class of conformal Majoron-like $\U{ }^\prime$ models that explain the totality of active neutrino oscillation data and produce a detectable stochastic gravitational wave background (SGWB) at LIGO, LISA and ET. We place constraints
on the $\U{ }^\prime$ breaking scale and gauge coupling using current LIGO-Virgo-Kagra data. We find that strong supercooling can be ruled out in large regions of parameter space if a SGWB  is not detected by these experiments. A null signal at LIGO and ET will disfavor a type-I seesaw scale above $10^{14}~\mathrm{GeV}$, while a positive signal is a signature of heavy right-handed neutrinos. 
On the other hand, LISA will be sensitive to seesaw scales as low as a TeV, and could detect a SGWB even if the right-handed neutrinos are decoupled. }

\begin{document}

\maketitle
\flushbottom

\section{Introduction}\label{Sect:intro}

Despite the tremendous success of the Standard Model (SM), there is overwhelming evidence for new physics. The detection of neutrino masses, the need for dark matter, and the inferred baryon asymmetry of the Universe are examples that motivate the search for new phenomena in both particle physics and cosmological data.
In the SM, baryogenesis requires electroweak symmetry breaking to be a first-order phase transition (FOPT).
However, both the quantum chromodynamic (QCD) and electroweak phase transitions are smooth crossovers~\cite{DOnofrio:2015gop,Fodor:2004nz}. It is conceivable that new physics allows for FOPTs in the early Universe, which in turn produce gravitational waves (GWs).

The detection of GWs~\cite{LIGOScientific:2016aoc} has opened a new avenue of fundamental physics exploration. In particular, the mounting evidence of a stochastic gravitational wave background (SGWB) from Pulsar Timing Arrays~\cite{NANOGrav:2023gor,Reardon:2023gzh,EPTA:2023fyk,Xu:2023wog} could mark the first direct measurements of the Universe prior to the Big Bang Nucleosynthesis (BBN) era, representing a breakthrough comparable to the discovery of the cosmic microwave background. 
The scale of new physics probed is well above that of TeV collider searches which have come up empty-handed so far.

This article focuses on the study of the SGWB produced during supercooled FOPTs, characteristic of classically scale-invariant models. Scale invariance is a symmetry of the classical action with respect to the simultaneous transformations, $\Phi \to \Phi^\prime = \rho^{-a} \Phi$ and $x \to x^\prime = \rho x$, where $\Phi$, $x$ and $\rho$ represent a generic field (boson or fermion), space-time coordinates, and the scale factor respectively. Here, $a=1$ for bosons and $3/2$ for fermions. This is a particular case of conformal symmetry~\cite{Shaposhnikov:2022zhj,Srivastava:1973iig}, and models incorporating it are typically referred to as conformal models. In Refs.~\cite{Coleman:1973jx,Gildener:1976ih}, the scale-invariant scalar sector is described by a purely quartic potential at tree level. Upon gauge-symmetry breaking, one of the scalars remains massless and becomes a pseudo-Goldstone boson known as the \textit{scalon}, emergent from spontaneous breaking of the continuous scale symmetry. Note that this is a classical symmetry that is explicitly broken by quantum corrections due to the non-polynomial nature of the Coleman-Weinberg (CW) potential \cite{Coleman:1973jx},
from which the scalon obtains its mass. 
 
A distinctive feature of classically conformal models is the substantial amount of gravitational radiation emitted during a FOPT. The potential barrier is absent at zero temperature and is induced by thermal corrections such that the FOPT persists for a relatively long time as the Universe cools down. As a result, the amount of released latent heat is much larger than in models without supercooling, due to a significant difference in potential energy between the true and false vacuum. Classical scale invariance is, therefore, an exceptional paradigm to be tested at gravitational interferometers, with a real possibility of excluding broad regions of the parameter space, or even leading to a breakthrough discovery.

Recent studies have explored classical scale invariance and its implications for the SGWB in the context of the real~\cite{Ahriche:2023jdq} and complex~\cite{Jaeckel:2016jlh} singlet extensions of the SM, the $\U{B-L}$ model~\cite{Marzo:2018nov,Ellis:2020nnr,Ellis:2019oqb,Jinno:2022mie}, and a non-Abelian $\SU{2}{}$ model of dark matter~\cite{Hambye:2018qjv,Baldes:2018emh,Marfatia:2020bcs,Kierkla:2022odc,Kierkla:2023von}. In this article, we study a wide class of locally scale-invariant Majoron-like models equipped with a type-I seesaw mechanism and generic 
$\mathrm{U(1)^\prime}$-gauge charge assignments. In these models, the generation of neutrino masses is driven by spontaneous symmetry breaking. If this breaking occurs via a FOPT, the GWs produced will carry information about the breaking mechanism and the scale of symmetry breaking associated with neutrino mass generation, as this scale is typically correlated with the peak frequency of the GW spectrum. As such, we begin with a comprehensive survey of all possible scale-invariant $\U{}^\prime$ models that explain the entirety of the neutrino oscillation data. Then, we identify which of these models can be tested at LIGO, the Einstein Telescope (ET), and LISA. Our goal is to determine which scales and parameter regions in this class of models can be studied with current and forthcoming GW data. 

This article is structured as follows: In \cref{Sect:model}, we introduce the $\U{}^\prime$ models of interest. In \cref{sec:FOPT_GWs}, we review the theory behind thermal FOPTs, focusing on the key elements necessary for our analysis, including the template functions for the SGWB spectrum. In \cref{sec:effective_pot}, we describe the renormalization Group (RG) improved thermal effective potential. In \cref{sec:num_results}, we present our results and summarize in \cref{sec:conclusions}.

\section{Generic scale-invariant Majoron-like models}\label{Sect:model}

We explore a class of generic $\U{}^\prime$ extensions of the SM, primarily designed to accommodate neutrino masses and mixing through the inclusion of three generations of right-handed neutrinos and a type-I seesaw mechanism. By imposing scale invariance, all tree-level dimensionful parameters of the theory are forbidden, thereby reducing the number of free parameters in the model Lagrangian. Specifically, the conventional Majorana mass term $M \bar{N}^c N$ is not allowed and must be replaced by a Yukawa term yielding the Majorana mass~$M\to y_\sigma\mean{\sigma}/\sqrt{2}$ through the introduction of a complex singlet scalar $\sigma$. In standard Majoron models~\cite{Chikashige:1980qk,Chikashige:1980ui,Gelmini:1980re}, the Majoron is identified with a pseudo Nambu-Goldstone boson resulting from the breaking of a global $\U{L}$ lepton number symmetry. In this study, however, neither is $\U{}^\prime$ global nor does a corresponding physical pseudo-Goldstone boson exist in the spectrum. Nevertheless, we refer to $\sigma$ as Majoron due to its role in generating Majorana masses for right-handed neutrinos, while the corresponding class of models will be called Majoron-like.

The field content, their quantum numbers and transformations under rescaling are shown in \cref{tab:charges}. The third column lists the anomaly-free $\U{}^\prime$ charges adopted from Ref.~\cite{Oda:2015gna}. It is important to note that in the current framework, which includes only one additional scalar $\sigma$, the anomaly-free conditions require flavor universality to describe neutrino oscillation data. In this context, the class of models presented is representative of all such scenarios, including the $\U{B-L}$ model~\cite{Iso:2009ss}  with $x_\mathcal{H}=0$ and $x_\sigma=2$ in \cref{tab:charges}.  We treat $x_\sigma$ and $x_\mathcal{H}$ as free parameters. The SM gauge group representations are shown in the last three columns. 
\begin{table}[t]
	\centering
	\begin{tabular}{c|c|c|c|c|c}
		\textbf{Field} & \textbf{Scale symmetry} & $\mathbf{U(1)^\prime}$ & $\mathbf{SU(3)_\text{C}}$ & $\mathbf{SU(2)_\text{L}}$ & $\mathbf{U(1)_\text{Y}}$  \\ \hline\Tstrut
		$Q$ & $\rho^{-3/2}$ & $\frac{1}{3} x_\mathcal{H} + \frac{1}{6} x_\sigma$ & $\bm{3}$ & $\bm{2}$ & $1/6$  \\[0.5em]
		$u_R$ & $\rho^{-3/2}$ & $\frac{4}{3} x_\mathcal{H} + \frac{1}{6} x_\sigma$ & $\bm{3}$ & $\bm{1}$ & $2/3$ \\[0.5em]
  		$d_R$ & $\rho^{-3/2}$ & $-\frac{2}{3} x_\mathcal{H} + \frac{1}{6} x_\sigma$ & $\bm{3}$ & $\bm{1}$ & $-1/3$ \\[0.5em]
		$L$ & $\rho^{-3/2}$ & $ -x_\mathcal{H} - \frac{1}{2} x_\sigma$ & $\bm{1}$ & $\bm{2}$ & $-1/2$ \\[0.5em]
        $e_R$ & $\rho^{-3/2}$ & $ -2x_\mathcal{H} - \frac{1}{2} x_\sigma$ & $\bm{1}$ & $\bm{1}$ & $-1$ \\[0.5em]
		$\mathcal{H}$ & $\rho^{-1}$ & $x_\mathcal{H}$ & $\bm{1}$ & $\bm{2}$ & $1/2$ \\[0.5em]
        $N $ & $\rho^{-3/2}$ & $-\frac{1}{2} x_\sigma$ & $\bm{1}$ & $\bm{1}$ & $0$ \\[0.5em]
        $\sigma$ & $\rho^{-1}$ & $x_\sigma$ & $\bm{1}$ & $\bm{1}$ & $0$ \\ [0.5em]
        \hline
	\end{tabular}
	\caption{\footnotesize Scaling and anomaly-free gauge quantum numbers of the field content in the class of scale-invariant Majoron-like models considered. The $\U{}^\prime$ charges are defined in terms of those of the SM Higgs doublet $\mathcal{H}$, $x_\mathcal{H}$, and the Majoron field $\sigma$, $x_\sigma$. The $\rho$ parameter denotes the scale factor for each field that also enters the coordinate transformation, $x \to x^\prime = \rho x$, as required by the scale symmetry.}
 \label{tab:charges}
\end{table}

\subsection{Yukawa sector}\label{SubSect:yukawas}

The neutrino sector Lagrangian of a classically scale invariant type-I seesaw mechanism with a Majoron reads~\cite{Chikashige:1980qk,Chikashige:1980ui,Gelmini:1980re}
\begin{equation}
    \mathcal{L}_\nu = y_\nu^{ij} \overline{N}_{i} \mathcal{H} L_j + y_\sigma^{ij} \Bar{N}_{i}^c N_{j} \sigma
    + \mathrm{h.c.}\,,
    \label{eq:Lnu}
\end{equation}
where
\begin{equation}
    L_i =
    \begin{pmatrix}
\nu_{\mathrm{L}i} \\
e_{\mathrm{L}i} 
\end{pmatrix}
\qquad
\textrm{and}
\qquad
\Tilde{\mathcal{H}} \equiv i \tau_2 \mathcal{H}^\dagger\,.
\end{equation} 
The mass matrix in the basis $\left\{\Bar{N}_{i},\Bar{N}_{i}^c\right\} \otimes \left\{\nu_{\mathrm{L}j},N_{j}\right\}$ is in a block compact form, 
\begin{equation}
\bm{M_\nu} = 
\left( \begin{array}{cc} 
0        &  \tfrac{v}{\sqrt{2}} \bm{y_\nu}^{\mathrm{T}} \\  
\tfrac{v}{\sqrt{2}} \bm{y_\nu}  & \tfrac{v_\sigma}{\sqrt{2}}  \bm{y_\sigma}
\end{array} \right) \,,
\label{eq:Mnu}
\end{equation}
where $v\simeq 246$ GeV and $v_\sigma$ represent the vacuum expectation values (VEVs) of the $\mathcal{H}$ and $\sigma$ fields, respectively.
The eigenvalues of $\bm{M_\nu}$ are the masses of the three active neutrinos, $m_1$, $m_2$ and $m_3$, and the three heavy neutrinos $M_{N_1}$, $M_{N_2}$ and $M_{N_3}$. The light neutrino mass matrix is
\begin{equation}
\bm{m}_\nu \approx \frac{1}{\sqrt{2}} \frac{v^2}{v_\sigma} \bm{y_\nu}^\mathrm{T} \bm{y_\sigma}^{-1} \bm{y_\nu} \,.
\label{eq:nu-light}
\end{equation} 
Similarly, the masses of the heavy neutrinos are given by
\begin{equation}\label{eq:heavyN}
\bm{M_{N}} \approx \frac{v_\sigma}{\sqrt{2}} \bm{y_\sigma} \,.
\end{equation}

The neutrino mass matrix can be inverted using the neutrino mass differences and the entries of the Pontecorvo–Maki–Nakagawa–Sakata matrix, $\bm{U}_\mathrm{PMNS}$, as input parameters. For a type-I seesaw mechanism with a diagonal $\bm{y_\sigma}$, we can write \cite{Cordero-Carrion:2019qtu}
\begin{equation}\label{eq:Ynu_matrix}
    \bm{y_\nu} = i \bm{\Sigma}^{-1/2} \bm{R} \bm{D}_{\sqrt{m}} \bm{U}^\dagger_\mathrm{PMNS} \,, 
\end{equation}
where $\bm{\Sigma}$ is a diagonal $3\times 3$ matrix whose entries are the singular values of $[(\sqrt{2} v^2)/(2 v_\sigma)] \bm{y^{-1}_\sigma}$, and $\bm{D}_{\sqrt{m}} = \mathrm{diag}(\sqrt{m_1}, \sqrt{m_2}, \sqrt{m_3})$. Here, $\bm{R}$ is a generic complex orthogonal $3\times 3$ matrix that satisfies~\cite{Casas:2001sr} 
\begin{equation}
   \bm{R}\bm{R}^T = \bm{R}^T\bm{R}=\bm{I}\,. 
\end{equation} 
For the numerical analysis, we used the latest neutrino oscillation data fit from the NuFIT collaboration \cite{Esteban:2020cvm}, assuming a normal mass hierarchy. Cosmological constraints on the neutrino mass sum are also taken into account, with an upper bound set at $\sum m_\nu < 0.12~\mathrm{eV}$  \cite{Planck:2018vyg}.

Note that current LVK data impose constraints on classically conformal Majoron models because GW interferometers operating in the $10 - 100~\mathrm{Hz}$ frequency range are sensitive to scales between $10^{12}~\mathrm{GeV}$ and $10^{17}~\mathrm{GeV}$, which are typically associated with heavy neutrino masses in a high-scale type-I seesaw mechanism. 

\subsection{Scalar potential}
\label{Sec:scalar_pot}

The tree-level scalar potential of a generic classically conformal Majoron model is given by
\begin{equation}\label{eq:tree_potential}
\begin{aligned}
V_{0}(\mathcal{H},\sigma) &= \lambda_h(\mathcal{H}^{\dagger}\mathcal{H})^2 + 
\lambda_{\sigma}(\sigma^{\dagger}\sigma)^2 + 
\lambda_{\sigma h}(\mathcal{H}^{\dagger}\mathcal{H})(\sigma^{\dagger}\sigma) \,.
\end{aligned}
\end{equation}
We can expand the Higgs doublet $\mathcal{H}$ and the Majoron $\sigma$ in terms of real-valued components as
\begin{equation}\label{eq:Hsigma}
\begin{aligned}
\mathcal{H} = \dfrac{1}{\sqrt{2}} 
\begin{pmatrix}
\omega_1 + i \omega_2  \\
\phi_h + h_r + i \eta
\end{pmatrix}\,,	
\qquad
\sigma = \dfrac{1}{\sqrt{2}} \left( \phi_\sigma + h^\prime_r + i J \right)\,,
\end{aligned}
\end{equation}	
where $h_r$ and $h^\prime_r$ represent radial quantum fluctuations about the classical field configurations $\phi_h$ and $\phi_\sigma$, while $\omega_{1,2}$, $\eta$, and $J$ are the Goldstone modes corresponding to the longitudinal degrees of freedom of the gauge bosons upon symmetry breaking. In terms of classical fields, the tree-level potential reads
\begin{equation}\label{eq:tree_potential_classival}
    V_{0}(\phi_h,\phi_\sigma) = \frac{1}{4} \Big( \lambda_h \phi^4_h + \lambda_\sigma \phi^4_\sigma + \lambda_{\sigma h} \phi^2_h \phi^2_\sigma \Big)\,.
\end{equation}
We can then express the field-dependent squared mass matrix as
\begin{equation}\label{eq:mass_matrix}
\bm{M^{2}_{(0)}} = \begin{pmatrix}
3 \phi_h^2 \lambda_h + \frac{1}{2}\lambda_{\sigma h}\phi^2_\sigma & \phi_h \phi_\sigma \lambda_{\sigma h} \\
\phi_h \phi_\sigma \lambda_{\sigma h} & 3 \phi_\sigma^2 \lambda_\sigma +  \frac{1}{2}\lambda_{\sigma h} \phi_h^2
\end{pmatrix}\,,
\end{equation}
with corresponding eigenvalues given by
\begin{equation}\label{eq:mass_values}
    \begin{aligned}
        M_{h_1}^2 = &\frac{1}{4} \Big(-\sqrt{2 \phi _h^2 \phi _{\sigma }^2 \left(6 \lambda _{\sigma h} \left(\lambda _h+\lambda _{\sigma }\right)-36 \lambda _h \lambda _{\sigma
   }+7 \lambda _{\sigma h}^2\right)+\phi _h^4 \left(\lambda _{\sigma h}-6 \lambda _h\right)^2+\phi _{\sigma }^4 \left(\lambda _{\sigma
   h}-6 \lambda _{\sigma }\right)^2} \\ 
   +&\phi _h^2 \left(6 \lambda _h+\lambda _{\sigma h}\right)+\lambda _{\sigma h} \phi _{\sigma }^2+6 \lambda_{\sigma } \phi _{\sigma }^2\Big) \,, \\
        M_{h_2}^2 = &\frac{1}{4} \Big(\sqrt{2 \phi _h^2 \phi _{\sigma }^2 \left(6 \lambda _{\sigma h} \left(\lambda _h+\lambda _{\sigma }\right)-36 \lambda _h \lambda _{\sigma
   }+7 \lambda _{\sigma h}^2\right)+\phi _h^4 \left(\lambda _{\sigma h}-6 \lambda _h\right){}^2+\phi _{\sigma }^4 \left(\lambda _{\sigma
   h}-6 \lambda _{\sigma }\right)^2} \\ 
   +&\phi _h^2 \left(6 \lambda _h+\lambda _{\sigma h}\right)+\lambda _{\sigma h} \phi _{\sigma }^2+6 \lambda
   _{\sigma } \phi _{\sigma }^2\Big)\,.
    \end{aligned}
\end{equation}
For the Goldstone bosons, the corresponding field-dependent masses are
\begin{equation}\label{eq:gold_masses}
\begin{aligned}
M_{G_1}^2 = \lambda_h \phi_h^2 + \frac{\lambda_{\sigma h}}{2}\phi^2_\sigma \,,\quad  M_{G_2}^2 = \lambda_\sigma \phi_\sigma^2 + \frac{\lambda_{\sigma h}}{2}\phi^2_h \,,\quad M_{G^\pm}^2 = \lambda_h \phi_h^2 + \frac{\lambda_{\sigma h}}{2}\phi^2_\sigma \,. \\
\end{aligned}
\end{equation}

The absence of quadratic terms at leading order results in a massless physical scalar, which will be identified with the $h_2$ boson, once the gauge and conformal symmetries are broken. As a pseudo-Goldstone boson of the scale symmetry, it acquires an explicit mass term due to the non-linear form of the CW potential \cite{Coleman:1973jx} that violates the scale symmetry of the Lagrangian. In the Landau gauge and assuming the $\overline{\mathrm{MS}}$ renormalization scheme, the CW potential takes the generic form,
\begin{equation}\label{eq:CW_potential}
    V_{\mathrm{CW}}(\phi_h, \phi_\sigma) = \frac{1}{64\pi^2} \sum_a n_a M_a^4(\phi_h, \phi_\sigma) \qty(\ln \frac{M_a^2(\phi_h, \phi_\sigma)}{\mu^2} - c_a), 
\end{equation}
where $a$ runs over all vector, scalar and fermionic degrees of freedom, while $M_a(\phi_h,\phi_\sigma)$ denotes the tree-level field-dependent mass for a field $a$, and $\mu$ is the renormalization scale. The $c_a$ factors are renormalization-dependent constants taking the values $c_a=3/2$ for fermions and scalars and $c_a=5/6$ for vectors in the $\overline{\mathrm{MS}}$ scheme. The pre-factor $n_a$ is given by
\begin{equation}\label{eq:pre-factor}
    n_a = (-1)^{2s_a} Q_a N_a (2s_a + 1)\,.
\end{equation}
Here, $s_a$ denotes the spin of particle $a$, $Q_a$ is 1 for uncharged particles and 2 for charged ones, whereas $N_a$ is 1 for uncolored particles and 3 for colored ones. 

The one-loop effective potential is 
\begin{equation}
V = V_0(\phi_h, \phi_\sigma) + V_{\mathrm{CW}}(\phi_h, \phi_\sigma) \,,
\end{equation}
which must be extremized in order to study the symmetry breaking patterns. In the zero temperature limit, the tadpole equations are
\begin{equation}\label{eq:tadpoles_1loop}
    \begin{aligned}
        & 0 = \lambda_h v^3 + \frac{1}{2}\lambda_{\sigma h} v v_\sigma^2 + \frac{\del V_{\mathrm{CW}}}{\del \phi_h}\Bigr|_{\substack{\phi_h=v, \phi_\sigma=v_\sigma}}\,, \\
        & 0 = \lambda_\sigma v_\sigma^3 + \frac{1}{2}\lambda_{\sigma h} v^2 v_\sigma + \frac{\del V_{\mathrm{CW}}}{\del \phi_\sigma}\Bigr|_{\substack{\phi_h=v, \phi_\sigma=v_\sigma}}\,. \\
    \end{aligned}
\end{equation}
 These equations can be used to fix the values of $\lambda_{\sigma h}$ and $\lambda_\sigma$ in our numerical analysis. The zero external momentum contribution to the scalar mass spectrum is derived by computing the eigenvalues of the Hessian matrix of the effective potential, while the momentum-dependent part is obtained from the self-energies $\Pi$. Separating the one-loop and tree-level parts, the mass matrix can  be written as~\cite{Costa:2014qga} 
\begin{equation}\label{eq:mass_matrix_1loop}
\bm{M^2}(p^2) = \bm{M^{2}_{(0)}} + \bm{\mathcal{M}}\,,
\end{equation}
with $\bm{\mathcal{M}} = \mathrm{Re}\{\Delta\bm{\Pi}(p^2)\} + \bm{\partial^2} V_\mathrm{CW}$, where $\bm{\partial^2} V_\mathrm{CW}$ is the Hessian matrix of the CW potential and $\Delta \bm{\Pi} = \bm{\Pi}(p^2 = M^2) - \bm{\Pi}(p^2 = 0)$. We then obtain the loop corrected masses,
\begin{equation}\label{eq:loop_corrected_masses}
    \begin{aligned}
        M_{h_1}^2 = \frac{1}{4} \Big(\Sigma - \Bigl\{32 \bm{\mathcal{M}}_{12} \phi _h \lambda _{\sigma h} \phi _{\sigma }+\phi _h^2 \left(6 \lambda _h-\lambda _{\sigma
   h}\right) \left(\phi _h^2 \left(6 \lambda _h-\lambda _{\sigma h}\right)+4 \left[ \bm{\mathcal{M}}_{11}-\bm{\mathcal{M}}_{22} \right] \right) \\
   +2 \phi _{\sigma }^2 \left(\Phi _h \phi _h^2 + \Phi _{\sigma} \right)+\phi _{\sigma }^4 \left(\lambda _{\sigma h}-6 \lambda _{\sigma }\right)^2+4 \left[(\bm{\mathcal{M}}_{11} - \bm{\mathcal{M}}_{22} )^2+4 \Pi^2_{h_1 h_2}\right] \Bigl\}^{1/2}\Big)\,, \\[1.0em]
        M_{h_2}^2 = \frac{1}{4} \Big(\Sigma + \Bigl\{32 \bm{\mathcal{M}}_{12}  \phi _h \lambda _{\sigma h} \phi _{\sigma }+\phi _h^2 \left(6 \lambda _h-\lambda _{\sigma
   h}\right) \left(\phi _h^2 \left(6 \lambda _h-\lambda _{\sigma h}\right)+ 4 \left[ \bm{\mathcal{M}}_{11}-\bm{\mathcal{M}}_{22} \right]  \right) \\
   +2 \phi _{\sigma }^2 \left(\Phi _h \phi _h^2+\Phi _{\sigma} \right)+\phi _{\sigma }^4 \left(\lambda _{\sigma h}-6 \lambda _{\sigma }\right)^2+4 \left[(\bm{\mathcal{M}}_{11} - \bm{\mathcal{M}}_{22} )^2+4 \Pi^2_{h_1 h_2}\right] \Bigl\}^{1/2}\Big)\,, \\[0.5em]
    \end{aligned}
\end{equation}
where 
\begin{equation}\label{eq:defs}
\begin{aligned}
&\Sigma = \phi _h^2 \left(6 \lambda _h+\lambda _{\sigma h}\right)+\lambda _{\sigma h} \phi _{\sigma }^2+6 \lambda _{\sigma } \phi _{\sigma }^2+2 \bm{\mathcal{M}}_{11} + 2 \bm{\mathcal{M}}_{22}  \,, \\
&\Phi_h = 6 \lambda _{\sigma h} \left(\lambda _h+\lambda _{\sigma }\right)-36 \lambda _h \lambda _{\sigma }+7 \lambda _{\sigma h}^2 \,, \\
&\Phi_\sigma = -2 (\bm{\mathcal{M}}_{11} - \bm{\mathcal{M}}_{22}) \left(6 \lambda _{\sigma }-\lambda _{\sigma h}\right) \,. \\
\end{aligned}
\end{equation}
Following symmetry breaking, we have a total of 7 parameters: three quartic couplings, two VEVs, and two physical scalar masses. Of them, the Higgs VEV and mass $M_{h_1}$ are fixed to their experimentally measured values, so we are left with $\lambda_h$, $\lambda_\sigma$, $\lambda_{\sigma h}$, $v_\sigma$, and $M_{h_2}$. The tadpole relations in \cref{eq:tadpoles_1loop} fix two of these parameters, which we choose to be $\lambda_\sigma$ and $\lambda_{\sigma h}$, while the one-loop corrected masses in \cref{eq:loop_corrected_masses} constrain two additional parameters, which we choose to be $v_\sigma$ and $\lambda_h$. Consequently, the only remaining free parameter is the mass of the additional Higgs, $M_{h_2}$.  The full set of one-loop diagrams and the corresponding expressions for the self-energies are provided in \cref{app:self_energy}. 

To simplify the numerical analysis and in the interest of  efficiency, we make the following approximations. First, we neglect mixing effects generated at one-loop, \textit{i.e.}~$\Pi_{h_1 h_2} = 0$, and instead utilize results from the tree level contribution. This approximation is not expected to significantly impact the final outcome, given that (a) experimental constraints already favor a relatively small Higgs mixing angle, and (b) we have found numerically that such an angle always remains small. Second, since the strength of phase transitions in classically conformal models with additional gauge groups is typically driven by gauge interactions, where the relation between scalar quartic $\lambda$ and gauge $g$ couplings is $\lambda \sim g^4$~\cite{Khoze:2014xha}, we anticipate that for $g \lesssim 1$, contributions from scalar fields to both one-loop masses and the effective potential are small. Indeed, we have numerically verified that for $\lambda_{\sigma} \sim \mathcal{O}(0.1)$, as obtained in our simulations, scalar corrections lead to changes in $\lambda_{\sigma}$ of about $6\%$ and in $\lambda_h$ of about $0.1\%$, with $v_{\sigma}$ and $\lambda_{\sigma h}$ unchanged. Consequently, we only include fermion and gauge boson contributions in the calculation of one-loop masses and the effective potential. This approximation is supported by arguments presented in \cite{Kierkla:2022odc}, particularly in the context of an $\SU{2}{}$ conformal model. The immediate advantage of these approximations is evident in the calculation of one-loop tadpole equations in \cref{eq:tadpoles_1loop}, which can be solved fully analytically (see \cref{sec:one_loop_min_masses} for the complete expressions). Furthermore, as the Higgs vacuum does not play a role in the FOPT, the relevant contributions at one loop are those from the $\mathrm{Z'}$ boson and the heavy neutrinos.

\subsection{Gauge sector}\label{sec:gauge_sector}

The presence of a new $\U{}^\prime$ gauge symmetry implies the existence of an additional heavy gauge boson $\mathrm{Z}^\prime$ that mixes with the SM photon and Z$^0$ boson. To illustrate this, consider the covariant derivatives which explicitly depend on the Higgs doublet and Majoron $\mathrm{U(1)^\prime}$ gauge charges:
\begin{equation}\label{eq:gauge_cov}
\begin{aligned}
    &D_\mu \mathcal{H} = \qty(\partial_\mu - \frac{i}{2} g_1 B_\mu - ig_2 \tau_aA^a_\mu - i x_\mathcal{H} g_L B^\prime_\mu - ig_{12} x_\mathcal{H} B_\mu - \frac{i}{2}g_{21} B^\prime_\mu  )\mathcal{H}\,,  \\
    &D_\mu \sigma = \left(\partial_\mu -ig_{21}x_{\sigma} B_\mu -ix_\sigma g_L B^\prime_\mu \right)\sigma\,.
\end{aligned}
\end{equation}
The usual $\U{Y}$ and $\SU{2}{L}$ gauge couplings are denoted as $g_1$ and $g_2$, respectively, while $g_L$ is the $\U{}^\prime$ gauge coupling. The $\U{Y}-\U{}^\prime$ kinetic mixing is generally described by two parameters, $g_{12}$ and $g_{21}$, although it is possible to rotate to a basis in which $g_{21} = 0$~\cite{delAguila:1988jz}. The $\U{Y}$, $\U{}^\prime$ and $\SU{2}{L}$ gauge fields are denoted as $B_\mu$, $B_\mu^\prime$ and $A_\mu^a$, respectively. By expanding the kinetic terms $\abs{D_\mu \mathcal{H}}^2$ and $\abs{D_\mu \sigma}^2$, one obtains the field-dependent mass matrix, which in the basis $\{A^1_\mu, A^2_\mu, B_\mu, A^3_\mu, B^\prime_\mu\} \otimes \{A^1_\mu, A^2_\mu, B_\mu, A^3_\mu, B^\prime_\mu\}$ is 
\begin{equation}
    \bm{M^2_V} = \begin{pmatrix}
          \dfrac{g_2^2 \phi^2_h}{4} & 0 & 0 & 0 & 0 \\
          0 & \dfrac{g_2^2 \phi^2_h}{4} & 0 & 0 & 0 \\
          0 & 0 & \dfrac{g_1^2 \phi_h^2}{4} & -\dfrac{1}{4}g_1 g_2 \phi_h^2 & \dfrac{1}{4} g_1 \qty(g_{12} + 2x_\mathcal{H}g_L)\phi_h^2 \\
          0 & 0 & -\dfrac{1}{4}g_1 g_2 \phi_h^2 & \dfrac{g_2^2 \phi_h^2}{4} & -\dfrac{1}{4}g_2 \qty(g_{12} + 2x_\mathcal{H}g_L)\phi_h^2 \\
          0 & 0 & \dfrac{1}{4}g_1\qty(g_{12} + 2g_L x_\mathcal{H})\phi_h^2 & -\dfrac{1}{4}g_2\qty(g_{12} + 2 x_\mathcal{H} g_L)\phi_h^2 & \dfrac{1}{4}\qty(g_{12} + 2x_\mathcal{H}g_L)^2\phi_h^2 + g_L^2 x_\sigma^2 \phi_\sigma^2
          \end{pmatrix}\,,
\end{equation}
and whose eigenvalues provide the vector bosons' field-dependent masses,
\begin{equation}\label{eq:masses_Z0Zp}
    \begin{aligned}
        &M^2_{\gamma} = 0 \,,\\
        &M^2_{W^\pm} = \frac{g_2^2 \phi_h^2}{4} \,,\\
        &M^2_{\mathrm{Z^0}} = \frac{1}{8} (G \phi_h^2+4 g_L^2 x_\sigma^2 \phi_\sigma ^2) \\
        &\hspace*{3em}- \sqrt{G^2 \phi_h^4+8 g_L^2 x_\sigma^2 \phi_h^2 \phi_\sigma^2 \left[(2 g_L x_\mathcal{H}+g_{12})^2-g_1^2-g_2^2\right]+16 g_L^4 x_\sigma^4 \phi_\sigma ^4} \,,\\
        &M^2_{\mathrm{Z^\prime}} = \frac{1}{8} (G \phi_h^2+4 g_L^2 x_\sigma^2 \phi_\sigma ^2)\\
        &\hspace*{3em}+ \sqrt{G^2 \phi_h^4+8 g_L^2 x_\sigma^2 \phi_h^2 \phi_\sigma^2 \left[(2 g_L x_\mathcal{H}+g_{12})^2-g_1^2-g_2^2\right]+16 g_L^4 x_\sigma^4 \phi_\sigma^4}\,,
    \end{aligned}
\end{equation}
where we define $G \equiv g_1^2 + g_2^2 + (g_{12}^2 + 2g_L x_\mathcal{H})^2$. 
Due to the flavor universality of the $\U{}^\prime$ charges in \cref{tab:charges}, the new $\mathrm{Z}^\prime$ boson couples to all SM fermions. Consequently, direct searches for $\mathrm{Z}^\prime$ bosons at the LHC impose stringent constraints on its mass, restricting it to be slightly above 
5~TeV~\cite{ATLAS:2019erb,ATLAS:2017eiz,ATLAS:2018tfk,ATLAS:2020lks}. Additionally, constraints from the Large Electron-Positron collider (LEP) \cite{ALEPH:2005ab} impose strict limitations on the allowed values of the kinetic mixing, parameterized here via $g_{12}$. Note that a heavy $\mathrm{Z'}$ boson implies a hierarchical relation among the VEVs, $v_\sigma \gg v$, from which we can derive approximate expressions for the masses of the $\mathrm{Z^0}$ and $\mathrm{Z'}$ as follows:
\begin{equation}\label{eq:Mzprime_Mz_masses}
M^2_{\mathrm{Z^0}} = \frac{\phi_h^2}{16}(g_1^2 + g_2^2)\qty(4 - \frac{(g_{12} + 2g_L x_\mathcal{H})\phi_h^2}{g_L^2 x_\sigma^2 \phi_\sigma^2} )\,, \quad
M^2_{\mathrm{Z^\prime}} = \frac{1}{4}\qty(g_{12} + 2g_L x_\mathcal{H})^2\phi_h^2 + g_L^2 x_\sigma^2 \phi_\sigma^2\,.
\end{equation}
While the kinetic mixing is expected to be small, the large value of $v_\sigma$ suppresses the $\mathrm{Z^0}-\mathrm{Z^\prime}$ mixing angle (see e.g. \cite{Coriano:2015sea}), proportional to $v^2/v_\sigma^2$, which in turn allows $g_{12}$ to be $\mathcal{O}(1)$. Indeed, previous studies \cite{Ellis:2020nnr, Marzo:2018nov} have shown that setting $g_{12}(\mu = M_{\mathrm{Z'}}) = -0.5$ contributes to the stabilization of the Higgs vacuum by ensuring that $\lambda_h$ remains positive up to the Planck scale. Although this conclusion was drawn within the context of the B$-$L model (equivalent to setting $x_\mathcal{H} = 0$ and $x_\sigma = 2$), we have observed that for other charge assignments, a nonzero value to $g_{12}$ at the $M_\mathrm{Z'}$ scale also aids in stabilizing the SM vacuum. For a cleaner analysis, we fix $g_{12} = 0$ at the electroweak scale, such that the only free parameters from the gauge sector are the charges $x_\mathcal{H}$ and $x_\sigma$ and the gauge coupling $g_L$. However, RG evolution regenerates a nonzero value of $g_{12}$  at the $M_{\mathrm{Z'}}$ scale so that the potential is bounded from below provided that $41 x_\mathcal{H} + 8 x_\sigma \neq 0$ (see $\beta^{(1)}(g_{12})$ in \cref{app:rges}).

\section{Gravitational waves from supercooled FOPTs}\label{sec:FOPT_GWs}

The dynamics of phase transitions is well-established, with its theoretical foundations laid out in past works \cite{Linde:1981zj,Coleman:1977py} (for a recent review see Ref.~\cite{Athron:2023xlk}). Here, we briefly outline the calculation of the GW spectrum from supercooled FOPTs in $\mathrm{U(1)}^\prime$ conformal Majoron models.

In a thermal bath, phase transitions are primarily driven by thermal fluctuations, with the decay rate given by
\begin{equation}\label{eq:decay_rateT}
\Gamma = A e^{-S_3/T}\,,
\end{equation}
where the prefactor $A$ is typically approximated as $A \sim T^4 \qty(S_3/2\pi T)^{3/2}$ in terms of the Euclidean action of the 3D theory \cite{Linde:1981zj,Coleman:1977py}: 
\begin{equation}\label{eq:euclidean_action}
S_3(T) = 4\pi \int_0^\infty dr~r^2 \qty[\frac{1}{2} \qty(\frac{d\phi_\sigma}{dr})^2 + V_\mathrm{eff}(\phi_\sigma,T)]\,.
\end{equation}
Here, $V_\mathrm{eff}$ is the thermal effective potential, and the bounce solution $\phi_\sigma$, is determined by solving the equation of motion, 
\begin{equation}
\frac{\partial^2 \phi_\sigma}{\partial r^2} + \frac{2}{r} \frac{\partial \phi_\sigma}{\partial r}  = \frac{\partial V_\mathrm{eff}}{\partial \phi_\sigma}\,,
\end{equation}
for the path that minimizes the energy of the scalar field.
We employ \texttt{CosmoTransitions} \cite{Wainwright:2011kj} as a bounce solver in our numerical analysis. We validate the results against our own algorithm.

\subsubsection*{Percolation and reheating temperatures}

As the Universe cools down from the symmetric phase, vacuum bubbles of the broken phase form. The critical temperature, $T_c$, is defined as the temperature at which the true and false vacuum are degenerate. Below $T_c$, thermal fluctuations can become significant enough to nucleate a true vacuum bubble per cosmological horizon. This defines the nucleation temperature, $T_n$, given by
\begin{equation}\label{eq:nucle}
    \int_{T_n}^{T_c} \frac{dT}{T} \frac{\Gamma(T)}{ H(T)^4} = 1\,.
\end{equation}
Here, $H(T)$ is the Hubble parameter which evolves as
\begin{equation}\label{eq:hubble_rate}
    H^2(T) = \frac{1}{3 \overline{M}^2_{\mathrm{Pl}}}\qty( \rho_R(T)+ \Delta V(T))\,,
\end{equation} 
where $\overline{M}_{\mathrm{Pl}} \approx 2.4 \times 10^{18}~\mathrm{GeV}$ is the reduced Planck mass, and 
$\Delta V(T)$ is the potential energy difference between the true and false vacuum at temperature $T$, \textit{i.e.}, $\Delta V(T) = V(T, 0) - V(T, v_\mathrm{True}(T))$, where $v_\mathrm{True}$ is the VEV of the true vacuum.
The radiation energy density is $\rho_R(T) = g_*(T)(\pi^2/30) T^4$, where $g_*(T)$ is the total number of SM and 
dark sector relativistic degrees of freedom (dof), which includes three right-handed neutrinos, a massive $\mathrm{Z^\prime}$ boson, and a massive scalar $h_2$. It is common to assume that $g_*$ is constant, given that for temperatures above $100~\mathrm{GeV}$, $g_*(T) \approx 100$. However, for temperatures just above the QCD scale, the number of dof drops by about 40~\cite{Husdal:2016haj}. 
This has a significant impact in conformal scenarios with phase transition temperatures close to the QCD scale. Additionally, while in the supercooled case, $\Delta V(T)$ provides the dominant contribution to $H(T)$, making the temperature dependence of $g_*(T)$ unimportant, for non-supercooled scenarios $\rho_R$ becomes comparable to 
$\Delta V(T)$, necessitating a proper accounting of the dof.  

We assume a dark sector above the electroweak scale that fully thermalizes with the SM sector. Then, constraints on the extra effective number of neutrino species, $\Delta N_\mathrm{eff}$, from BBN and cosmic microwave background data are easily satisfied. 
Right-handed neutrinos can thermalize with the SM through the Yukawa interactions, $y_\nu^{ij} \overline{N}_{i} \mathcal{H} L_j$, and reach thermal equilibrium at a temperature~\cite{diBari:2021dri,DiBari:2019zcc}
\begin{equation}\label{eq:thermal_eq_temp}
    T^\mathrm{eq}_i \approx 0.2 \frac{(\bm{y_\nu} \bm{y_\nu}^\dagger)_{ii} v^2}{m_\mathrm{eq}} = 0.2 M_{N_i} K_i\,,
\end{equation}
where the effective equilibrium neutrino mass, $m_\mathrm{eq} \approx 1.1~\mathrm{meV} \sqrt{g_*/g_*^\mathrm{SM}}$, with $g_*^\mathrm{SM}$ the relativistic dof in the SM sector. Both $g_*$ and $g_*^\mathrm{SM}$ are evaluated at the temperature of the phase transition. Here, $M_{N_i}$ are the masses of the three physical right-handed neutrinos, and their decay parameters are
\begin{equation}\label{eq:thermaleq}
    K_i = \frac{(\bm{y_\nu} \bm{y_\nu}^\dagger)_{ii} v^2}{M_{N_i} m_\mathrm{eq}}\,.
\end{equation}
Neutrinos reach thermal equilibrium before the onset of the phase transition if $M_{N_i} K_i \gtrsim 5 T_c$~\cite{diBari:2021dri}. Since the right-handed neutrinos couple to the Majoron $\sigma$ via the Yukawa interactions $y_\sigma^{ij} \Bar{N}_{i}^c N_{j}\sigma$, the Majoron $\sigma$ also thermalizes if $y_\sigma$ is sufficiently large. We also expect $\sigma$ to thermalize via its interactions with a thermalized 
$\mathrm{Z^\prime}$ through the gauge coupling $g_L$ which is $\mathcal{O}(0.1)$.\footnote{The portal coupling $\lambda_{\sigma h} \sim -v^2/v^2_\sigma$ also plays a role, though it becomes increasingly subdominant for larger Majoron masses; see \cref{sec:one_loop_min_masses}.} Thermalization of the $\mathrm{Z^\prime}$ occurs through direct processes like $\bar{f}_{\mathrm{SM}} + f_{\mathrm{SM}} \leftrightarrow \mathrm{Z^\prime}$, where $f_{\mathrm{SM}}$ are SM fermions, or indirectly by first thermalizing with the right-handed neutrinos through $\bar{N}_{i} + N_{i} \leftrightarrow \mathrm{Z^\prime}$, which is also mediated by the gauge coupling. 

As the vacuum bubbles expand and occupy $34\%$ of the Universe's volume, they become causally connected, preventing the Universe from reverting to its initial symmetric phase. This defines the percolation temperature, $T_p$, and corresponds to the cosmological epoch at which the SGWB is generated. Quantitatively, the fraction of space in the false vacuum is $\mathcal{P}(T) = e^{-I(T)}$, where $I(T)$ corresponds to the true vacuum volume per unit comoving volume:
\begin{equation}\label{eq:It_percolation}
    I(T) = \frac{4\pi v^3_w}{3} \int_T^{T_c} dT^\prime \frac{\Gamma(T^\prime)}{T^{\prime 4} H(T^\prime)} \qty(\int_T^{T^\prime} \frac{d\tilde{T}}{H(\tilde{T})})^3\,,
\end{equation}
where $v_w=1$ is the bubble wall velocity.
The percolation temperature is then calculated from \cref{eq:It_percolation} by requiring that $I(T_p) = 0.34$ is satisfied, or equivalently, $\mathcal{P}(T_p) = 0.7$. 
To confirm that percolation indeed takes place, we ensure that the false vacuum volume is decreasing near $T_p$ by requiring
\begin{equation}\label{eq:perc_condition}
H(T)\left(3 + T\frac{dI}{dT}\right)\Biggr|_{\substack{T = T_p}} < 0\,.
\end{equation}
 Note that this condition may become valid at a temperature below percolation. In fact, we find a number of scenarios in which this condition is not valid at $T_p$ but is satisfied at a lower temperature. Then, it is unclear whether percolation is guaranteed~\cite{Ellis:2018mja}.

As the FOPT takes place, the energy released to the surrounding plasma reheats the Universe back to a higher temperature $T_\mathrm{RH}$. This is particularly relevant in the case of supercooling due to the substantial amount of latent heat released. Consequently, immediately after percolation, the heavy physical scalar field $h_2$ will begin to oscillate around the true vacuum and eventually decay away. If its decay rate $\Gamma_{h_2} > H(T_p)$, then reheating is almost instantaneous, and the Universe immediately enters a period of radiation domination. However, if $\Gamma_{h_2} < H(T_p)$, an interim period of matter domination occurs until the heavy scalar has decayed away~\cite{Ellis:2019oqb}.  With this in mind, the reheating temperature can be written as\footnote{A more accurate estimate of $T_\mathrm{RH}$ can be obtained from energy conservation by matching the energy density before and after the transition: $\rho(\phi(\phi_\mathrm{False}(T_p), T_p) = \rho(\phi(\phi_\mathrm{True}(T_\mathrm{RH}), T_\mathrm{RH})$ \cite{Athron:2023mer}.}
\begin{equation}\label{eq:reheat}
    \begin{aligned}
        T_\mathrm{RH} &\approx  \qty(\frac{\Gamma_{h_2}}{H(T_p)})^{1/2} T_p [1 + \alpha(T_p)]^{1/4}\,, \ \ \ \ \Gamma_{h_2} < H(T_p)\,, \\
                &\approx  T_p [1 + \alpha(T_p)]^{1/4}\,, \ \ \ \ \ \ \ \ \  \ \ \ \ \ \ \ \ \ \ \, \ \ \Gamma_{h_2} > H(T_p)\,.
    \end{aligned}
\end{equation}
Only after reheating does the Universe enter a period of radiation domination. In this case, the temperature at which the phase transition ends  should be taken to be $T_\mathrm{RH}$. However, the remaining {\it thermodynamic parameters}, like $\alpha$ and $\beta/H(T_p)$ (discussed below), are evaluated at $T_p$ \cite{Athron:2023xlk}. In the absence of supercooling, \textit{i.e.,}~$\alpha \ll 1$, one can approximate $T_\mathrm{RH} \approx T_p$ as long as the Universe immediately enters the radiation dominated era.

\subsubsection*{Strength of the phase transition $\bm{\alpha}$}

The strength of the phase transition, $\alpha$, is defined as the ratio of the latent heat released during the phase transition to the total radiation energy density. It can be expressed in terms of $\Delta V(T)$ as follows:
\begin{equation}\label{eq:alpha_param}
\alpha = \frac{\Delta V}{\rho_R}\Biggr|_{\substack{T = T_p}} - \frac{T}{\rho_R} \frac{\partial \Delta V}{\partial T} \Biggr|_{\substack{T = T_p}}\,,
\end{equation}
where the second term on the right-hand side encodes entropy density variation. In the case of supercooling, $\Delta V$ dominates the radiation energy density, \textit{i.e.},~$\Delta V \gg \rho_R$, leading to $\alpha \gg 1$.

\subsubsection*{Inverse time duration $\bm{\beta/H(T_p)}$}

The duration of the phase transition can be calculated using the false vacuum decay rate expressed as a function of time, $\Gamma(\tau) \sim e^{\beta \tau}$. By comparing this with \cref{eq:decay_rateT}, we obtain $\beta = -\big(\tfrac{d}{d\tau} \tfrac{S_3}{T}\big)|_{\tau_p}$. Using $\tfrac{dT}{d\tau} = -T H$, 
\begin{equation} \label{eq:beta}   
\frac{\beta}{H(T_p)} = T_p\frac{d (S_3/T)}{dT} \Biggr|_{\substack{T = T_p}}\,.
\end{equation}
This quantity can also be expressed in terms of the characteristic length scale $R_*$ corresponding to the average size of the bubble~\cite{Caprini:2019egz},
\begin{equation}
\frac{\beta}{H(T_p)} = (8\pi)^{1/3}{\frac{\mathrm{max}(v_w,c_s)}{H(T_p)R_*}}\,,
\end{equation}
where $c_s = 1/\sqrt{3}$ is the sound speed in the plasma and~\cite{Turner:1992tz,Enqvist:1991xw} 
\begin{equation}\label{eq:radius_Tp}
    R_* = \qty[T^3_p \int^{T_c}_{T_p} \frac{dT^\prime}{T^{\prime 4}} \frac{\Gamma(T^\prime)}{H(T^\prime)} e^{-I(T^\prime)}]^{-1/3} \,.
\end{equation}
The templates describing the SGWB spectrum are expressed in terms of $R_*$. 

\subsubsection*{Spectral templates for the SGWB}

We use the latest templates for the stochastic GW background spectrum characterized by the amplitude 
$\Omega_\GW$ and frequency $f$ as provided by the LISA Cosmology Working Group~\cite{Caprini:2024hue}. The SGWB gets contributions from three main sources: sound waves~\cite{Hindmarsh:2013xza, Hindmarsh:2015qta, Hindmarsh:2017gnf}, bubble wall collisions~\cite{Kosowsky:1991ua, Kosowsky:1992vn, Kamionkowski:1993fg}, and magneto-hydrodynamics turbulence in the plasma~\cite{Kamionkowski:1993fg, Caprini:2009yp}. Sound waves typically provide a dominant contribution. However, in the presence of supercooling the bubbles undergo unbounded expansion, making wall collisions efficient in producing gravitational radiation. Turbulence effects are still not well understood and remain largely uncertain compared to the other sources. Since the contribution from turbulence is expected to be subdominant, we neglect it in what follows.

For bubble collisions and highly relativistic fluid shells, the spectrum admits a broken power law that can be expressed as \cite{Caprini:2024hue}
\begin{equation}
	\Omega_\GW^{\rm BC}(f, \Omega_\mathrm{GW}^\mathrm{peak}, f_\mathrm{peak}) = \Omega_\mathrm{GW}^\mathrm{peak} \frac{\left(n_1 - n_2\right)^{\frac{n_1 - n_2}{a_1}}}{\left[-n_2 \left(\frac{f}{f_\mathrm{peak}}\right)^{-\frac{n_1 a_1}{n_1 - n_2}} + n_1 \left(\frac{f}{f_\mathrm{peak}}\right)^{-\frac{n_2 a_1}{n_1 - n_2}}\right]^{\frac{n_1 - n_2}{a_1}}} \,,
 \label{eq:BPLalt}
\end{equation}
where $\Omega_\mathrm{GW}^\mathrm{peak}$ and $f_\mathrm{peak}$, which we call {\it geometric parameters},  correspond to the peak energy density amplitude and frequency, respectively. The $n_i$ and $a_i$ parameters result from a fit to numerical simulations and are given as $n_1 = 2.4$, $n_2 = -2.4$ and $a_1 = 1.2$ \cite{Caprini:2024hue}. We can relate the geometric and thermodynamics parameters as follows
\begin{equation}\label{eq:amp_and_freq_strongPT}
\begin{aligned}
h^2 \Omega_\mathrm{GW}^\mathrm{peak}  &= h^2 F_{\GW,0} \,A_\text{str} \,\tilde{K}^2\,\left(\frac{\beta}{H(T_p)}\right)^{-2} \,, 
\quad
f_\mathrm{peak} &\simeq 0.11\,H_{*,0}\, \frac{\beta}{H(T_p)}\,,
\end{aligned}
\end{equation}
where $\tilde{K} \equiv \kappa_{\mathrm{BC}}[\alpha/(1 + \alpha)]$ is the fractional energy density and $A_{\mathrm{str}} \simeq 0.05$ \cite{Lewicki:2022pdb}. The parameters $F_{\GW,0}$ and $H_{*,0}$ account for the redshift as follows:
\begin{equation}\label{eq:redshit_H_FGW0}
    \begin{aligned}
        &H_{*,0} \simeq 1.65\times 10^{-5}~\mathrm{Hz}\,\left(\frac{g_*}{100}\right)^{1/6} \left(\frac{T_\mathrm{RH}}{\mathrm{GeV}}\right) \qty(\frac{\Gamma_{h_2}}{H(T_p)})^{-1/3} \,, \\
        &h^2 \, F_{\GW,0}  \simeq 1.65\times 10^{-5} \left(\frac{100}{g_{*}}\right)^{1/3} \qty(\frac{\Gamma_{h_2}}{H(T_p)})^{2/3}\,,
    \end{aligned}
\end{equation}
with $H_0 = 100h~\mathrm{km/s/Mpc}$. The $\Gamma_{h_2}/{H(T_p)}$ factors are taken from Ref.~\cite{Ellis:2019oqb}. In the case of radiation domination at percolation, $\Gamma_{h_2}/H(T_p)=1$. 

The efficiency factor $\kappa_{\mathrm{BC}}$ is model dependent. Whether the bubble wall reaches a terminal velocity depends on the pressure exerted by the plasma on the walls. Two contributions apply: a leading-order (LO) contribution due to $1\rightarrow 1$ scattering~\cite{Bodeker:2009qy} and a next-to-leading order (NLO) one from $1 \rightarrow N$ splittings~\cite{Bodeker:2017cim}: 
\begin{equation}\label{eq:pressures_LO_NLO}
    P_\mathrm{LO} = \sum_{a=\mathrm{Z}^\prime, N_i} k_a c_a \frac{\Delta m^2_a T^2_p}{24} \quad \mathrm{and} \quad
    P_\mathrm{NLO} = k_{\mathrm{Z^\prime}} g^2_L \Delta m_\mathrm{Z^\prime} T^3_p \,,
\end{equation}
where $c_a = 1~(1/2)$ for bosons (fermions), $k_a$ denote the corresponding degrees of freedom, $\Delta m_a^2$ is the squared mass difference of the particles in the false and true vacuum, $\Delta m_\mathrm{Z^\prime}$ is the mass difference of the $\mathrm{Z^\prime}$ boson  in the two vacua, and $g_L$ is the U(1)$'$ gauge coupling. Defining~\cite{Ellis:2019oqb}
\begin{equation}
    \alpha_\mathrm{eq} = \frac{P_\mathrm{NLO}}{\rho_R} \quad \mathrm{and} \quad \alpha_{\infty} = \frac{P_\mathrm{LO}}{\rho_R}\,,
\end{equation}
the equilibrium Lorentz factor (corresponding to the pressure terms $P_\mathrm{LO}$ and $P_\mathrm{NLO}$ being balanced by the potential energy difference 
$\Delta V$) is
\begin{equation}
    \gamma_{\mathrm{eq}} = \frac{\alpha - \alpha_{\infty}}{\alpha_\mathrm{eq}}\,.
\end{equation}
Neglecting plasma effects {\it{i.e.}}, $P_{\mathrm{NLO}}$, as the bubble grows, its Lorentz factor can be approximated as~\cite{Ellis:2019oqb}
\begin{equation}
    \tilde{\gamma}_* \approx \frac{2}{3}\frac{R_*}{R_0}\,,
\end{equation}
where $R_0$ is the initial bubble radius,
\begin{equation}
    R_0 = \qty(\frac{3 S_3(T_p)}{4\pi\Delta V(T_p)})^{1/3}\,.
\end{equation}
The efficiency factor can be estimated as the ratio of the energy of the bubble wall to the total energy released~\cite{Ellis:2019oqb}:
\begin{equation}
    \kappa_{\mathrm{BC}} = \begin{cases}
        & \qty(1 - \dfrac{1}{3}\qty(\dfrac{\tilde{\gamma}_*}{\gamma_{\mathrm{eq}}})^2)\qty(1- \dfrac{\alpha_{\infty}}{\alpha}) \quad \mathrm{for} \quad  \tilde{\gamma}_* < \gamma_{\mathrm{eq}} \,, \\[1.5em]
        &\dfrac{2}{3} \dfrac{\gamma_{\mathrm{eq}}}{\tilde{\gamma}_*}\qty(1 - \dfrac{\alpha_\infty}{\alpha}) \qquad \mathrm{otherwise}.
    \end{cases}
\end{equation}

For the sound wave contribution, the SGWB template is described by the following double broken power law \cite{Caprini:2024hue}
\begin{align}
&\Omega^{\rm SW}_{\GW}(f, \Omega_2, f_1, f_2
) =
\Omega_\text{int} \times S(f) \,, \label{eq:DBPL}
\\
S(f) &= N \left( \frac{f}{f_1} \right)^{n_1}
\left[
1 + \left( \frac{f}{f_1} \right)^{a_1}
\right]^{\frac{- n_1 + n_2}{a_1}}
\left[
1 + \left( \frac{f}{f_2} \right)^{a_2}
\right]^{\frac{- n_2 + n_3}{a_2}} \,, \nonumber
\end{align}
where the fit parameters are $n_1 = 3$, $n_2 = 1$, $n_3 = -3$, $a_1 = 2$ and $a_2 = 4$. Here, $N$ is a normalization factor that is determined by $\int^{+\infty}_{-\infty} S(f) d\ln f = 1$. The geometric parameters $f_1$ and $f_2$ are given by 
\begin{align}\label{eq:geom_soundwave}
& f_1 \simeq 0.2 \, H_{*,0} \, (H(T_p) R_*)^{-1} \,, \\
& f_2 \simeq 0.5 \, H_{*,0} \, \Delta_w^{-1}~(H(T_p) R_*)^{-1} \,,
\label{eq:sw_shape}
\end{align}
where $\Delta_w = v_\mathrm{shell}/\mathrm{max}(v_w,c_s)$ with $v_\mathrm{shell} = |v_w - c_s|$ the dimensionless sound shell thickness. A definition for $H(T_p) R_*$ was given below Eq.~\eqref{eq:radius_Tp}. The integrated amplitude $\Omega_{\rm int}$ obeys the relation~\cite{Jinno:2022mie},
\begin{align}
h^2 \Omega_\mathrm{int}
&=
0.11 h^2 F_{\GW,0} \, K^2 \left( H(T_p) \tau_{\rm SW} \right) \left( H(T_p)R_* \right) \,,
\label{eq:sw_amplitude}
\end{align}
where the lifetime of sound waves in units of Hubble time is 
 $H(T_p) \tau_{\mathrm{SW}} = \mathrm{min} (2H(T_p) R_*/\sqrt{3K}, 1)$, and $K = 0.6\kappa_{\mathrm{SW}}\alpha/(1+\alpha)$ is the fractional kinetic energy converted into sound waves. The efficiency factor is~\cite{Ellis:2020nnr}
\begin{equation}\label{eq:efficiency_SW}
    \kappa_{\mathrm{SW}} = \frac{\alpha_{\mathrm{eff}}}{\alpha} \frac{\alpha_\mathrm{eff}}{0.73 + 0.083 \sqrt{\alpha_\mathrm{eff}} + \alpha_\mathrm{eff}}\,, \quad \alpha_\mathrm{eff} = \alpha(1 - \kappa_{\mathrm{BC}}) \,.
\end{equation}

\section{Effective thermal potential}\label{sec:effective_pot}

It is frequently asserted that small theoretical inaccuracies in the thermodynamic parameters of phase transitions can result in significant variations, spanning several orders of magnitude, in the predicted SGWB. We assess the generality of this statement and argue that it depends on the nature of the FOPT. The peak amplitudes for the sound wave and bubble collision contributions, 
$\Omega_\mathrm{SW}^\mathrm{peak}$ and $\Omega_\mathrm{BC}^\mathrm{peak}$ respectively, and their peak frequency, $f_\mathrm{peak}$, scale with the phase transition parameters according to
\begin{equation}\label{Omegaf_approx}
\Omega_\mathrm{SW}^\mathrm{peak} \propto \left(\frac{\kappa_\mathrm{SW} \alpha}{1+\alpha}\right)^2 \left( \frac{\beta}{H(T_p)} \right)^{-1}\,, \quad \Omega_\mathrm{BC}^\mathrm{peak} \propto \left(\frac{\kappa_\mathrm{BC} \alpha}{1+\alpha}\right)^2 \left( \frac{\beta}{H(T_p)} \right)^{-2}\,, \quad f_\mathrm{peak} \propto \frac{\beta}{H(T_p)}\,.
\end{equation}
 The impact of uncertainties in the efficiency factors is illustrated in \cref{fig:GW_uncer}. 
 In the limit of strong supercooling, $\alpha \gg 1$, 
 according to Fig.~6 of Ref.~\cite{Ellis:2020awk},
\begin{equation}\label{eq:approx_beta}
     \dfrac{\beta}{H(T_p)} \approx \text{constant}\,.
\end{equation}
  For simplicity, taking $\kappa_\mathrm{BC} = 1$ (for bubble collisions) and $\kappa_\mathrm{BC} = 0$ (for sound waves), and using \cref{eq:efficiency_SW}, we estimate
\begin{equation}\label{eq:approx_Om}
    \Omega_\mathrm{SW,BC}^\mathrm{peak} \approx  \text{constant}\,.   
\end{equation}
Likewise, for the peak frequency, 
\begin{equation}\label{eq:approx_fpeak}
    f_\mathrm{peak} \approx \text{constant}\,.
\end{equation}
Although a different choice of the renormalization scale implies a change in $T_p$,
for strong supercooling, \cref{eq:approx_Om,eq:approx_fpeak} suggest that neither 
$\Omega_\mathrm{SW,BC}^\mathrm{peak}$ nor $f_\mathrm{peak}$ are expected to be significantly altered.
In contrast, for $\alpha < 1$, a small change in $T_p$ and $\Delta V(T_p)$ can be amplified at least by the second power of $\alpha$ in \cref{Omegaf_approx}, $i.e.,$ $(\Delta V)^2 /T_p^8$, which mostly impacts the peak amplitude.

Various methods have been proposed to mitigate theoretical uncertainties. These include constructing the RG-improved potential, where each coupling and field are evolved by means of their RG equations~\cite{Kierkla:2022odc, Chataignier:2018kay, Chataignier:2018aud, Ellis:2020nnr}, which is particularly relevant for supercooling. Another method involves dimensional reduction of the original 4D theory into a 3D effective field theory (EFT)~\cite{Gould:2022ran, Kainulainen:2019kyp, Schicho:2021gca, Niemi:2021qvp, Schicho:2022wty, Croon:2020cgk, Hirvonen:2021zej}. However, based 
on~\cref{eq:approx_beta,eq:approx_Om,eq:approx_fpeak}, we argue that the advantage of using dimensional reduction for the study of classically scale-invariant models is questionable. Furthermore, the supercooling effect for low temperatures invalidates the high-temperature approximation for most field values.  Recent work~\cite{Kierkla:2023von} has demonstrated that the 3D EFT approach is valid only for small field values. At NNLO precision, the Euclidean action \cref{eq:euclidean_action} is corrected with an additional factor $Z(\phi)$ in the kinetic term as $S = 4\pi \int dr r^2 Z(\phi) (\partial \phi)^2 + V_{\mathrm{eff}}$. In general, $Z(\phi)$ scales as $1/\phi$ as $\phi\to 0$ which diverges in the symmetric vacuum, implying a breakdown of the derivative expansion. Since Ref.~\cite{Kierkla:2023von} has numerically verified that this correction is responsible for the observed differences between the 3D and 4D approaches, it remains unclear to us whether such an effect is physical or merely a consequence of operating in a regime where the derivative expansion may not be valid.  Therefore, we subscribe to the 4D RG-improved effective potential.

\subsection{Renormalization-group improved thermal potential: a 4D effective theory}\label{subsec:RG_improvement}

The RG-improved effective potential at zero temperature for classically scale invariant models can be formulated as~\cite{Meissner:2008uw} 
\begin{equation}
    V_\mathrm{eff}(\phi_\sigma, \lambda, t) = \lambda(t)\phi_\sigma^4 \exp{\frac{1}{4}\int_0^t dt \gamma[\lambda(t)]}\,,
\end{equation}
where $\lambda$ denotes a set of couplings, $\gamma$ is the anomalous dimension, and $t = \ln(\mu/\mu_{\mathrm{ref}})$, with $\mu$ being the RG scale and $\mu_{\mathrm{ref}}$ -- a reference scale. In our numerical analysis, we set $\mu_{\mathrm{ref}}$ equal to the mass of the $\mathrm{Z^0}$ boson. The choice of the reference scale is arbitrary; however, we have verified that a factor of two variation in $\mu_\mathrm{ref}$ results in a deviation of only 0.1\% in $V_\mathrm{eff}$. In practice, RG improvement entails rescaling the couplings and fields according to the following transformations:
\begin{equation}\label{eq:rescale}
    \begin{aligned}
        & \lambda \rightarrow \lambda(t)\,, \\
        & \phi_\sigma^2  \rightarrow \frac{\phi_\sigma^2}{2} \exp{\int_0^t dt^\prime \gamma[\lambda(t^\prime)]}\,,
    \end{aligned}
\end{equation}
which are applied to the tree-level, one-loop and thermal potentials. The beta functions, provided in \cref{app:rges}, determine the evolution of the couplings, while the $\gamma$ functions, provided in \cref{app:anom_dim}, control the field's evolution. For simplicity, we omit the explicit $t$ dependence of the fields and couplings throughout, unless necessary.

The choice of the renormalization scale must take into account its field-dependent nature and an additional scale introduced by the temperature: 
\begin{equation}\label{eq:renorm_scale}
\mu = \mathrm{max}[M_{\mathrm{Z^\prime}}(\phi_\sigma), \kappa T] \,,
\end{equation}
where $M_\mathrm{Z^\prime}(\phi_\sigma)$ is the field-dependent mass of the $\mathrm{Z^\prime}$ boson in \cref{eq:masses_Z0Zp}, with its couplings evaluated at this mass, and $\kappa$ is an arbitrary factor. Setting $\kappa = 4 \pi e^{-\gamma_E}$ ensures an exact cancellation between the logarithmic terms of the CW potential and the high-temperature potential $V_T$. However, any $\kappa$ value close to this is perfectly acceptable. In our calculations, we set $\kappa = \pi$ and find that varying it by a factor of 5 does not significantly affect the results.

The thermal corrections are described by the following two contributions as detailed in Refs.~\cite{Quiros:1999jp,Athron:2023xlk}: 
\begin{equation}\label{eq:thermal_potential}
V_\mathrm{th}(\phi_\sigma, T) = V_T(\phi_\sigma, T) + V_\mathrm{Daisy}(\phi_\sigma, T)\,,
\end{equation}
where the one-loop thermal potential is
\begin{equation}\label{eq:thermal_cont}
V_T(\phi_\sigma, T) = \frac{T^4}{2\pi^2} \left[3 J_B\qty(\frac{M_{\mathrm{Z^\prime}}^2(\phi_\sigma)}{T^2}) + \sum_f n_f J_F\qty(\frac{M_f^2(\phi_\sigma)}{T^2}) \right]\,.
\end{equation}
The fermionic contributions arise from the heavy neutrinos $f = N_1, N_2, N_3$, with the number of d.o.f.'s $n_{N_i} = 2$. The thermal functions are given by
\begin{equation}\label{eq:thermal_int}
J_{B,F}(y) = \int_0^\infty dx\, x^2 \ln\Big[1 \mp e^{\sqrt{x^2 + y}}\Big]\,.
\end{equation}
While no closed-form expressions are available, analytical expressions can be derived in the high-temperature ($M_i^2/T^2 \ll 1$) and low-temperature ($M_i^2/T^2 \gg 1$) regimes. Specifically~\cite{Quiros:1999jp},
\begin{equation}\label{eq:approx_Js}
\begin{aligned}
    &(M_i^2/T^2 \ll 1):~\begin{cases}
    & J_B(y) = -\dfrac{\pi^4}{45} + \dfrac{\pi^2}{12}y - \dfrac{\pi}{6}y^{3/2} - \dfrac{y^2}{32}\mathrm{ln}\dfrac{y}{c_B} + \dots\\
    & J_F(y) = \dfrac{7\pi^4}{360} - \dfrac{\pi^2}{24}y - \dfrac{y^2}{32} \mathrm{ln}\dfrac{y}{c_F} +\dots
    \end{cases} \\
    &(M_i^2/T^2 \gg 1):~ J_B(y) = J_F(y) = -\sqrt{\frac{\pi}{2}}y^{3/4} e^{-\sqrt{y}}\qty(1 + \frac{15}{8}y^{-1/2} + \frac{105}{128}y^{-1}) + \dots
\end{aligned}    
\end{equation}
where the ellipses represent subleading terms, $c_B = 16\pi^2\exp(3/2 - \gamma_E)$ and $c_F = \pi^2\exp(3/2 - 2\gamma_E)$, with $\gamma_E \approx 0.5772$ the Euler-Mascheroni constant. While we use Eq.~\eqref{eq:thermal_int} for the numerical evaluation of the finite temperature effective potential, the $J_B(y)$ function in the high-$T$ limit in Eq.~\eqref{eq:approx_Js} is employed to compute the thermal masses of the scalar fields below.

Symmetry restoration due to $T^2$-terms in the effective potential typically leads to the breakdown of perturbation theory near the critical temperature. Consequently, an all-order resummation of higher-order contributions, known as Daisy diagrams, is required \cite{Gross:1980br,Parwani:1991gq,Arnold:1992rz,Espinosa:1992kf}. We use the Arnold-Espinosa method, where the Daisy resummation is expressed as~\cite{Arnold:1992rz} 
\begin{equation}\label{eq:daisy}
V_\mathrm{Daisy}(\phi_\sigma, T) = -\frac{T}{12\pi} n_{\mathrm{Z^\prime}} \qty[\overline{M}_{\mathrm{Z^\prime}}^3(\phi_\sigma, T) - M_{\mathrm{Z^\prime}}^3(\phi_\sigma)]\,.
\end{equation}
 Fermions do not contribute because the Matsubara summation for fermions lack zero-frequency
modes~\cite{Espinosa:1992kf}. We define $\overline{M}_\mathrm{Z^\prime}$ to incorporate thermal mass corrections to the $\mathrm{Z^\prime}$ boson and express it as 
\begin{equation}\label{eq:thermal_mass}
\overline{M}_{\mathrm{Z^\prime}}(\phi_\sigma, T) = M_\mathrm{Z^\prime}(\phi_\sigma) + m_{D,\mathrm{Z^\prime}}(T)\,,
\end{equation}
where $m_{D,\mathrm{Z^\prime}}(T)$ is the Debye mass. For vector field contributions, only the longitudinal modes acquire thermal masses, which are evaluated through the one-loop self-energies of the gauge bosons~\cite{Croon:2020cgk,Laine:2016hma,Brauner:2016fla}. We utilize the \texttt{DRAlgo} package~\cite{Ekstedt:2022bff} to obtain the temperature-dependent $\mathrm{Z^\prime}$ boson mass for a generic charge assignment: 
\begin{equation}\label{eq:thermal_masses_vectors}
\begin{aligned}
    & m_{D,\mathrm{Z^\prime}}(T) = \frac{g_L^2 T^2}{3}  (22 x_\mathcal{H}^2 + 8 x_\mathcal{H} x_\sigma + 3 x_\sigma^2)\,.
\end{aligned}
\end{equation}

\begin{figure*}[t]
	\centering
	\subfloat{\includegraphics[width=\textwidth]{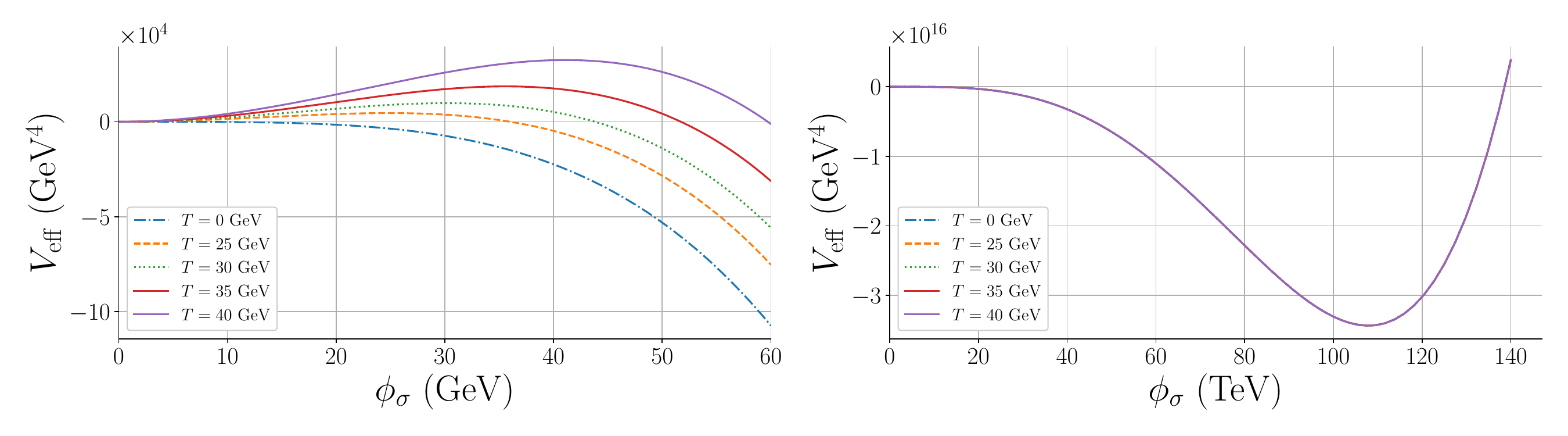}}  
    \caption{\footnotesize Snapshots of the effective potential for several values of the temperature for field values near the barrier (left panel) and near the true vacuum (right panel) for benchmark point (BP) (a) of the conformal $\U{B-L}$ model with $x_\mathcal{H} = 0$, $x_\sigma=2$; see \cref{tab:bench_HEP}. All the curves are overlapping in the right panel. }
	\label{fig:potential_vs_temperature}
\end{figure*}

With all the components in place, the complete RG-improved effective potential is the sum of Eqs.~\eqref{eq:tree_potential}, \eqref{eq:CW_potential}, and \eqref{eq:thermal_cont}. Here, all couplings and fields are rescaled according to Eq.~\eqref{eq:rescale}. Based on the approximations discussed in the last paragraph of \cref{Sec:scalar_pot}, contributions from scalar fields are neglected in one-loop computations. Given the expected suppression of scalar mixing and the hierarchy $v_\sigma \gg v$, the phase transition is governed by a single field $\phi_\sigma$. However, it is important to note that the SM sector indirectly affects the FOPT through the RG evolution of the couplings. Then, the full effective potential can be explicitly written as
\begin{equation}\label{eq:full_4DEFT}
\begin{aligned}
    V_{\mathrm{eff}} = &~\frac{1}{4} \lambda_\sigma(t) Z^2_{\sigma}(t) \phi_\sigma^4 \\
                        &+\frac{1}{64\pi^2} \sum_{a=\mathrm{Z^\prime},N_1,N_2,N_3} n_a M_a^4\qty(\sqrt{Z_\sigma(t)} \phi_\sigma) \qty(\ln \frac{M_a^2\qty(\sqrt{Z_\sigma(t)} \phi_\sigma)}{\mu^2} - c_a) \\
                        &+3 J_B\qty(\frac{M_{\mathrm{Z^\prime}}^2\qty(\sqrt{Z_\sigma(t)}\phi_\sigma)}{T^2}) + \sum_{f=N_1,N_2,N_3} J_F\qty(\frac{M_f^2\qty(\sqrt{Z_\sigma(t)} \phi_\sigma)}{T^2}) \\
                        &-\frac{T}{12\pi} \qty[\overline{M_\mathrm{Z^\prime}}^3(\sqrt{Z_\sigma(t)}\phi_\sigma, T) - M_{\mathrm{Z^\prime}}^3(\sqrt{Z_\sigma(t)}\phi_\sigma)]\,.
\end{aligned}
\end{equation}
In this expression, $Z_\sigma(t) = (1/2) \exp{\int_0^t dt^\prime \gamma(\lambda(t^\prime))}$ is the wave function renormalization, with $\gamma$ defined in Eq.~\eqref{eq:anom_dim}. To understand how the model parameters affect the shape of the potential -- specifically, the behavior of the potential barrier and the true vacuum with respect to variations of the couplings -- it is useful to simplify \cref{eq:full_4DEFT}. First, to assess the behavior near the barrier, we expand this expression in the high-temperature (HT) regime, valid for low field values. For clarity, we disregard the RG dependence of the couplings and fields, \textit{i.e.},~$\lambda(t) \rightarrow \lambda$ and $Z_\sigma(t) \rightarrow 1$. We also fix the charges to $x_\sigma = 2$ and $x_\mathcal{H} = 0$, and set the renormalization scale to $\mu = \pi T$.
 Note that the scale must remain 
proportional to $T$ to ensure that the logarithmic terms from the CW potential cancel with those arising from the high-$T$ expansion of the thermal functions. With this in mind, we expand up to fourth-order in the fields and obtain
\begin{equation}\label{eq:potential_HighT}
    \begin{aligned}
    V_{\mathrm{eff}}^{\mathrm{HT}} =~& \phi_\sigma ^4 \left(\frac{g_L^4(1 - 3\gamma_E + 6\,\mathrm{ln}2) }{2 \pi ^2}-\frac{g_L^3}{2 \sqrt{2} \pi}+\frac{\lambda_\sigma }{4}+\frac{\gamma_E {\rm Tr}(\bm{y_\sigma}^4)}{64 \pi ^2}\right) - \phi_\sigma^3\frac{4 g_L^3 T}{3 \pi }  \\
    &+\phi_\sigma ^2 \left(\frac{g_L^2 T^2}{2}-\frac{g_L^3 T^2}{\sqrt{2} \pi}+\frac{T^2}{48}{\rm Tr}(\bm{y_\sigma}^2) \right)\,.
   \end{aligned}
\end{equation}
We observe that quadratic and cubic terms in $\phi_\sigma$ are generated at finite temperature, and vanish as $T \to 0$. The negative sign of the latter indicates that a potential barrier between the true and false vacuum is induced.
This characteristic of classically conformal models is depicted in the left panel of \cref{fig:potential_vs_temperature}, which shows that the potential barrier is absent at $T=0$ and grows with temperature.
A key feature of this class of models is that the gauge sector is responsible for the FOPT at finite temperatures. This necessitates a nonzero $g_L$ coupling to generate a cubic term, and consequently, a potential barrier. In contrast, 
$\lambda_\sigma$ and the heavy neutrino Yukawa couplings $\bm{y_\sigma}$ determine the location of the true minimum where the quartic term becomes relevant. However, \cref{eq:potential_HighT} is not adequate for analyzing the behavior of the true vacuum, as the high-$T$ expansion fails at large field values for which $\phi_\sigma^4$ term dominates.  While the emergence of a potential barrier is a purely thermal effect, the zero-temperature theory fixes the minimum of the effective potential. Although finite temperature corrections contribute to the position and depth of the true vacuum, we have numerically confirmed that the thermal potential has a minimal impact, thereby validating the zero-temperature approximation. This can be seen in the right panel of \cref{fig:potential_vs_temperature} where the position of the true vacuum is insensitive to temperature corrections. While a complete minimization must include the CW potential, the minimum of the potential essentially arises
from the tree-level RG-improved contribution, $V_0(t)=\lambda_\sigma(t) Z^2_\sigma(t) \phi_\sigma^4/4$.  This suffices for the analytical expressions discussed in this section.
 In the RG-improved approach, the renormalization scale depends on the field value, causing the magnitude and sign of the quartic coupling to vary across the potential. Since $Z_\sigma(t)$ is an order $\mathcal{O}(1)$ parameter that does not change sign, the location of the minimum is governed by the RG evolution of $\lambda_\sigma(t)$, which must be negative for low field values and become positive for higher field values. This ensures the existence of a nonzero minimum and guarantees that the potential remains bounded from below.

\begin{figure*}[t]
	\centering
    \subfloat{\includegraphics[width=\textwidth]{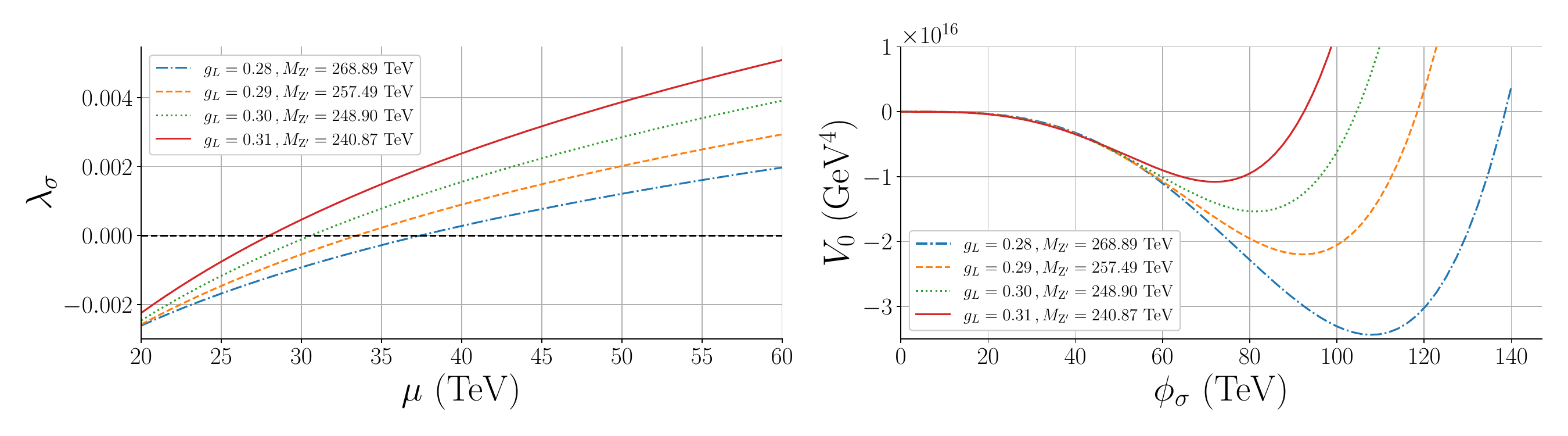}}
    \caption{\footnotesize 
    \textbf{Left panel:} RG evolution of $\lambda_\sigma$ as a function of the renormalization scale $\mu = M_{Z^0} e^t$ for several values of $g_L$ and the resulting $M_{\mathrm{Z}^\prime}$. The blue dashed curve corresponds to BP(a) of the $\U{B-L}$ model, and the other curves illustrate the dependence on $g_L$. At the minimum of the potential the value of $\lambda_\sigma$ depends on $g_L$ via \cref{eq:lambda_relation}. \textbf{Right panel:} Leading-order contribution to the potential, $V_0(t) = \lambda_\sigma(t) Z_\sigma^2(t) \phi_\sigma^4/4$. 
    }
	\label{fig:potential_running}
\end{figure*}

In the left panel of \cref{fig:potential_running}, we 
show the RG evolution of $\lambda_\sigma$ for four values of 
the gauge coupling and corresponding $\mathrm{Z^\prime}$ mass.
Observe that the slopes of the curves increase for larger $g_L$, so that the transition from negative to positive 
$\lambda_\sigma$ occurs at a lower value of $\mu$ as $g_L$ increases. This results in the generation of a minimum at lower field values given that we set the scale in \cref{eq:renorm_scale} to that of the field-dependent $\mathrm{Z^\prime}$ mass in \cref{eq:masses_Z0Zp}.
This behavior follows from the $\lambda_\sigma$ beta-function in \cref{eq:lsigma_betafunction}, where the leading contribution is positive and scales as $6 g^4_L x^4_\sigma$. Thus, increasing $g_L$ leads to faster running and, consequently a sign flip at lower scales. Since the false vacuum is at the origin, the potential energy difference $\Delta V$ during the FOPT is larger for smaller values of $g_L$, as is evident from the right panel of \cref{fig:potential_running}. We also observe that a larger $\mathrm{Z^\prime}$ mass yields a larger field value for the true vacuum. In general, $M_\mathrm{Z^\prime} \sim \phi_\sigma$ which has implications for the peak SGWB frequency as we discuss in \cref{sec:BL_charges}.

\begin{figure*}[t]
	\centering
    \subfloat{\includegraphics[width=\textwidth]{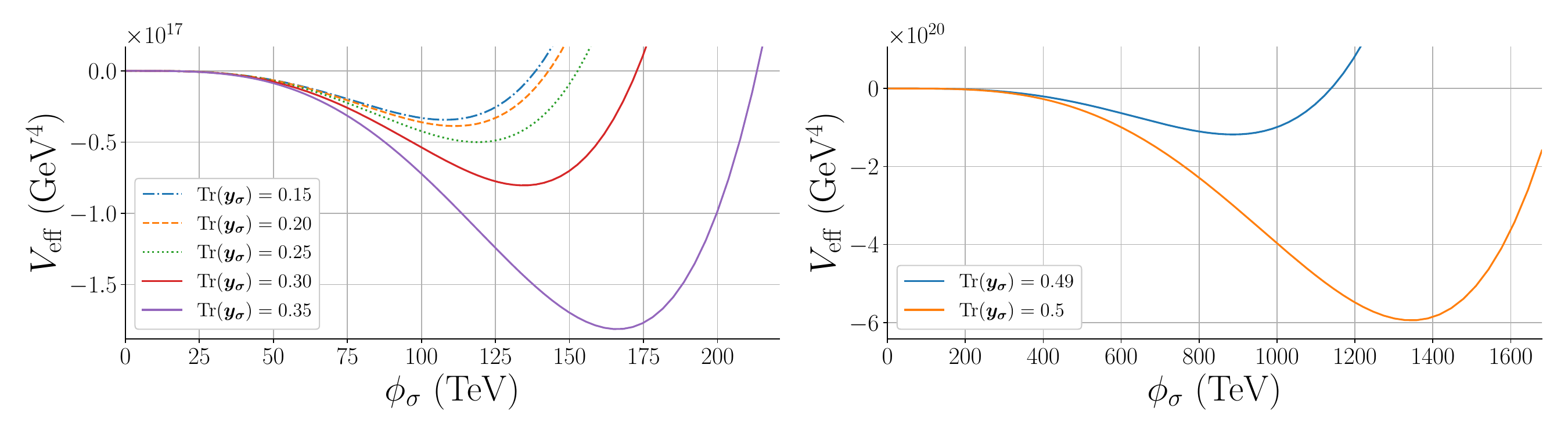}}
    \caption{\footnotesize The effective potential for different values of $\mathrm{Tr}(\bm{y_\sigma})$ for the $\U{B-L}$ model. The other parameters are that of BP(a). }
	\label{fig:Potential_TrueFalse_Yukawas}
\end{figure*}

Yukawa couplings contribute to \cref{eq:lsigma_betafunction} with a negative sign and can dominate the RG evolution if $\mathrm{Tr}(\bm{y_\sigma}) \gtrsim g_L$. Therefore, 
larger $\bm{y_\sigma}$ values push the sign flip in $\lambda_\sigma$ to higher scales.
This is illustrated in \cref{fig:Potential_TrueFalse_Yukawas}, which shows the effective potential for different values of $\mathrm{Tr}(\bm{y_\sigma})$. Both the depth of the effective potential and its minimum are strongly affected.
%

The sensitivity of the true vacuum to $g_L$ and $\bm{y_\sigma}$ can also be assessed by minimizing the one-loop potential. Suppressing RG factors, the zero-temperature potential at the minimum, $\mu = v_\sigma$, is given by
\begin{equation}\label{eq:effective_potential_mZp}
    \begin{aligned}
    V_{\mathrm{eff}} = \frac{\lambda_\sigma  \phi_\sigma ^4}{4} + \frac{3 g_L^4 \phi_\sigma ^4}{4 \pi^2} \left[\mathrm{ln}\left(\frac{4 g_L^2 \phi_\sigma ^2}{v_\sigma^2}\right)-\frac{5}{6}\right]
    - \sum_{i=1}^3 \frac{[\bm{y_\sigma}^4]_{ii} \phi_\sigma ^4}{128 \pi ^2} \left(\mathrm{ln}\left(\frac{[\bm{y_\sigma}^2]_{ii}\phi_\sigma ^2}{2 v_\sigma^2}\right)-\frac{3}{2}\right) \,.
    \end{aligned}
\end{equation}
 Minimizing this expression with respect to $\phi_\sigma$, leads to the following relation for $\lambda_\sigma$:
\begin{equation}\label{eq:lambda_relation}
    \begin{aligned}
    \lambda_\sigma = \frac{1}{32 \pi ^2} \Bigg[32 g_L^4-96 g_L^4 \ln \left(\frac{4 g_L^2 \phi_\sigma ^2}{v_\sigma ^2}\right) - \sum_{i=1}^3 \qty([\bm{y_\sigma}^4]_{ii} + [\bm{y_\sigma}^4]_{ii} \mathrm{ln} \left(  \frac{[\bm{y_\sigma}^2]_{ii} \phi_\sigma^2}{2v_\sigma^2} \right)) \Bigg]\,.
    \end{aligned}
\end{equation}
In the limit that the Yukawa couplings vanish, this relation reduces to 
\begin{equation}
    \lambda_\sigma = \frac{g_L^4}{\pi ^2}\left[1-3\,\mathrm{ln} \left(\frac{4 g_L^2 \phi^2_\sigma}{v_\sigma^2}\right)\right]\,,
\end{equation}
which is the standard relation between $\lambda_\sigma$ and the fourth power of the gauge coupling in conformal models~\cite{Khoze:2014xha}. This relation indicates that varying $g_L$ affects not only the RG evolution of $\lambda_\sigma$ as discussed above, but also its initial value 
at the minimization scale through the tadpole relations in \cref{eq:tadpoles_1loop}. Plugging \cref{eq:lambda_relation} into \cref{eq:effective_potential_mZp} leads to
\begin{equation}\label{eq:potential_at_VEV}
    V_{\mathrm{min}} = \frac{v_\sigma^4}{256 \pi ^2} \Big[-96 g_L^4 + {\rm Tr}(\bm{y_\sigma}^4)\Big]\,,
\end{equation}
which underscores the dependence of the scalar potential on the fourth power of $g_L$, $\bm{y_\sigma}$, and $v_\sigma$, which affects both the thermodynamic and geometric parameters of the SGWB. \cref{eq:effective_potential_mZp,eq:lambda_relation,eq:potential_at_VEV} also highlight how the neutrino sector affects the $\U{}^\prime$ phase transition. While \cref{eq:potential_at_VEV} suggests that for a fixed value of $g_L$, the minimum should get shallower with increasing $\mathrm{Tr}(\bm{y_\sigma})$, in seeming contradiction with Fig.~\ref{fig:Potential_TrueFalse_Yukawas},
the effect of increasing $\mathrm{Tr}(\bm{y_\sigma})$ is much stronger on the RG evolution of $\lambda_\sigma$, and causes the minimum to get deeper.
\subsection{Theoretical uncertainties}\label{subsec:theor_errors}

\begin{figure*}[t]
    \centering
    \hspace*{-3em}
    \subfloat[]{\includegraphics[width=0.52\textwidth]{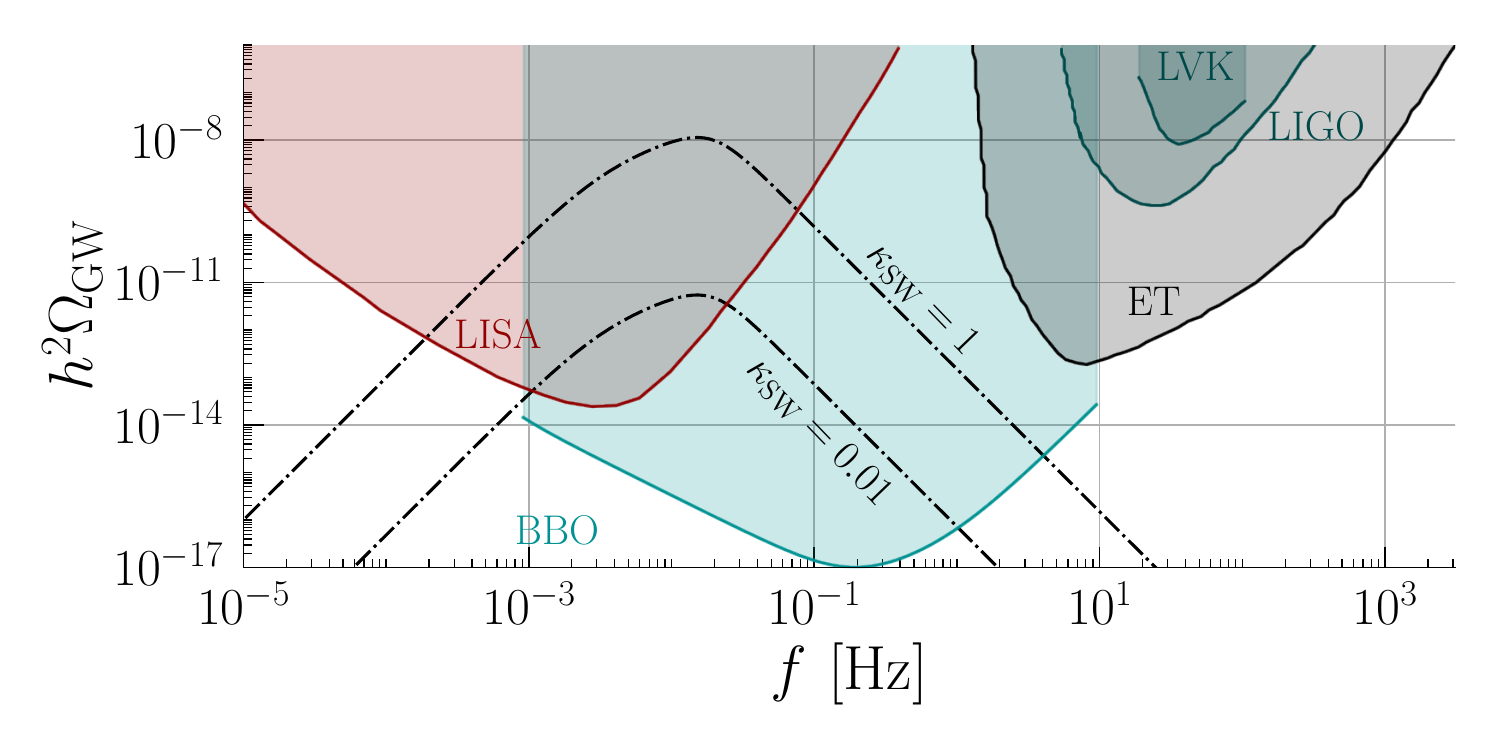}} 
    \subfloat[]{\includegraphics[width=0.52\textwidth]{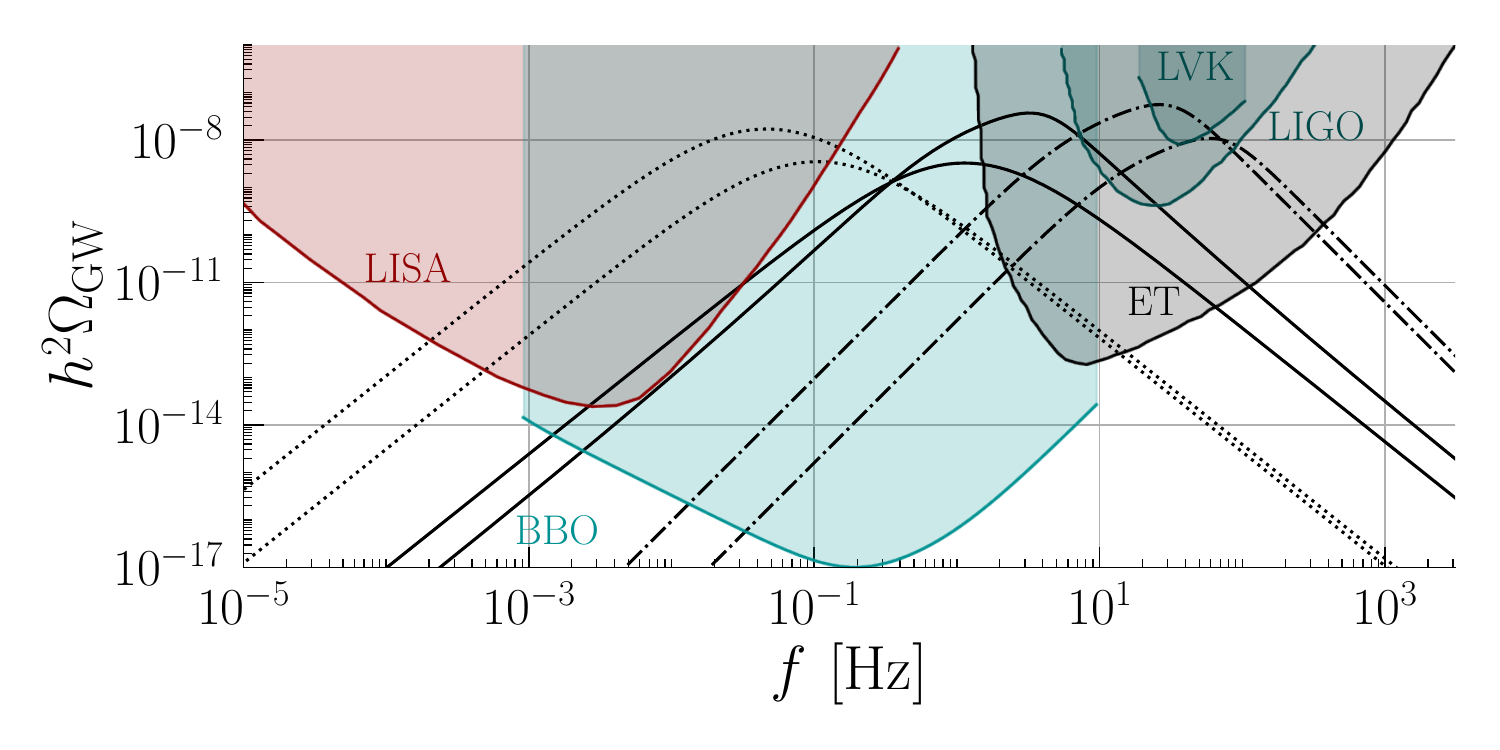}} \\
    \hspace*{-3em}
    \subfloat[]{\includegraphics[width=0.52\textwidth]{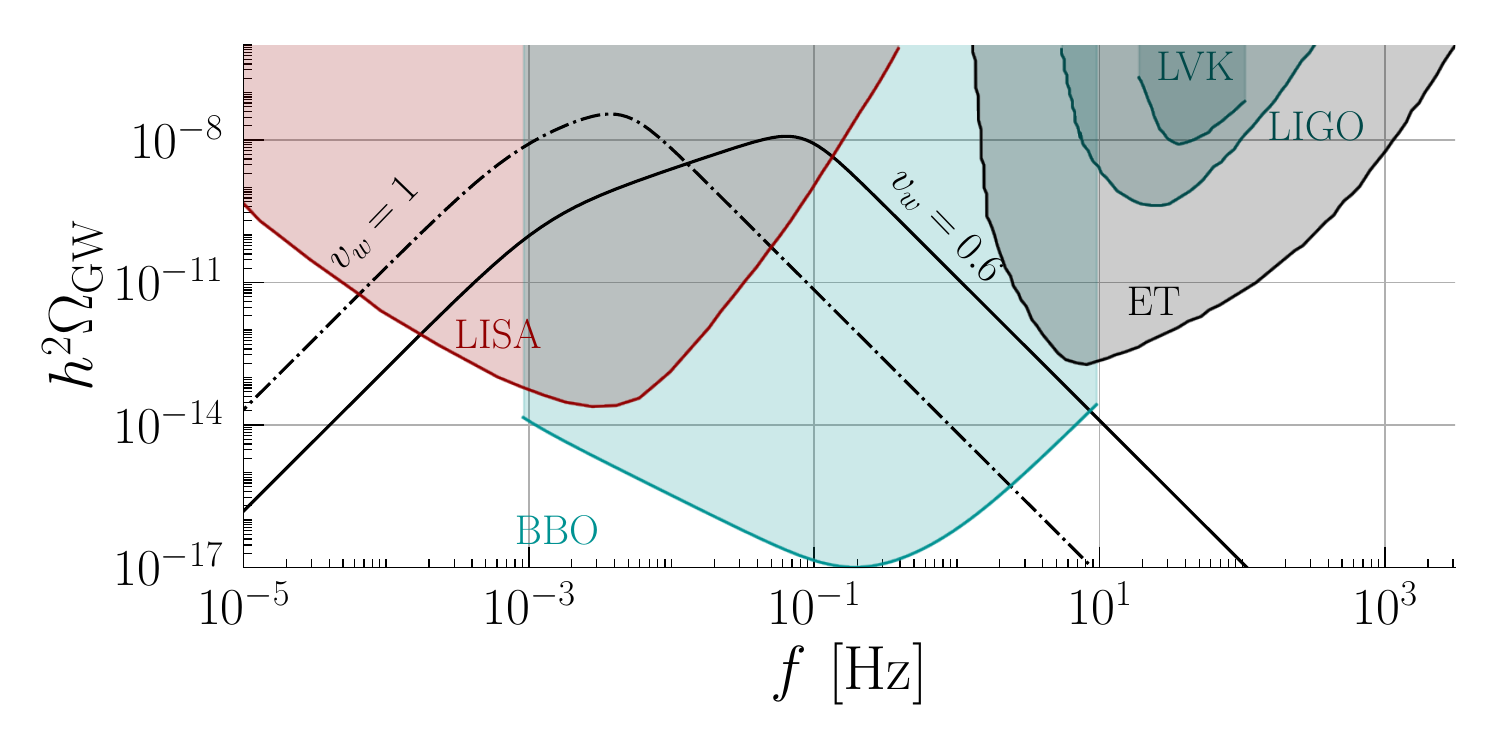}}
    \subfloat[]{\includegraphics[width=0.52\textwidth]{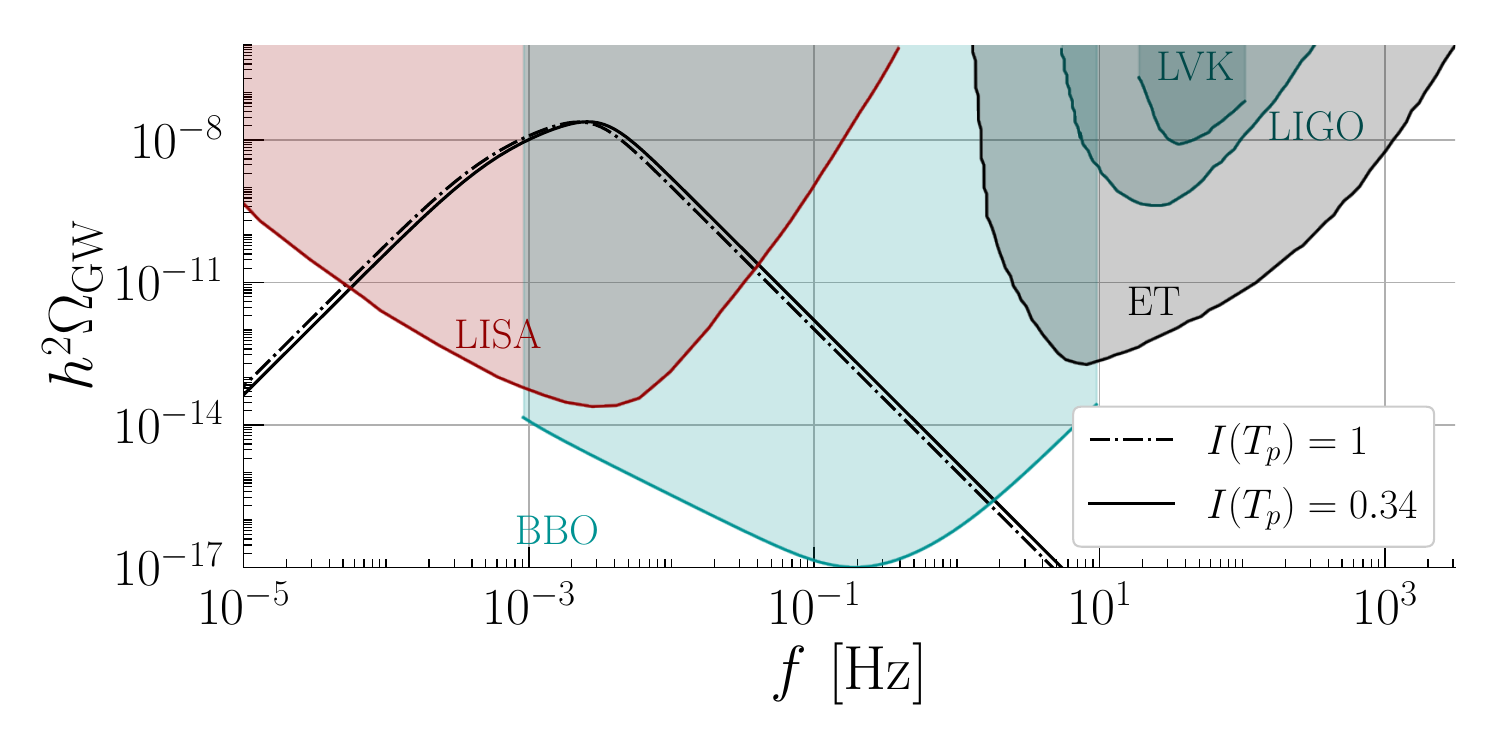}}
    \caption{\footnotesize \textbf{Panel (a):} SGWB spectrum from sound waves for the efficiency factor $\kappa_{\mathrm{SW}}$ between 0.01 and 1, with $v_w = 1$, $\mathrm{log_{10}}~\alpha = 5.7$, $\beta/H(T_p) = 21$ and $T_\mathrm{RH} = 1.36\times 10^{4}~\mathrm{GeV}$. \textbf{Panel (b):} Impact of the bubble radius distribution on the SGWB spectrum for bubble collisions (dotted curves), sound waves (dot-dashed curves) and for both sources (solid curves), with $v_w=1$. For the sound-wave source, we fix $\mathrm{log_{10}}~ \alpha = 16.2$, $\beta/H(T_p) = 7$ and $T_\mathrm{RH} = 33.5~\mathrm{GeV}$, for the bubble-collision source we fix  $\mathrm{log_{10}}~ \alpha = 9.3$, $\beta/H(T_p) = 9.5$ and $T_\mathrm{RH} = 250~\mathrm{TeV}$, and for the mixed source we fix $\mathrm{log_{10}}~ \alpha = 14.6$, $\beta/H(T_p) = 8$ and $T_\mathrm{RH} = 1.9~\mathrm{TeV}$. The curves peaked at a higher (lower) frequency correspond to a monochromatic (extended) radius distribution. \textbf{Panel (c):} SGWB spectrum for two values of the bubble wall velocity with $\mathrm{log_{10}}~\alpha = 11.9$, $\beta/H(T_p) = 12$ and $T_\mathrm{RH} = 5.9~\mathrm{TeV}$.
      \textbf{Panel (d):} SGWB spectrum for two choices of the percolation condition for $v_w = 1$, $\mathrm{log_{10}}~\alpha = 8.6$, $\beta/H(T_p) =14.5$ and $T_\mathrm{RH} = 2.9~\mathrm{TeV}$. Also shown are the LVK bound~\cite{KAGRA:2021kbb} and sensitivity curves for LISA~\cite{LISA:2017pwj}, BBO~\cite{Yagi:2011wg}, LIGO O5~\cite{LIGOScientific:2014pky} and ET~\cite{Punturo:2010zz}.  }
    \label{fig:GW_uncer}
\end{figure*}

Current analytical template functions of the SGWB spectrum rely on efficiency factors that introduce theoretical uncertainties into the SGWB. We quantify the uncertainty due to $\kappa_{\mathrm{SW}}$ by treating it as a free parameter ranging from $0.01$ to $1$. Panel (a) of \cref{fig:GW_uncer} demonstrates that a two order-of-magnitude uncertainty in the efficiency factor results in approximately a three order-of-magnitude uncertainty in the SGWB.

Another commonly used approximation is to fix the radius of the expanding true vacuum bubbles to an average radius $R_*$. A more realistic treatment should consider extended bubble radius distributions. In Ref.~\cite{Marfatia:2024cac}, this effect was analyzed in the context of non-supercooled FOPTs in which the dominant source of GWs is sound waves. A broadening of the spectrum below the peak frequency was noted. Following Ref.~\cite{Marfatia:2024cac}, with the radius distribution of Ref.~\cite{Lu:2022paj}, we show in panel (b) of \cref{fig:GW_uncer} how the radius distribution impacts a pure bubble-collision spectrum (dotted curves with $\kappa_{\mathrm{BC}} = 1$) and a pure sound-wave spectrum (dot-dashed curves with $\kappa_{\mathrm{SW}} = 1$) in the case of supercooling. The spectral broadening found in Ref.~\cite{Marfatia:2024cac} is applicable for both bubble collision and sound wave sources. If both sources contribute (solid curves with $\kappa_{\mathrm{BC}} = \kappa_\mathrm{SW} = 0.5$), spectral broadening occurs at both ends of the spectrum, with a greater impact at higher frequencies. In each case, the curve peaked at a higher (lower) frequency corresponds to a monochromatic (extended) radius distribution.

Uncertainties can also arise from the modelling of the phase transition dynamics, particularly the choice of percolation condition and the determination of the bubble wall velocity. For the former, we have previously defined the percolation temperature via $I(T_p) = 0.34$, which is supported by studies of the percolation of uniformly nucleated bubbles \cite{10.1063/1.1338506,LIN2018299,LI2020112815}. However, percolation can alternatively be defined by requiring $I(T_p) = 1$ (or equivalently, that the probability is given by $\mathcal{P}(T_p) = 1/e$), based on the requirement that the comoving volume equals the volume of true vacuum bubbles \cite{Guo:2021qcq}. 

\begin{figure*}[t]
    \centering
    \includegraphics[width=\textwidth]{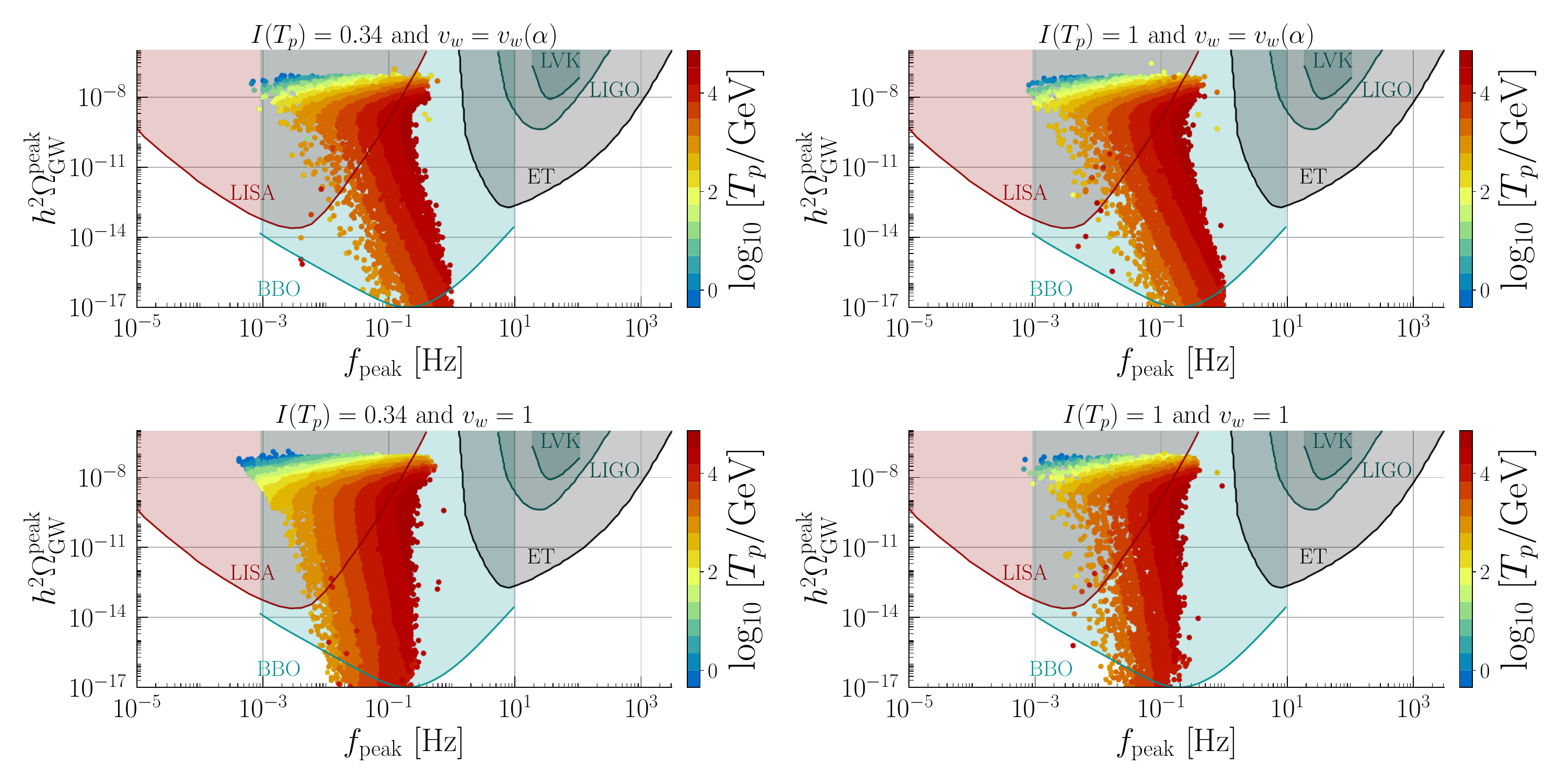} \\
    \caption{\footnotesize Scatter plots of the SGWB peak amplitude $h^2 \Omega^\mathrm{peak}_\mathrm{GW}$ as a function of the peak frequency 
    for the four possible combinations of the percolation condition, $I(T_p) = 0.34$ or $1$, and $v_w=1$ or the Chapman-Jouguet velocity. } 
    \label{fig:ITP_comparision_Omf}
\end{figure*}

Although for supercooled FOPTs it is safe to assume that the bubbles approach the speed of light ($v_w = 1$), in non-supercooled scenarios, bubbles may acquire a subluminal terminal velocity. While recent hydrodynamic simulations suggest that in most cases $v_w = 1$ \cite{Krajewski:2024gma}, analytical estimates indicate otherwise \cite{Ai:2023see}. To assess the impact of $v_w$, we adopt the Chapman-Jouguet velocity~\cite{Steinhardt:1981ct},
\begin{equation}
    v_w(\alpha) = \frac{\sqrt{1/3} + \sqrt{\alpha^2 + 2\alpha/3}}{1+\alpha}\,,
    \label{Chapman}
\end{equation}
as a crude estimate. We perform a scan in the parameter space of the B$-$L model in the mass range $m_{h_2} = [10^3, 10^8]~\mathrm{GeV}$ and gauge coupling range $g_L = [0.26, 0.62]$. The results in the $(f_\mathrm{peak}, h^2 \Omega^{\mathrm{peak}}_{\mathrm{GW}})$ plane are shown in Fig.~\ref{fig:ITP_comparision_Omf}.
Note that employing either $I(T_p) = 0.34$ or $I(T_p) = 1$ has virtually no impact on the peak amplitude and frequency of the SGWB. The bubble wall velocity significantly affects the spectrum, shifting weaker signals (lower $\alpha$) to higher frequencies if wall velocities are determined by \cref{Chapman}, as can be seen from  \cref{fig:ITP_comparision_Omf}. These features are also evident in panels (c) and (d) of Fig.~\ref{fig:GW_uncer}.

\section{Numerical results}\label{sec:num_results}

The potential barrier between the true and false vacuum, which is absent at leading order, is a quantum thermal effect that typically persists for a long time as the Universe cools down. This can extend to temperatures below 0.17~GeV, at which the QCD phase transition occurs, thus entering a non-perturbative regime where our calculations become unreliable. This is especially true in scenarios in which the quark Yukawa interactions are relevant in the effective potential. Although within a local B$-$L framework, a modified cosmology can assist the QCD phase transition through the Higgs portal, allowing for reliable perturbative calculations~\cite{Ellis:2020nnr,Iso:2017uuu,Marzo:2018nov}, we focus on temperatures above the QCD scale.

We perform a numerical scan with parameters in the ranges displayed in \cref{tab:num_ranges_scan}. The bounce action is computed using \texttt{CosmoTransitions}~\cite{Wainwright:2011kj} 
and validated against our own numerical implementation. Further details are provided in \cref{app:numeric_procedure}.  To reduce numerical uncertainties in the calculation of the action due to discretization in temperature, we perform a spline fit to it. Based on the discussion in \cref{subsec:RG_improvement}, we split our parameter space into two parts: one with points that have ${\rm Tr}(\bm{y_\sigma}) < g_L$, and the other with ${\rm Tr}(\bm{y_\sigma}) > g_L$.  Note that in the first dataset, the hierarchy between the heavy neutrinos and the $\U{}^\prime$ breaking scale can be rather large if $(\bm{y_\sigma})_{ii} \ll g_L$, whereas in the second dataset, the heavy neutrino masses are always close to $v_\sigma$, and hence to $M_\mathrm{Z^\prime}$ and $M_{h_2}$.

In our simulations, we only consider FOPTs with $h^2 \Omega_\mathrm{GW}^\mathrm{peak} > 10^{-17}$ that can be probed by current experiments (LIGO), near-future experiments (LISA and ET), or planned future initiatives (BBO).
\begin{table}[t]
	\centering
    \captionsetup{justification=raggedright}
    \begin{tabular}{c|c|c|c|c|c|c|c}
		\toprule
		$M_{h_2}$ (GeV) & $g_{L}$ & $x_\mathcal{H}$ & $x_\sigma$ & $(\bm{y_\sigma})_{ii}$ & $\lambda_\sigma\,, \lambda_{\sigma h}$ & $\lambda_h\,, v_\sigma$ & $M_\mathrm{Z^\prime}$\\
		\midrule
		$  \left[ 150 , 10^{18} \right]$ & $ \left[0.20, 1\right]$ & $\left[-2, 2\right]$ & $\left[0, 5\right]$ & $\left[10^{-10}, 1\right]$  &Eq.~\eqref{eq:tadpoles_1loop} & Eq.~\eqref{eq:loop_corrected_masses} & Eq.~\eqref{eq:Mzprime_Mz_masses} 
		\\
		\bottomrule
	\end{tabular}%
	\caption{\footnotesize Input parameter ranges (defined at $\mu = M_\mathrm{Z^0}$) used in our numerical analysis. We sample $M_{h_2}$ and $(\bm{y_\sigma})_{ii}$ logarithmically and the other parameters linearly. The gauge charges $x_\mathcal{H}$ and $x_\sigma$ admit only rational values. In the last three columns, we refer to the equations used to calculate the quartic couplings, $v_\sigma$ and the $\mathrm{Z^\prime}$ mass.}
	\label{tab:num_ranges_scan}
\end{table}

\subsection{\texorpdfstring{$\U{B-L}$}{sb} scenario \texorpdfstring{$(x_\mathcal{H},x_\sigma) = (0,2)$}{xS}}\label{sec:BL_charges}

\begin{figure*}[t]
	\centering
    \subfloat{\includegraphics[width=\textwidth]{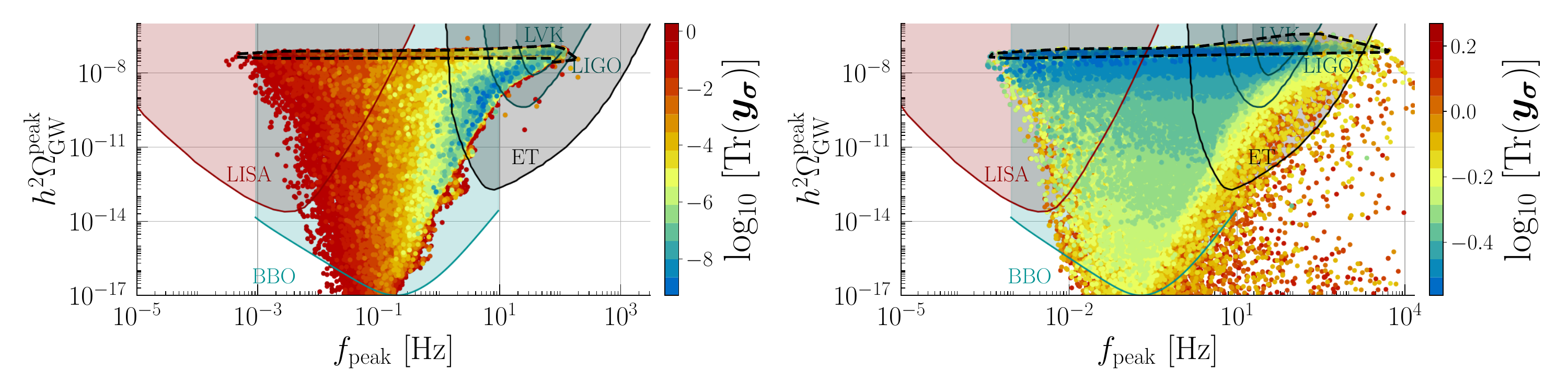}} \\
	\subfloat{\includegraphics[width=\textwidth]{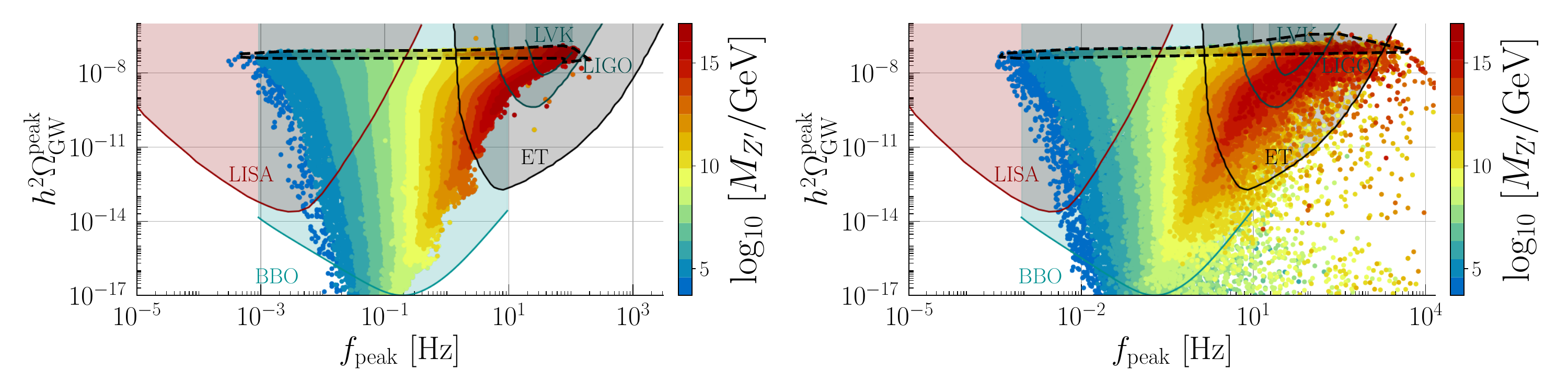}} \\
	\subfloat{\includegraphics[width=\textwidth]{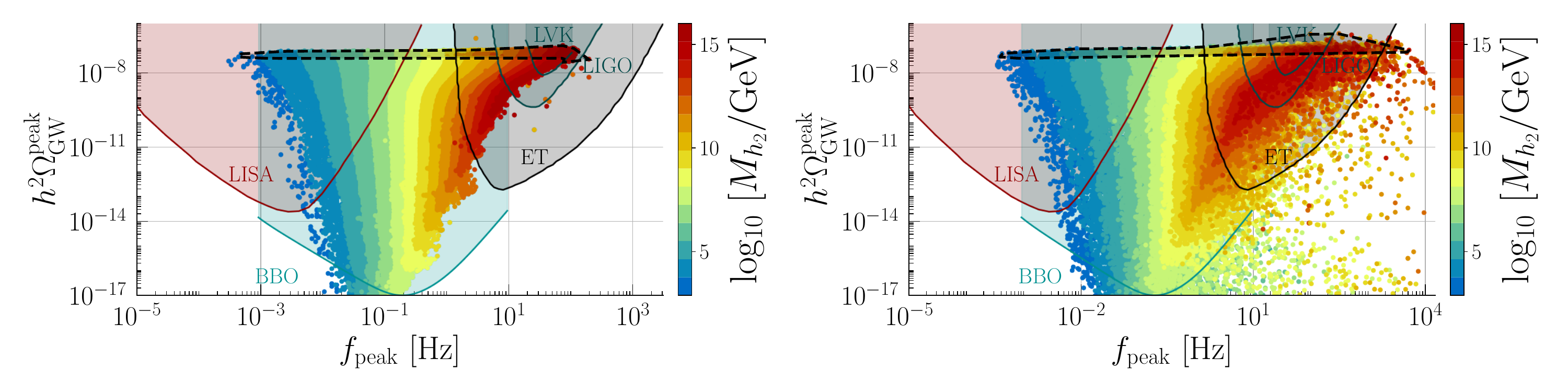}} \\
    \subfloat{\includegraphics[width=\textwidth]{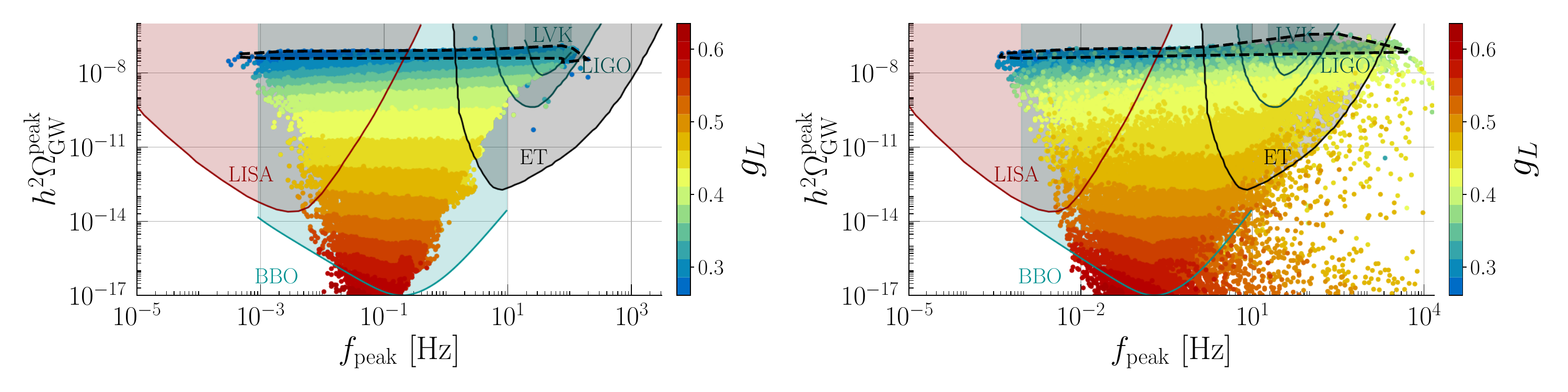}} \\
    \subfloat{\includegraphics[width=\textwidth]{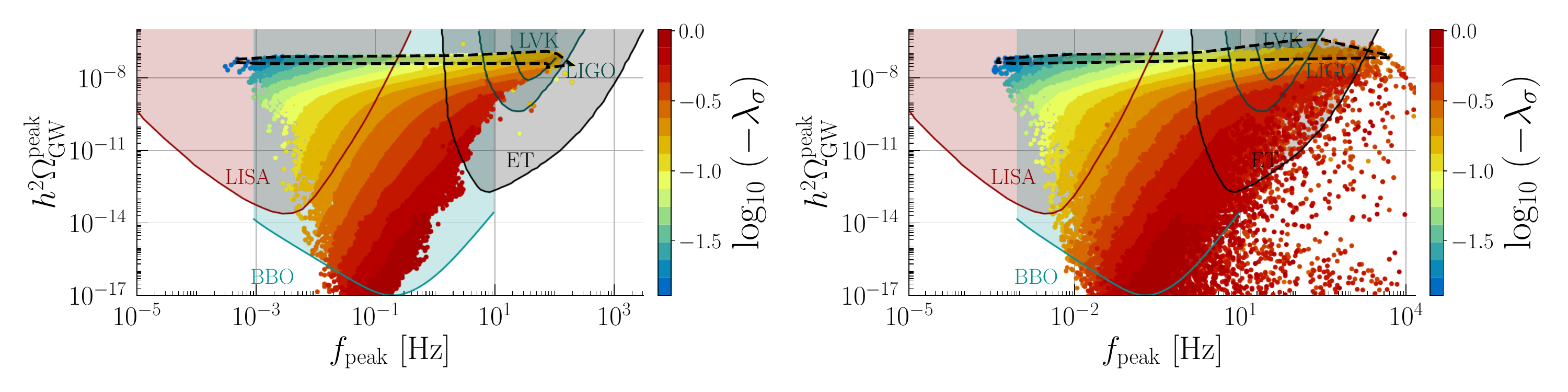}} 
    \caption{\footnotesize Scatter plots of the SGWB peak amplitude $h^2\Omega_\mathrm{GW}^\mathrm{peak}$ as a function of the peak frequency $f_{\mathrm{peak}}$ for the $\U{B-L}$ model. The color scales represent $\mathrm{Tr}(\bm{y_\sigma})$
    (first row), the $\mathrm{Z^\prime}$ boson mass (second row), the heavy scalar mass $M_{h_2}$ (third row), the gauge coupling $g_L$ (fourth row), and the quartic self-coupling of the Majoron $\lambda_\sigma$ (fifth row). In the regions enclosed by a dashed contour, percolation is not assured as prescribed by \cref{eq:perc_condition}, but may occur at a temperature below $T_p$. The left panels show points with $\mathrm{Tr}(\bm{y_\sigma}) < g_L$, and the right panels shows points with $\mathrm{Tr}(\bm{y_\sigma}) > g_L$.  }
	\label{fig:hGW_fpeak_proj_LowHigh_HEP}
\end{figure*}

\begin{figure*}[t]
	\centering
    \subfloat{\includegraphics[width=\textwidth]{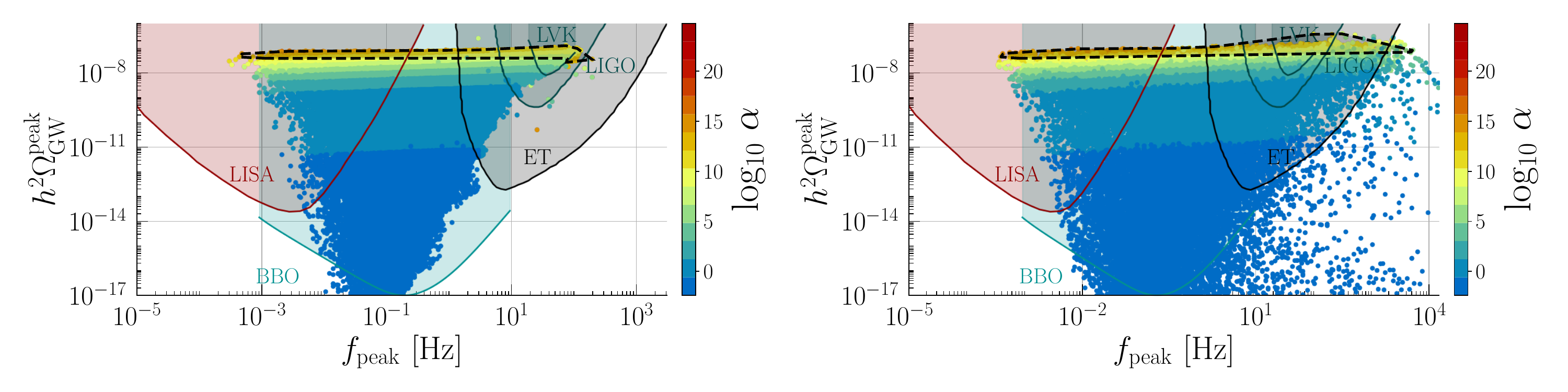}} \\
    \subfloat{\includegraphics[width=\textwidth]{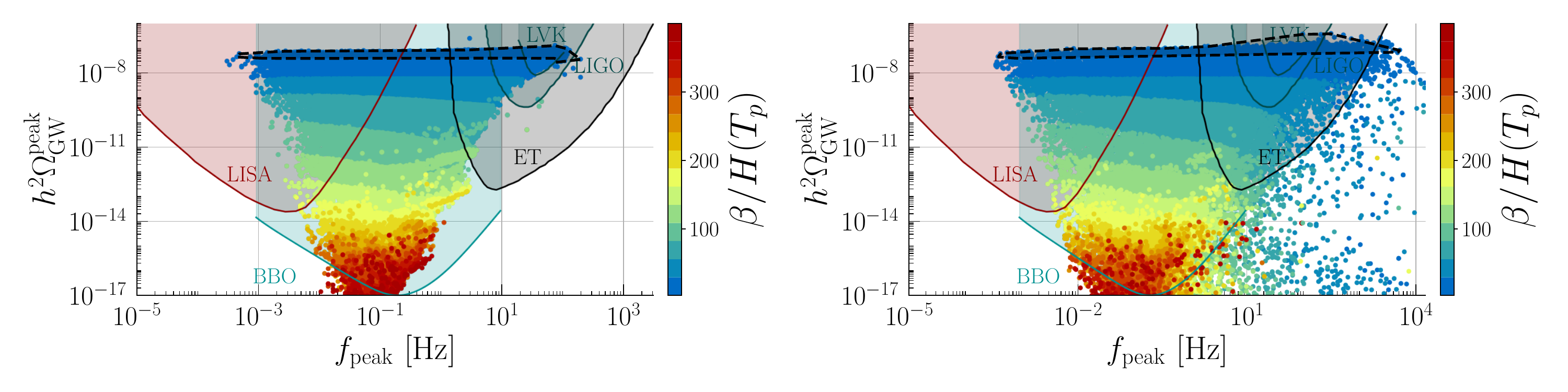}} \\
    \subfloat{\includegraphics[width=\textwidth]{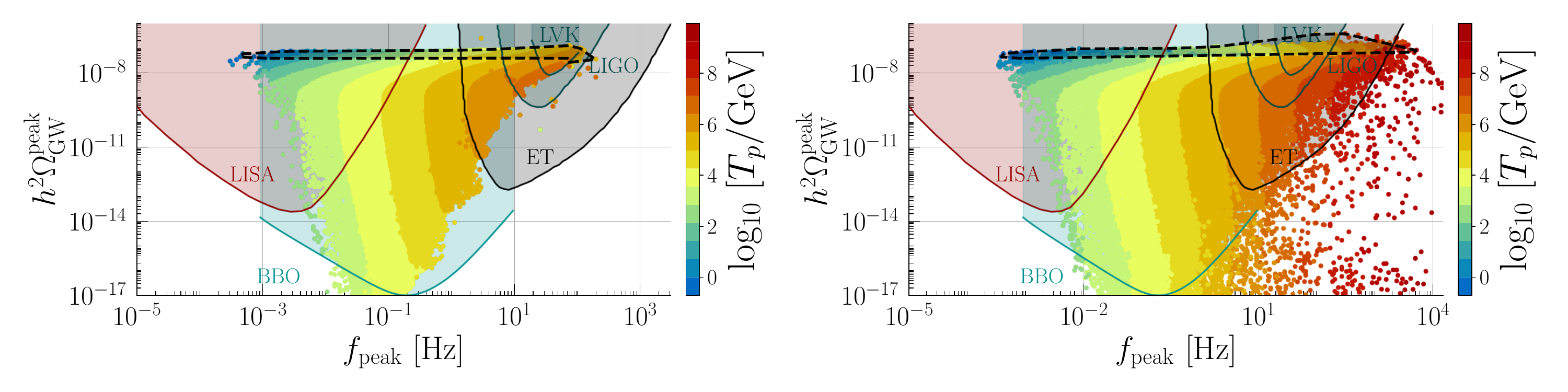}} \\
    \subfloat{\includegraphics[width=\textwidth]{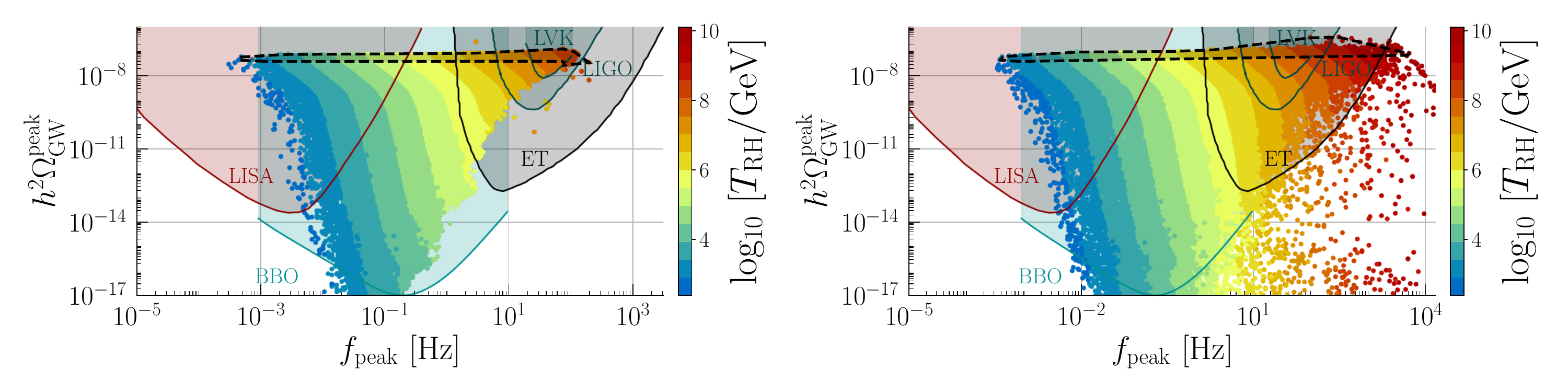}}
    \caption{\footnotesize 
    Similar to \cref{fig:hGW_fpeak_proj_LowHigh_HEP}, but the
    color scales indicate the phase transition strength $\alpha$ (first row), its inverse time duration $\beta/H(T_p)$ (second row), the percolation temperature $T_p$ (third row), and the reheating temperature $T_\mathrm{RH}$ (fourth row).}
	\label{fig:hGW_fpeak_proj_LowHigh_Th} 
\end{figure*}

First, we fix $x_\mathcal{H} = 0$ and $x_\sigma = 2$ and study the classically scale-invariant $\U{B-L}$ scenario. The remaining free parameters are set according to the ranges in \cref{tab:num_ranges_scan}. In \cref{fig:hGW_fpeak_proj_LowHigh_HEP}, we present predictions for the SGWB geometric parameters with respect to $\mathrm{Tr}(\bm{y_\sigma})$ (first row) $M_\mathrm{Z^\prime}$ (second row), the heavy Higgs mass (third row), the $\U{B-L}$ gauge coupling (fourth row), and the quartic coupling $\lambda_\sigma$ (fifth row).
The left (right) panels correspond to the dataset with 
$\mathrm{Tr} (\bm{y_\sigma}) < g_L$ ($\mathrm{Tr} (\bm{y_\sigma}) > g_L$). Note that if we bifurcate the dataset according to $\mathrm{Tr} (\bm{y_\sigma}) < (1-\delta) g_L$ and $\mathrm{Tr} (\bm{y_\sigma}) > (1-\delta) g_L$, where $\delta>0$, then the red points along the right edge of the colored region in the top-left panel migrate to the top-right panel.
The region enclosed by the black dashed contour does not satisfy the criterion for percolation at $T_p$ in \cref{eq:perc_condition}, but is fulfilled at some temperature below $T_p$. 
 In \cref{fig:hGW_fpeak_proj_LowHigh_Th}, we present similar scatter plots, but with the color gradient representing the thermodynamic parameters $\alpha$ (first row), $\beta/H(T_p)$ (second row), $T_p$ (third row), and $T_{\rm RH}$ (fourth row).
In \cref{fig:hGW_fpeak_proj_LowHigh_HEP,fig:hGW_fpeak_proj_LowHigh_Th}, $M_{h_2} \approx M_\mathrm{Z^\prime} \approx v_\sigma$, which explains the similarity between the second and third rows in \cref{fig:hGW_fpeak_proj_LowHigh_HEP}.  A closer inspection reveals a hierarchy of approximately one order of magnitude between $M_{h_2}$ and $M_\mathrm{Z^\prime}$ because $M_{h_2}$ is generated at one loop.

\subsubsection{Impact of the heavy bosons on the peak frequency of GWs}\label{sec:MZp_impact}

The color gradation in the second and third rows of \cref{fig:hGW_fpeak_proj_LowHigh_HEP} indicates that the peak frequency is governed by the $\U{B-L}$ breaking scale, represented by the $\mathrm{Z^\prime}$ and $h_2$ masses. As shown in \cref{fig:potential_running}, a larger value of $M_\mathrm{Z^\prime}$ results in a higher value of the field $\phi_\sigma$ in the true vacuum, thereby increasing the FOPT temperatures, as can be seen from the last two rows of \cref{fig:hGW_fpeak_proj_LowHigh_Th}. According to \cref{eq:amp_and_freq_strongPT,eq:redshit_H_FGW0}, the peak frequency is linearly dependent on the reheating temperature $T_{\rm RH}$ and thus scales with $M_\mathrm{Z^\prime}$ or, equivalently, with $M_{h_2}$.

The substantial difference between the critical and percolation temperatures highlights the degree of supercooling at play. As discussed in \cref{eq:reheat}, $\alpha \gg 1$ implies that $T_p \ll T_\mathrm{RH} < T_c$, which we observe across the entire parameter space. As long as the phase transition completes, the strong supercooling redshifts the SGWB to much higher frequencies 
as required by energy conservation. The disparity between $T_\mathrm{RH}$ and $T_p$ underscores the importance of  calculating the SGWB spectrum at the correct temperature in classically conformal models. Indeed, the dependence of the peak frequency on $T_\mathrm{RH}$ is stronger than on $T_p$.
Redshifting from $T_p$ would lead to the incorrect conclusion that points in the low mass edge of our scatter plots would populate the region probed by Pulsar Timing Arrays~\cite{Athron:2023mer}.

\subsubsection{Impact of the $\U{B-L}$ gauge coupling on the peak amplitude of GWs}
\label{sec:gL_impact}

The fourth row of \cref{fig:hGW_fpeak_proj_LowHigh_HEP} shows that increasing $g_L$ decreases the amplitude of the SGWB spectrum. As indicated in \cref{fig:potential_running}, 
$\Delta V$ decreases with larger $g_L$. Recall that $\Omega_\mathrm{GW}^\mathrm{peak}$ scales with $\Delta V$. From \cref{fig:potential_running}, we see that a 10\% increase in $g_L$ results in a factor of 1.5 decrease in $v_\sigma$, making the true vacuum shallower. This is a general feature of the potential for small $\mathrm{Tr} (\bm{y_\sigma})$ in most of the parameter space, and explains the color gradation in the fourth row of \cref{fig:hGW_fpeak_proj_LowHigh_HEP}; specifically, smaller $g_L$ values favor larger $h^2 \Omega_\mathrm{GW}^\mathrm{peak}$. A close correspondence between the $g_L$ color gradient and those of $\alpha$ and $\beta/H(T_p)$ is evident in the left panels of \cref{fig:hGW_fpeak_proj_LowHigh_Th}. It is worth mentioning that the SGWB peak amplitude tends to plateau at $h^2\Omega_\mathrm{GW}^\mathrm{peak} \approx 10^{-7}$ even for large values of $\alpha$. This is due to the asymptotically constant behavior of $\Omega_\mathrm{SW,BC}^\mathrm{peak}$ in the limit of strong supercooling, as in \cref{eq:approx_Om}.

 For a large part of the parameter space in \cref{fig:hGW_fpeak_proj_LowHigh_Th}, $1 \lesssim \alpha \lesssim 10^{20}$ and $10 \lesssim \beta/H(T_p) \lesssim 150$, defining supercooled and long-lasting FOPTs.  This entire parameter space falls within the detection capabilities of various ongoing and planned experiments. LISA will probe $\U{B-L}$ breaking scales ranging from tens of TeV, similar to the LHC reach, up to about $10^{8}~\mathrm{GeV}$, while the sensitivity of LIGO and ET extends to the GUT scale. We also find FOPTs with $\alpha < 1$ corresponding to $g_L \gtrsim 0.4$. These non-supercooled transitions predict a SGWB with peak frequencies within $10~\mathrm{mHz} \lesssim f_\mathrm{peak} \lesssim 1~\mathrm{Hz}$ and peak amplitudes $h^2 \Omega_\mathrm{GW}^\mathrm{peak} \lesssim 10^{-11}$, most of which are well below LISA sensitivity, but may be probed by BBO.

Since we restrict $h^2 \Omega_\mathrm{GW}^\mathrm{peak} > 10^{-17}$, the gauge coupling $g_L$ lies in the interval, $0.26 \lesssim g_L \lesssim 0.62$. However, even for FOPTs with $h^2 \Omega_\mathrm{GW}^\mathrm{peak} < 10^{-17}$, the upper limit in our scan, $g_L^\mathrm{max} = 1$, is never reached, as large values $g_L$ violate perturbativity at the $M_\mathrm{Z^\prime}$ scale. As demonstrated previously, the effective potential is highly sensitive to small variations in $g_L$, which partially explains the relatively narrow band in $g_L$. While the upper bound on $g_L$ corresponds to lower amplitudes, a couple of constraints restrict $g_L$ from below. Firstly, as the peak amplitude increases with decreasing $g_L$, the total integrated SGWB energy density must not exceed the amount of dark radiation allowed by BBN. This imposes the constraint $h^2 \Omega_{\mathrm{GW}} < 5.6 \times 10^{-6} \Delta N_{\mathrm{eff}}$~\cite{Pagano:2015hma}, which translates into an upper bound of $h^2 \Omega_{\mathrm{GW}} < 2.8 \times 10^{-6}$ for $\Delta N_{\mathrm{eff}} < 0.5$. Secondly, for small values of $g_L$, it is questionable whether percolation occurs, as Eq.~\eqref{eq:perc_condition} is only satisfied for $T < T_p$. 

At low peak frequencies, the percolation temperature approaches the QCD scale $\sim 0.17~\mathrm{GeV}$, which we do not consider in our analysis. Conversely, at high peak frequencies, the percolation temperature can reach up to $10^{10}~\mathrm{GeV}$ for $\mathrm{Tr} (\bm{y_\sigma}) > g_L$. This contrasts with SU(2) conformal models, for which the percolation temperature remains below $300~\mathrm{GeV}$~\cite{Kierkla:2023von}. This difference underscores a key distinction between Abelian and non-Abelian scenarios. In our case, the RG evolution of $g_L$  is asymptotically safe. However, for $\SU{N}{}$ models, the gauge coupling runs to non-perturbative values, constraining the percolation temperature from above~\cite{Kierkla:2022odc,Kierkla:2023von}.

\subsubsection{Role of the neutrino sector}
\label{sec:yN_impact}

The right panels in \cref{fig:hGW_fpeak_proj_LowHigh_HEP,fig:hGW_fpeak_proj_LowHigh_Th} correspond to scenarios in which the neutrino sector 
affects the running of $\lambda_\sigma$ and the minimum of the potential because $\mathrm{Tr}(\bm{y_\sigma}) > g_L$. As illustrated in \cref{fig:Potential_TrueFalse_Yukawas}, for increasing Yukawa couplings the minimum becomes deeper and shifts towards larger VEVs, augmenting the potential energy difference between the true and false vacuum, $\Delta V$. This effect allows for larger values of $g_L$ populating a greater area in the right panel of the fourth row of \cref{fig:hGW_fpeak_proj_LowHigh_HEP}. The shift towards larger $v_\sigma$ driven by the neutrino sector extends the frequency range up to about $10~\mathrm{kHz}$ and raises the critical and percolation temperatures by approximately five orders of magnitude. This effect is more pronounced at the high-frequency end of the last two rows of \cref{fig:hGW_fpeak_proj_LowHigh_Th}.

While the interplay between $g_L$ and $\mathrm{Tr} (\bm{y_\sigma})$ affects RG evolution of $\lambda_\sigma$, the shape of the potential is primarily controlled by the LO contribution, $V_0=\lambda_\sigma(t) Z^2_{\sigma}(t) \phi_\sigma^4/4$. Consequently, the correlation between $\lambda_\sigma$ and the peak amplitude/frequency is expected to be the same for $\mathrm{Tr}(\bm{y_\sigma}) < g_L$ and $\mathrm{Tr}(\bm{y_\sigma}) > g_L$, barring the spread in points to higher frequencies in the bottom-right panel of \cref{fig:hGW_fpeak_proj_LowHigh_HEP}. Said differently,  
for fixed values of $h^2 \Omega_\mathrm{GW}^\mathrm{peak}$ and $f_\mathrm{peak}$, the value of
$\lambda_\sigma$, and hence $V_0$ is the same whether  
$\mathrm{Tr}(\bm{y_\sigma}) < g_L$ or $> g_L$. This explains the smooth variation in $\lambda_\sigma$ in the bottom panels of \cref{fig:hGW_fpeak_proj_LowHigh_HEP}. The other parameters exhibit a certain degree of overlap of the different colors in the right panels.  

\begin{figure*}[t]
	\centering
	\subfloat{\includegraphics[width=\textwidth]{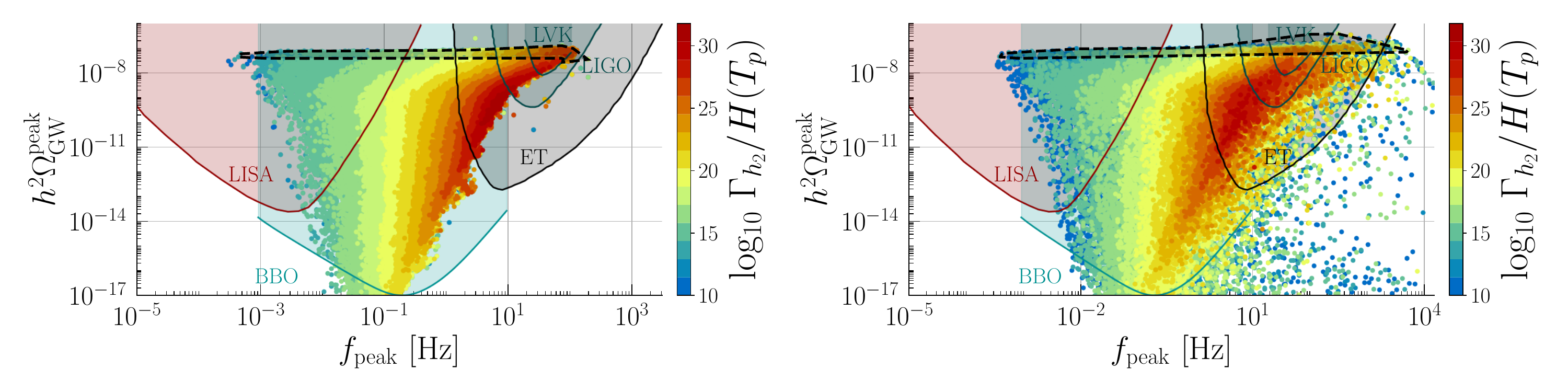}} \\
    \caption{\footnotesize 
    Similar to \cref{fig:hGW_fpeak_proj_LowHigh_HEP}, but the
    color scale indicates the ratio of the $h_2$ decay rate to the Hubble rate at the percolation temperature. }
	\label{fig:GW_GammaH}
\end{figure*}

Depending on the heavy Higgs decay rate $\Gamma_{h_2}$, the Universe will either immediately enter a radiation-dominated epoch if $\Gamma_{h_2} > H(T_p)$, or it will first pass through a period of matter domination if $\Gamma_{h_2} < H(T_p)$ until $\Gamma_{h_2} \simeq H$. To distinguish between these scenarios, we must compute the decay rate of $h_2$ and compare it with the Hubble rate at percolation. $h_2$ can decay through the neutrino channel $h_2 \rightarrow \bar{N}_i N_i$ via Yukawa interactions, followed by the decay of the right-handed neutrinos into SM particles, or directly into SM particles mediated by the mixing of $h_1$ and $h_2$. For the decay rates of the right-handed neutrinos, we consider all dominant two- and three-body 
channels~\cite{Helo:2010cw}. We also cross-checked the decay rates with \texttt{MadGraph}~\cite{Alwall:2014bza} and found agreement within 10\%. For the decays of $h_2$, we consider the two-body processes,
\begin{equation}\label{eq:decay_rates}
    \begin{aligned}
        & \Gamma_{h_2 \to h_1 h_1} = \frac{\lambda^2_{\sigma h} v^2_\sigma}{32\pi M_{h_2}} \sqrt{1 - \dfrac{M^2_{h_1}}{M^2_{h_2}} } \,, \\
        & \Gamma_{h_2 \to \bar{f}_\mathrm{SM} f_\mathrm{SM}} = \frac{M_{h_2} \sin^2\theta}{16\pi v^2} \sum_{f} M^2_f  \sqrt{1 - \frac{4 M^2_f}{M^2_{h_2}}} \,, \\
        & \Gamma_{h_2 \to VV} = \frac{C_V M_{h_2}^3 \sin^2\theta}{16\pi v^2} \sqrt{ 1 - \frac{4 M_V^2}{M_{h_2}^2}} \qty(1 - \frac{4 M_V^2}{M_{h_2}^2} + \frac{12M_V^4}{M_{h_2}^4}) \,, \\
        & \Gamma_{h_2 \to \bar{N}_{i} N_{i}} = \frac{M_{h_2}}{16\pi v^2_\sigma} \sum_{i=1}^3 M^2_{N_{i}}  \sqrt{1 - \frac{4 M^2_{N_{i}}}{M^2_{h_2}}} \,,
    \end{aligned}
\end{equation}
where $C_V=1,2$ for $V=\mathrm{Z^0}, W^\pm$,  and the scalar mixing angle $\theta$ is defined by
\begin{equation}\label{eq:mixing_scalar}
    \sin 2\theta = \frac{2v v_\sigma \lambda_{\sigma h}}{M_{h_1}^2 - M_{h_2}^2}\,.
\end{equation}
The total decay rate into SM final states is given by $\Gamma_{h_2} = \Gamma_{h_2 \rightarrow p_\mathrm{SM} p^*_\mathrm{SM}} + \Gamma_{h_2 \rightarrow \bar{N_i} N_i}\Gamma_{N_i \rightarrow p_\mathrm{SM} p^*_\mathrm{SM}}$, where $p_\mathrm{SM}$ denotes all SM particles. This is used to redshift the SGWB spectra according to \cref{eq:reheat,eq:redshit_H_FGW0}. For $\Gamma_{h_2} < H(T_p)$, the peak amplitude and peak frequency, are suppressed according to $\left(\Gamma_{h_2}/H(T_p)\right)^{1/6}$ and $\left(\Gamma_{h_2}/H(T_p)\right)^{2/3}$, respectively. The first three channels in \cref{eq:decay_rates} are suppressed due to the small scalar mixing that scales as $|\lambda_{\sigma h}| \sim v^2/v_\sigma^2 \ll 1$ (see \cref{app:numeric_procedure}), leaving $\Gamma_{h_2 \rightarrow \bar{N}_{i} N_{i}}$ as the dominant contribution to the decay width. 
\begin{figure*}[t]
	\centering
	\subfloat{\includegraphics[width=\textwidth]{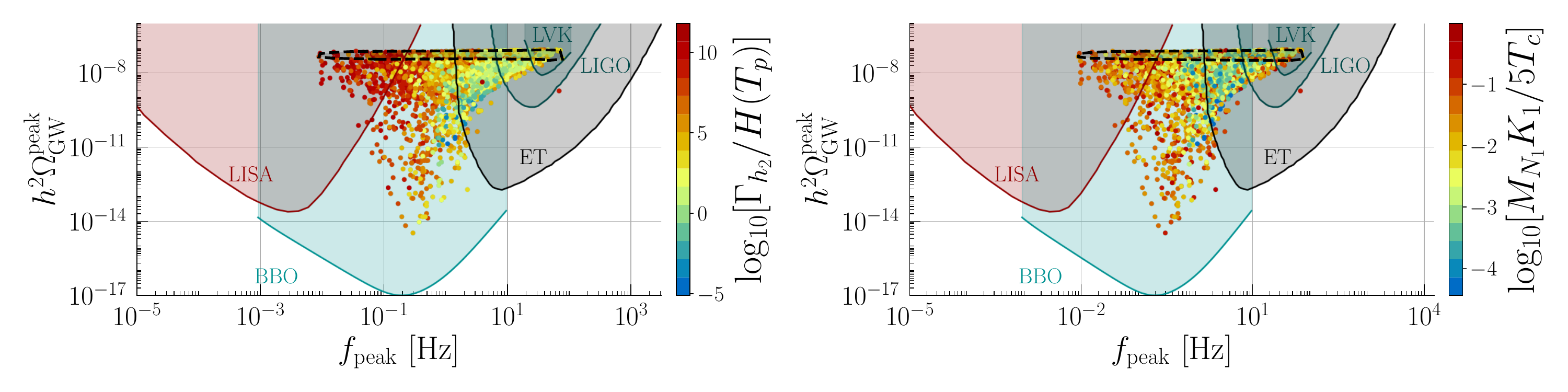}} \\
    \subfloat{\includegraphics[width=0.5\textwidth]{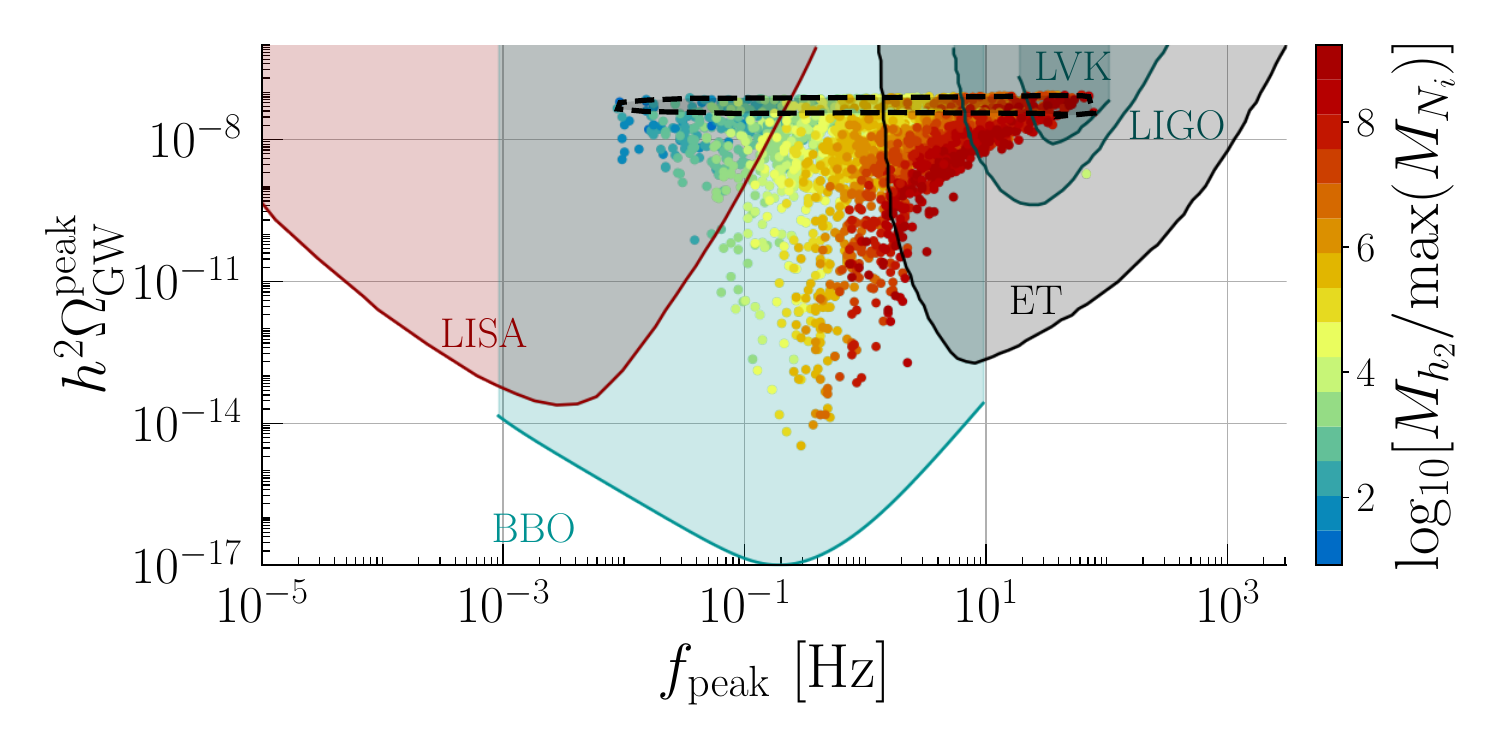}}
    \caption{\footnotesize Scatter plots for the $\U{B-L}$ model with points for which $\mathrm{Tr}(\bm{y_\sigma}) < g_L$ and the right-handed neutrinos do not thermalize, {\it i.e.,} $M_{N_i} K_i/(5 T_c) < 1$. In the top-left panel, the color scale represents the ratio of the heavy Higgs decay rate to the Hubble rate at the percolation temperature, in the top-right panel, the corresponding value of $M_{N_i} K_i/(5 T_c)$, and in the bottom panel, the mass ratio of the heavy Higgs and the heaviest right-handed neutrino. The black dashed contour has the same meaning as in \cref{fig:hGW_fpeak_proj_LowHigh_HEP}. }
	\label{fig:GW_neutrinohiggs}
\end{figure*}
From \cref{fig:GW_GammaH}, we conclude that the Universe promptly enters the radiation dominated era after percolation because \mbox{$\Gamma_{h_2}/H(T_p) > 10^{10}$}. This lower limit arises from the thermalization condition $M_{N_i} K_i \gtrsim 5 T_c$ which excludes all scenarios with feeble couplings to the SM. This is evident in \cref{fig:GW_neutrinohiggs}, which shows that $M_{N_i} K_i < 5 T_c$ (top-right panel) yields $\Gamma_{h_2}/H(T_p) \lesssim 10^{12}$ (top-left panel). This bound also establishes a maximum allowable hierarchy between the heavy neutrino masses and the $h_2$ mass, limiting it to eight orders of magnitude, as shown in the bottom panel. Furthermore, the cluster of blue points in the top-left panel corresponds to scenarios where the early matter-dominated period is long-lasting and the hierarchy between $M_{N_i}$ and $M_{h_2}$ is maximal. 

The impact of the neutrino sector on the FOPT becomes significant when the magnitudes of $\bm{y_\sigma}$ and $g_L$ are comparable; see \cref{eq:lsigma_betafunction}. However, even if the Yukawa sector does not contribute to the minimization, it strongly affects the SGWB at high frequencies. To quantify this, we consider a subset of our data with $\mathrm{Tr}(\bm{y_\sigma}) < 10^{-8}$. Then, $\Gamma_{h_2 \rightarrow \bar{N_i} N_i}$ and the contribution of $\mathrm{Tr}(\bm{y_\sigma}^4)$ to $V_\mathrm{min}$ are negligible, so that the right-handed neutrinos are effectively decoupled {and the Universe enters an era of matter domination after percolation.
In this scenario, $h_2$ decays only to SM particles with a rate $\Gamma_{h_2}^\mathrm{SM}$ that is strongly suppressed. We find that $H(T_p) \approx [5\times 10^{-16}~\mathrm{GeV}^{0.04}] M_{h_2}^{0.96}$ and $\Gamma_{h_2}^\mathrm{SM} \approx [10^{5.5}~\mathrm{GeV}^4]M_{h_2}^{-3}$, with $M_{h_2}$ in units of $\mathrm{GeV}$. Then, $\Gamma_{h_2}^\mathrm{SM}/H(T_p) \approx [6.32 \times 10^{20}~\mathrm{GeV}^{3.96}] M_{h_2}^{-3.96}$, which implies that $\Gamma_{h_2}^\mathrm{SM}/H(T_p) > 1$ for $M_{h_2} \gtrsim 1.8\times 10^{5}~\mathrm{GeV}$, as shown in \cref{fig:hGW_fpeak_fake}.  As $M_{h_2}$ increases, $\Gamma_{h_2}^\mathrm{SM}/H(T_p)$ rapidly decreases and significantly suppresses the peak frequency of the SGWB below $0.1~\mathrm{Hz}$, and the peak amplitude is reduced so that LISA is sensitive to scales below $10^{7}~\mathrm{GeV}$. Therefore, the SGWB accessible by LIGO and ET is a distinctive signature of the neutrino sector. 
\begin{figure*}[t]
	\centering
    \hspace*{-0.5em}
    \subfloat{\includegraphics[width=\textwidth]{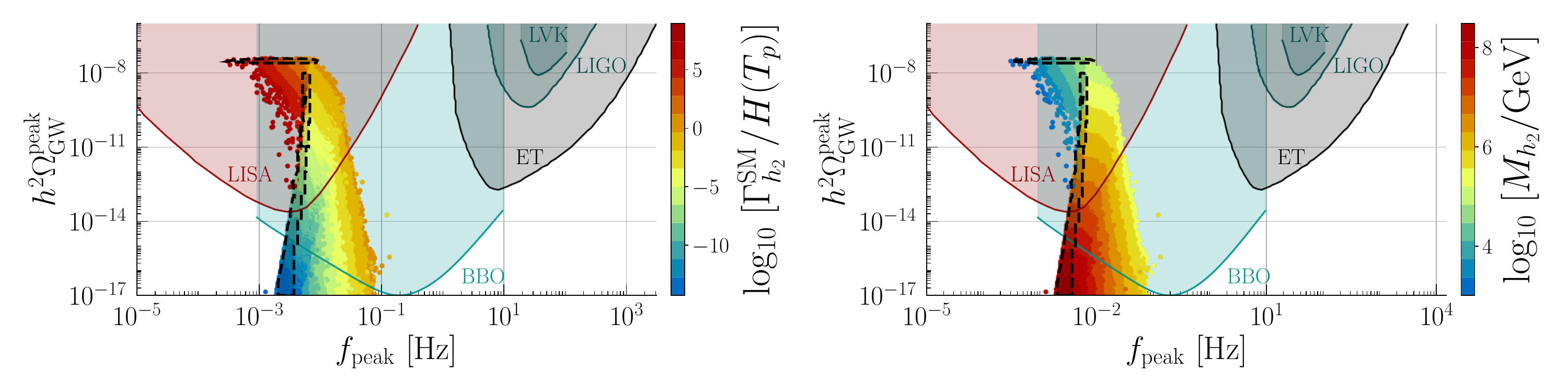}}
    \caption{\footnotesize 
    Similar to \cref{fig:hGW_fpeak_proj_LowHigh_HEP},
    but for a scenario with decoupled right-handed neutrinos. In the left panel, $\Gamma_{h_2}^\mathrm{SM}$ is the direct decay width of $h_2$ to SM particles.  }
	\label{fig:hGW_fpeak_fake}
\end{figure*}

The top and bottom rows of \cref{fig:hGW_fpeak_Yukawas} quantify the dependence of the SGWB on the size of the $\bm{y_\nu}$ Yukawa couplings, and the mass of the heaviest right-handed neutrino, respectively. We observe a clear correlation between the peak frequency and the magnitude of $\bm{y_\nu}$ and $\mathrm{max}(M_{N_i})$. This correlation arises from the type-I seesaw mechanism, and establishes a direct connection between GW physics and neutrino physics. Specifically, for interferometers operating in the Hz to kHz range, such as LIGO and ET, the observation of a SGWB would imply $\bm{y_\nu} \sim \mathcal{O}(1)$ and a neutrino mass scale between $10^{10}$ and $10^{15}$~GeV. In contrast, LISA can probe $\bm{y_\nu}$ ranging from approximately $10^{-6}$ (blue points) to $10^{-3}$ (yellow points), corresponding to a seesaw scale between $10^4$~GeV and $10^8$~GeV.
\begin{figure*}[t]
	\centering
    \subfloat{\includegraphics[width=\textwidth]{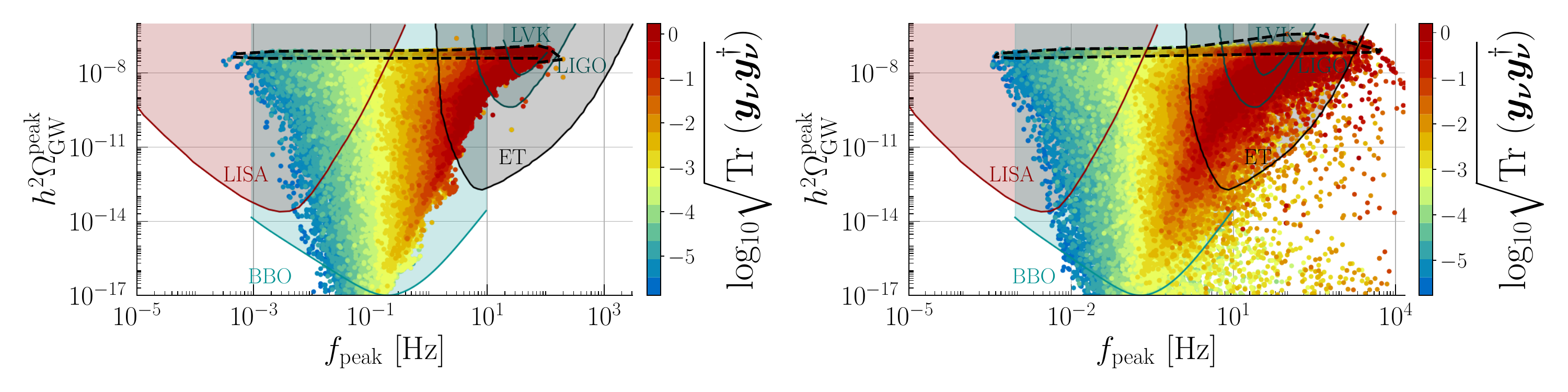}} \\
    \subfloat{\includegraphics[width=\textwidth]{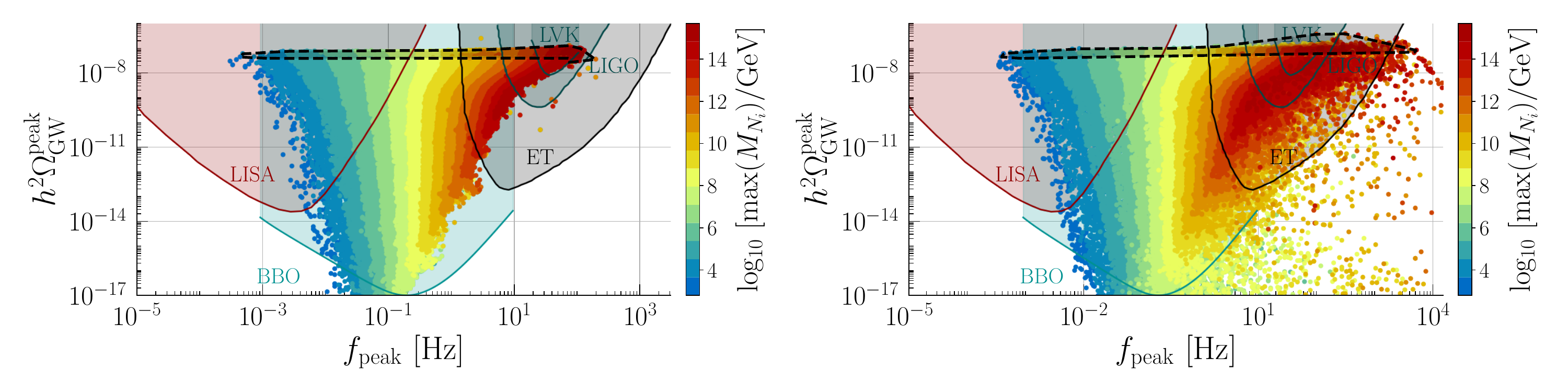}}
    \caption{\footnotesize 
    Similar to \cref{fig:hGW_fpeak_proj_LowHigh_HEP}, but the
    color scales indicate the magnitude of the Dirac Yukawa couplings, $\sqrt{\mathrm{Tr} (\bm{y}_\nu \bm{y}_\nu^\dagger)}$ (top row), and mass of the heaviest right-handed neutrino (bottom row). 
}\label{fig:hGW_fpeak_Yukawas}
\end{figure*}
For $\mathrm{Tr}(\bm{y_\sigma}) > g_L$ (right panels), the correlation is not as clean due to the competition between $g_L$ and $(\bm{y_\sigma})_{ii}$. However, at LIGO and ET frequencies, the magnitude of the Dirac neutrino Yukawa is typically of order one. This results from the fact that the 
$\U{B-L}$ breaking scale, which characterizes the scale of neutrino mass generation, approaches the GUT scale, so that $v^2/v_\sigma$ in \cref{eq:nu-light} is suppressed. In particular, $v^2/v_\sigma \lesssim 0.1~\mathrm{eV}$ for $v_\sigma \gtrsim 10^{14}~\mathrm{GeV}$. At lower frequencies, $v^2/v_\sigma$ is larger, necessitating smaller $\bm{y_\nu}$ to produce sub-eV neutrino masses.

\subsubsection{Constraining the parameter space with GW data}
\label{sec:SNR}

A $\mathrm{Z^\prime}$ boson that couples to electrons and muons has been constrained by the LHC to be heavier than approximately $5~\mathrm{TeV}$. GW experiments afford a completely different approach to constrain the model parameter space. Data from LIGO, ET and LISA will cover a mass range from the TeV scale all the way up to the GUT scale. To quantify the detection prospects for a given GW experiment, we calculate the signal-to-noise ratio,
\begin{equation}\label{eq:SNR_calc}
    \mathrm{SNR} = \sqrt{ \mathcal{T} \int df~ \frac{\Omega_{\mathrm{GW}}^2(f)}{ \Omega_{\mathrm{Sens}}^2(f)} } \,,
\end{equation}
where $h^2 \Omega_{\mathrm{GW}}(f)$ is the predicted GW spectrum and $h^2 \Omega_{\mathrm{Sens}}(f)$ is the expected experimental sensitivity. Except for the LVK bound, we set the observation time to $\mathcal{T} = 4~\mathrm{years}$ for all experiments. A signal is considered detectable if $\mathrm{SNR} > 10$. 

In \cref{fig:Spectrum_ng5Ligo} we illustrate three benchmark scenarios whose parameters are provided in \cref{tab:bench_HEP}.
\begin{itemize}
    \item[BP(a)] corresponds to physics at the $10-100$~TeV scale with a GW spectrum that peaks in the mHz regime, well within the reach of LISA. The large SNR qualifies this as an early discovery/exclusion benchmark for LISA. 
    \item[BP(b)] represents physics at a scale of approximately $10^{11}~\mathrm{GeV}$ with a GW spectrum that peaks in the Hz regime, well within the reach of ET. However, LIGO-O5 is sensitive to its high-frequency tail with an SNR of approximately 29. This allows for its discovery or exclusion during the LIGO-O5 run, well before ET comes online. 
    \item[BP(c)] is a scenario that can also be tested at LIGO-O5. The GW spectrum peaks at tens of Hz and features the highest $\U{B-L}$ breaking scale of the three BPs. An observation at LIGO would imply a strong confirmation at ET with an SNR of $10^6$. 
\end{itemize}
\begin{figure*}[t]
	\centering
    \subfloat{\includegraphics[width=0.8\textwidth]{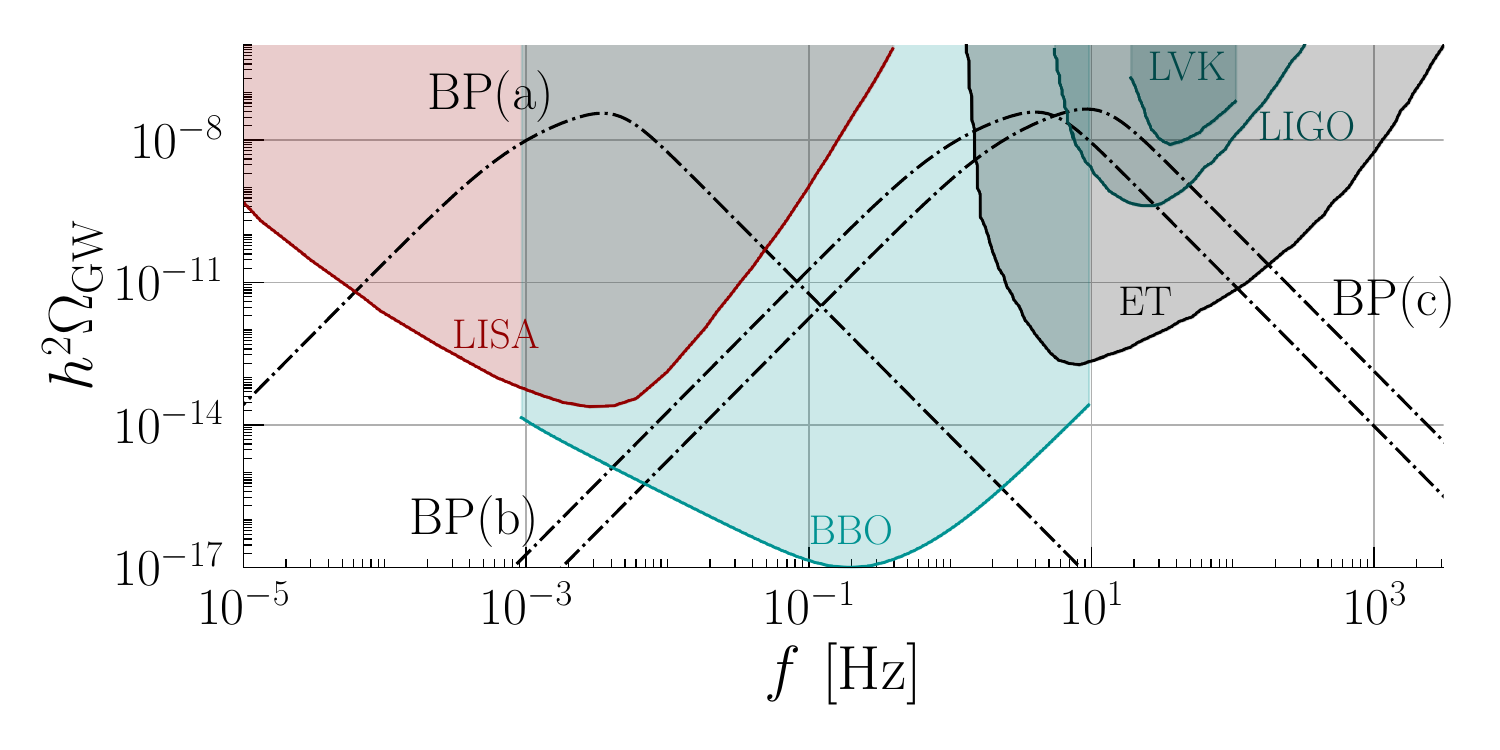}}
    \caption{\footnotesize SGWB spectra for three benchmark points (BPs) of the $\U{B-L}$ model whose GW spectra can be easily detected at LISA, ET and LIGO. The complete set of parameters for each BP can be found in \cref{tab:bench_HEP}. 
    }
	\label{fig:Spectrum_ng5Ligo}
\end{figure*}
All three BPs in \cref{fig:Spectrum_ng5Ligo} can be tested at  BBO.
\begin{table*}[h!]
\centering
\begin{tabular}{@{}rc|cccr@{}}
& \multicolumn{4}{c}{} \\
\hline
\hline
& & BP(a) & BP(b) & BP(c) &\\ \hline\hline
& $M_{h_2}$ & $2.91\times 10^{4}$ & $6.25\times10^{10}$ & $9.45\times10^{11}$ &\\
& $M_\mathrm{Z^\prime}$ & $2.69 \times 10^5$ & $5.24\times 10^{11}$ & $7.92\times 10^{12}$ &\\
& $M_{N_1}$ & $5.62\times 10^3$ & $4.11\times 10^9$ & $1.60\times 10^5$ &\\
& $M_{N_2}$ & $1.19 \times 10^4$ & $1.47\times 10^{10}$ & $1.989\times 10^7$ &\\
& $M_{N_3}$ & $3.65\times 10^4$ & $1.94\times 10^{10}$ & $2.25\times 10^9$ &\\
& $v_\sigma$ & $4.84\times 10^5$ & $8.57\times 10^{11}$ & $1.29\times 10^{13}$ &\\
& $g_L$ & $0.28$ & $0.31$ & $0.31$ & \\
& $\mathrm{Tr}(\bm{y_\sigma})$ & $0.16$ & $0.063$ & $2.5\times10^{-4}$ &\\
& $\sqrt{\mathrm{Tr}(\bm{y_\nu} \bm{y_\nu}^\dagger)}$ & $1.28\times 10^{-5}$ & $0.010$ & $0.0025$ &\\
& $\lambda_\sigma$ & $-0.025$ & $-0.11$ & $-0.13$ &\\
& $\lambda_{\sigma h}$ & $-7.62\times 10^{-8}$ & $-2.43\times 10^{-20}$ & $-1.07\times 10^{-22}$ &\\
\hline
& $\alpha$ & $5.37\times 10^{11}$ & $2.69\times10^{8}$ & $4.79\times10^{8}$ &\\
& $\beta/H(T_p)$ & $11.7$ & $11.4$ & $10.6$ &\\
& $T_p$ & $6.52$ & $5.19\times10^{4}$ & $1.11\times10^{5}$ &\\
& $T_\mathrm{RH}$ & $5594.42$ & $6.63\times 10^6$ & $1.64\times 10^7$ &\\
& $T_c$ & $2.02\times 10^4$ & $2.45\times 10^7$ & $6.11\times 10^7$ &\\
\hline
& SNR$^\mathrm{LIGO}$ & $1.89\times 10^{-8}$ & $29.4$ & $283.84$ &\\
& SNR$^\mathrm{ET}$ & $5.2\times 10^{-4}$ & $4.3\times 10^5$ & $1.11\times10^{6}$ &\\
& SNR$^\mathrm{LISA}$ & $2.15\times10^{5}$ & $0.06$ & $5.7\times 10^{-3}$ &\\
\hline\hline
\end{tabular}
\caption{ \footnotesize Model parameters at $\mu = M_{\mathrm{Z}}$, thermodynamic parameters, and SNR for the BPs in \cref{fig:Spectrum_ng5Ligo}. The $\U{B-L}$ breaking VEV $v_\sigma$, physical masses and temperatures are in units of GeV. 
}
\label{tab:bench_HEP}
\end{table*}

Current LVK data do not show evidence for a SGWB, either of cosmological or astrophysical origin. As shown in the scatter plots, numerous points fall within the LVK excluded region. In \cref{fig:GW_SNR_LVK}, we provide an estimate of the excluded region in the $(M_\mathrm{Z^\prime},g_L)$ plane (top row) and the $(M_{h_2},\lambda_\sigma)$ plane (bottom row).
\begin{figure*}[t]
	\centering
    \subfloat{\includegraphics[width=\textwidth]{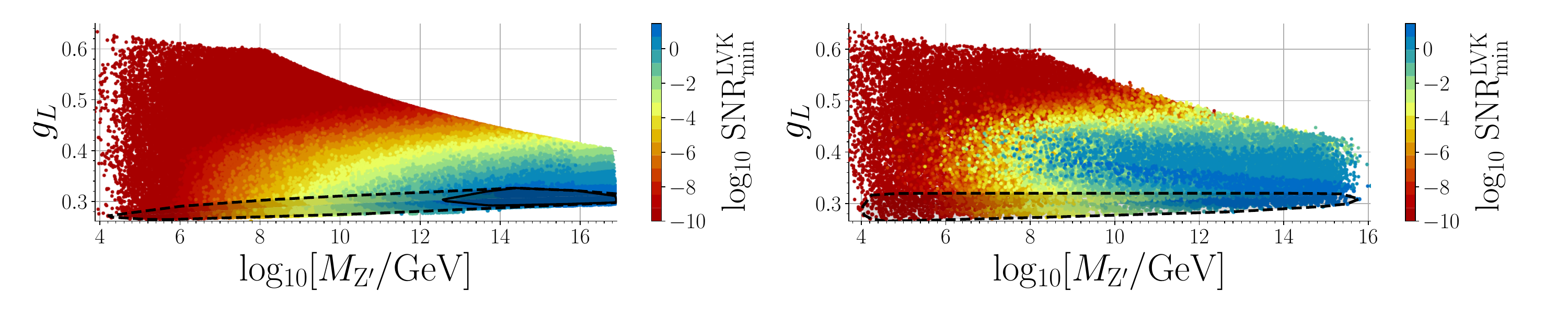}} \\[0.5em]
    \subfloat{\includegraphics[width=\textwidth]{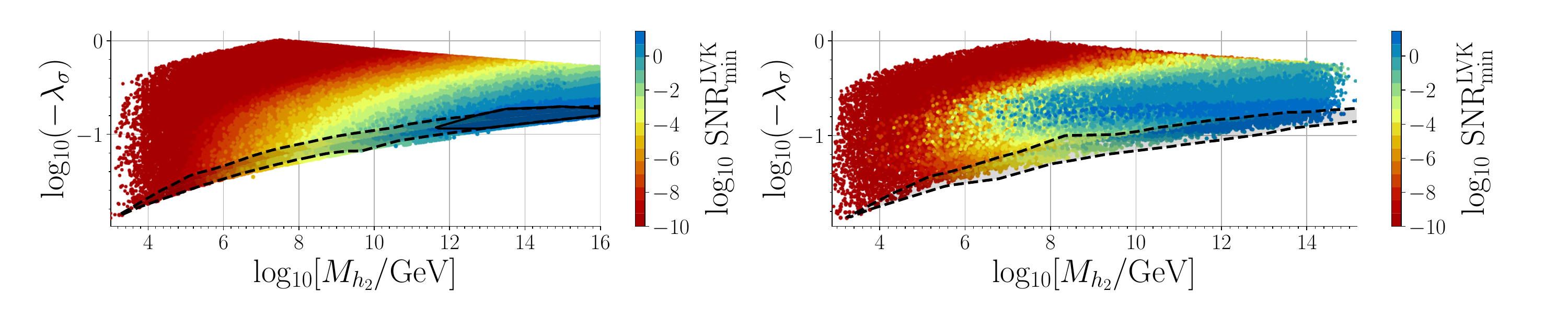}} \\[0.5em]
    \caption{\footnotesize Scatter plots for the $\U{B-L}$ model in the
    $(M_\mathrm{Z^\prime},g_L)$ and $(M_{h_2},\lambda_\sigma)$ planes for $\mathrm{Tr}(\bm{y_\sigma}) < g_L$ (left panels) and $\mathrm{Tr}(\bm{y_\sigma}) > g_L$ (right panels). The color scale represents the  minimum SNR at LVK. The area enclosed by the solid black contour is excluded by LVK since $\mathrm{SNR_{min}} > 10$. No regions are excluded by LVK in the right panels. For points enclosed by the dashed contour, percolation may occur at a temperature below $T_p$. }
	\label{fig:GW_SNR_LVK}
\end{figure*}
The shaded area within the solid black contour represents signals at LVK with a minimum SNR greater than 10. The region within the dashed contour has the same meaning as in \cref{fig:hGW_fpeak_proj_LowHigh_HEP}. Given the multi-dimensional nature of the parameter space, we consider a point in the $(M_\mathrm{Z^\prime}, g_L)$ and $(M_{h_2}, \lambda_\sigma)$ planes to be excluded if for all combinations of the other model parameters, the SNR is greater than 10. 
For instance, in the right panels, large Yukawa couplings lead to points of a fixed SNR spreading out. This results in the coexistence of, $e.g.$, blue and green points, with SNR values greater or smaller than 10 in the same region, which consequently, cannot be excluded.
In the right panels, the excluded region disappears because of the increased freedom provided by the large Yukawa couplings.

With the LIGO-O5 observation run it will be possible to test a broader region of the parameter space, as illustrated by the  area enclosed by the solid black curve in \cref{fig:GW_SNR_LIGO}. This run will explore scenarios compatible with $\mathrm{Z}^\prime$ masses down to $10^{11}~\mathrm{GeV}$ and $h_2$ masses down to $10^{10}~\mathrm{GeV}$ for $g_L \approx 0.3$ and $\lambda_\sigma \approx -0.1$. The high mass edge of the parameter space aligns with the LVK exclusion but can extend to $g_L \approx 0.4$ and $\lambda_\sigma \approx -0.5$.
\begin{figure*}[t]
	\centering
    \subfloat{\includegraphics[width=\textwidth]{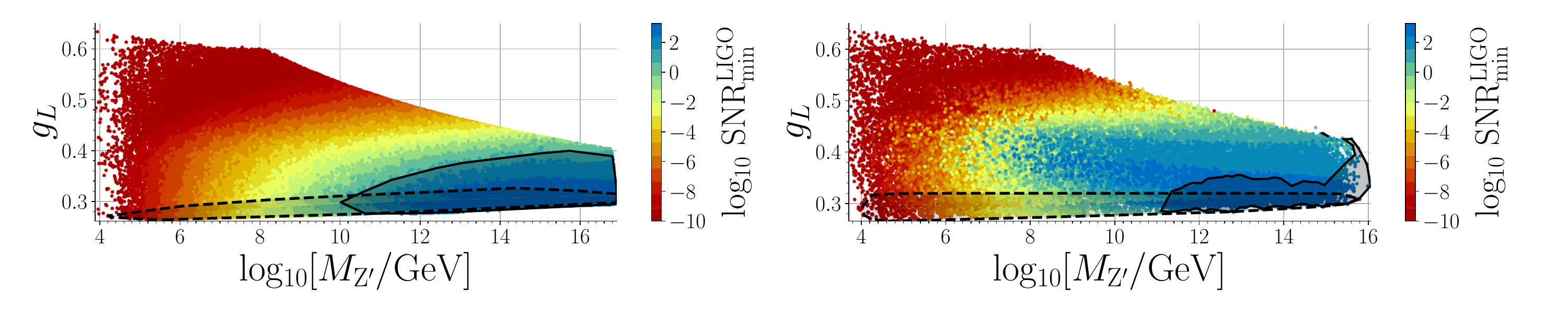}} \\[0.5em]
    \subfloat{\includegraphics[width=\textwidth]{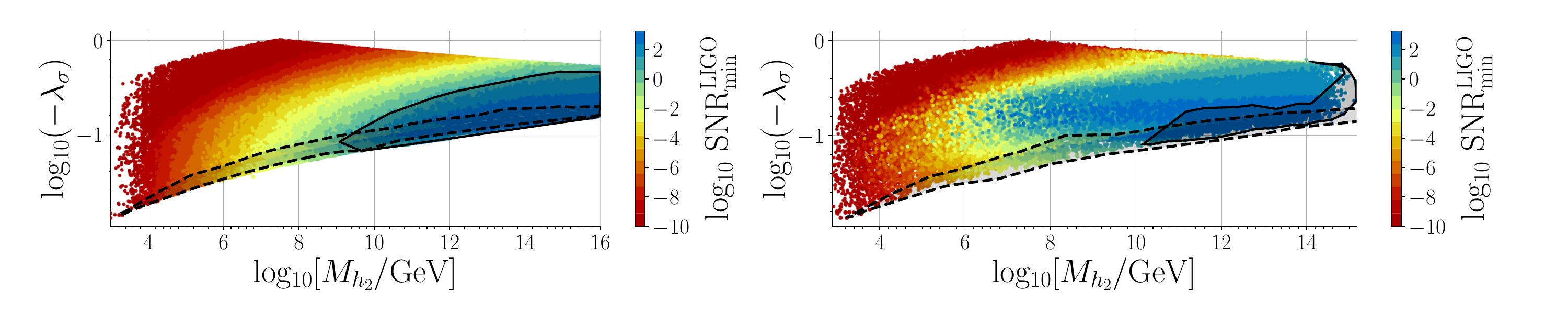}} \\[0.5em]
    \caption{\footnotesize 
    Similar to \cref{fig:GW_SNR_LVK}, but for a 4-year LIGO-O5 run. }
	\label{fig:GW_SNR_LIGO}
\end{figure*}

In the longer term, ET will explore a significantly larger region of parameter space, as shown in \cref{fig:GW_SNR_ET}. Specifically, we project sensitivity to a $\mathrm{Z}^\prime$ as light as $10~\mathrm{PeV}$ for $g_L \approx 0.3$, and to ${h_2}$ as light as 1~PeV for $\lambda_\sigma \approx -0.05$.
\begin{figure*}[t]
	\centering
    \subfloat{\includegraphics[width=\textwidth]{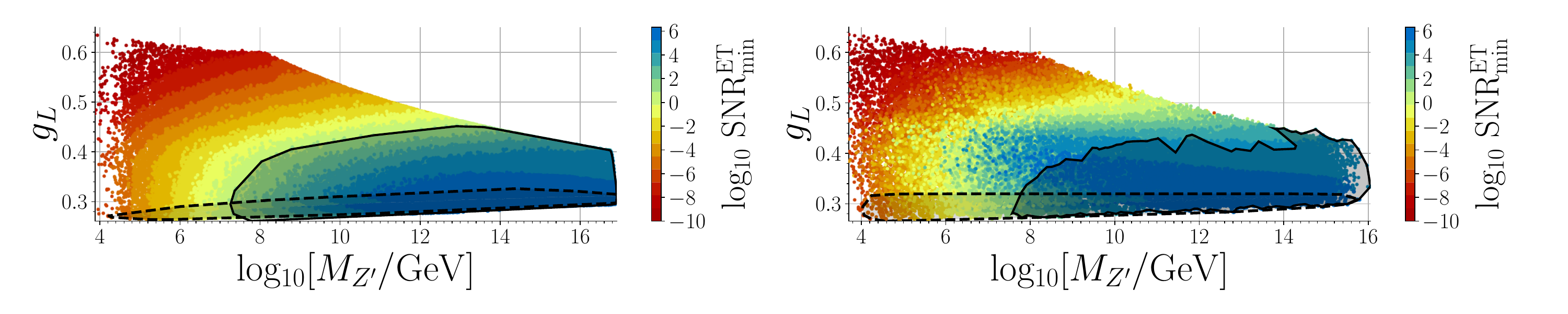}} \\[0.5em]
    \subfloat{\includegraphics[width=\textwidth]{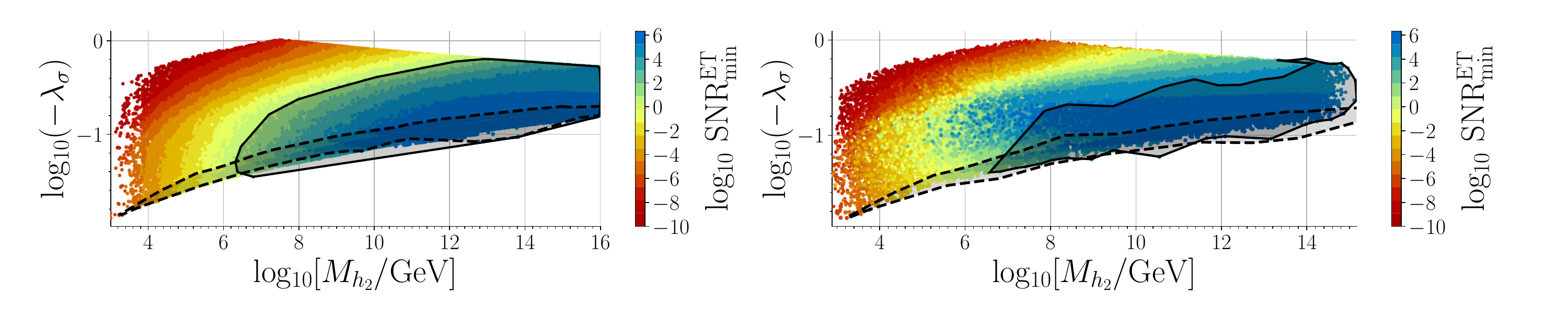}} \\[0.5em]
    \caption{\footnotesize 
    Similar to \cref{fig:GW_SNR_LVK}, but for ET.   }
	\label{fig:GW_SNR_ET}
\end{figure*}

LISA is complementary to LIGO and ET in that it will test the low mass edge of the parameter space, corresponding  to the smallest values of $\lambda_\sigma$. As shown in \cref{fig:GW_SNR_LISA}, a four-year exposure can probe $\mathrm{Z}^\prime$ masses from $10~\mathrm{TeV}$ to $10^{10}~\mathrm{GeV}$, and heavy Higgs boson masses from $1~\mathrm{TeV}$ to $10^{9}~\mathrm{GeV}$. 
\begin{figure*}[t]
	\centering
	\subfloat{\includegraphics[width=\textwidth]{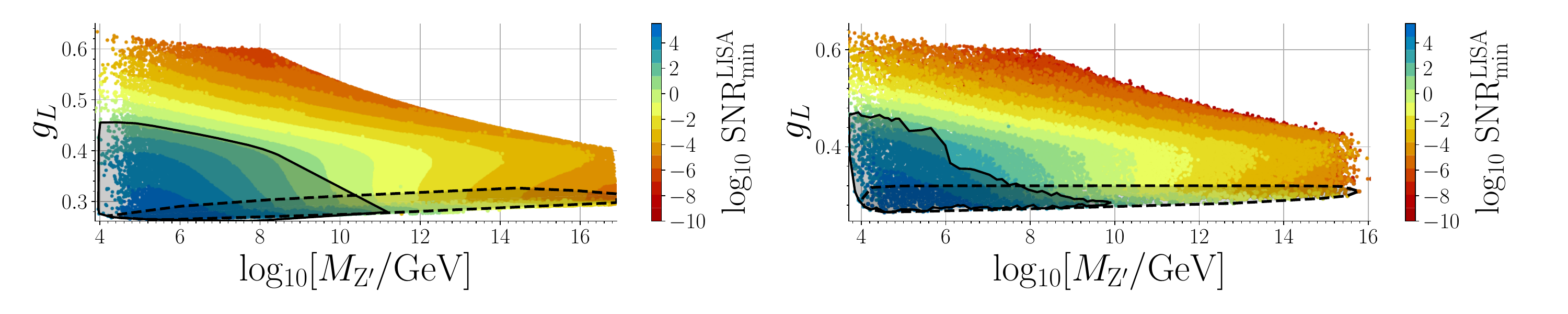}} \\[0.5em]
    \subfloat{\includegraphics[width=\textwidth]{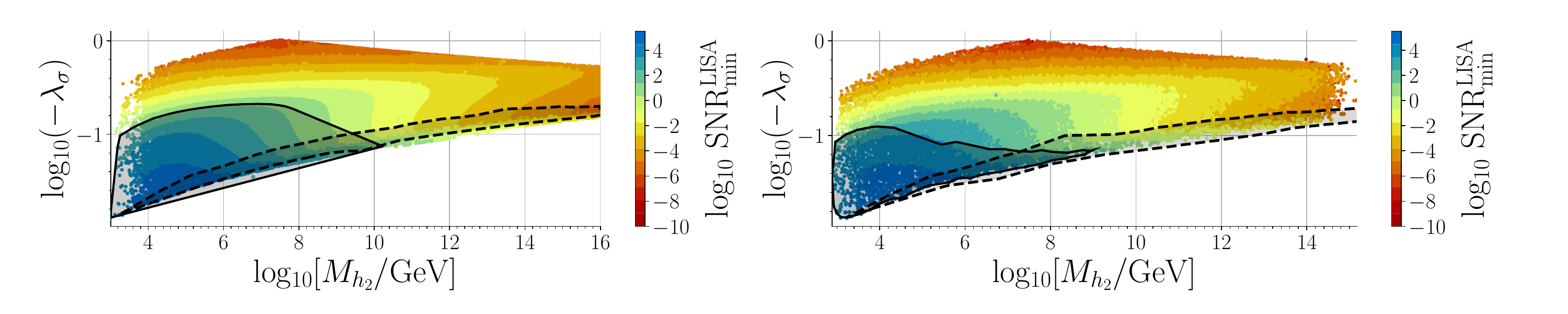}} \\[0.5em]
    \caption{\footnotesize 
    Similar to \cref{fig:GW_SNR_LVK}, but for LISA.  }
	\label{fig:GW_SNR_LISA}
\end{figure*}
Although LISA and ET do not operate in the same frequency band, we find many SGWB signals that peak in one experiment but with tails extending into the other, achieving SNR values above 10 in both experiments. Indeed, from \cref{fig:GW_SNR_ET,fig:GW_SNR_LISA} we observe overlapping sensitivity in the mass range, $10^{8}~\mathrm{GeV}$ to $10^{10}~\mathrm{GeV}$ for $M_{\mathrm{Z}^\prime}$, and $10^{7}~\mathrm{GeV}$ to $10^{9}~\mathrm{GeV}$ for $M_{h_2}$. 

We summarize the results of this subsection in Fig.~\ref{fig:GW_SNR_LVK_B-L}.
The color scales in the $(M_\mathrm{Z^\prime}, g_L)$ and $(\mathrm{max}[M_{N_i}], \mathrm{Tr}(\bm{y_\sigma}))$ planes represent the heavy Higgs mass, and the size of  $\bm{y_\nu}$, respectively. 
The top-left panel indicates that the $\U{B-L}$ model is excluded by the LVK bound for $M_\mathrm{Z^\prime} \approx 10 M_{h_2} > 10^{13}~\mathrm{GeV}$, with $g_L \approx 0.3$ and $\mathrm{Tr}(\bm{y_\sigma}) < \mathcal{O}(0.1)$. It also confirms that a wide range of $\mathrm{Z^\prime}$ masses can be tested for small $\mathrm{Tr}(\bm{y_\sigma})$.
In the bottom-left panel, there is no region with SNR$_{\rm {min}}> 10$ because the Yukawa couplings play a subdominant role in the phase transition. However, if no evidence for a SGWB with $\rm{SNR} >10$ is found at LISA, LIGO, and ET, strong supercooling defined by $g_L \lesssim 0.4$, will be disfavored for $\mathrm{Tr}(\bm{y_\sigma}) < \mathcal{O}(0.1)$. At LIGO and ET, strong supercooling will be tested for $\mathrm{Z}^\prime$ masses above $10^{9}$~GeV, and at LISA for masses below 100~TeV. Furthermore, the model can be fully excluded for $M_\mathrm{Z^\prime} \approx 10 M_{h_2} > 10^{14}$~GeV, as can be seen from the complete overlap of the dark purple contour and the colored region above $M_{\mathrm{Z}^\prime}=10^{14}$~GeV in the top panels. Correspondingly, the bottom-right panel shows that a high-scale seesaw mechanism characterized by right-handed neutrinos heavier than $10^{14.5}$~GeV and Yukawa couplings of $\mathcal{O}(1)$, can be excluded.  Note that for $\mathrm{Tr}(\bm{y_\sigma}) \gtrsim 0.4$, the LVK data do not exclude any parameter space, and that a nonobservation of a GW signal will exclude $\mathrm{Tr}(\bm{y_\sigma}) \approx 0.45$ and $g_L \approx 0.3$ in the entire mass range. 
\begin{figure*}[t]
	\centering
    \subfloat{\includegraphics[width=\textwidth]{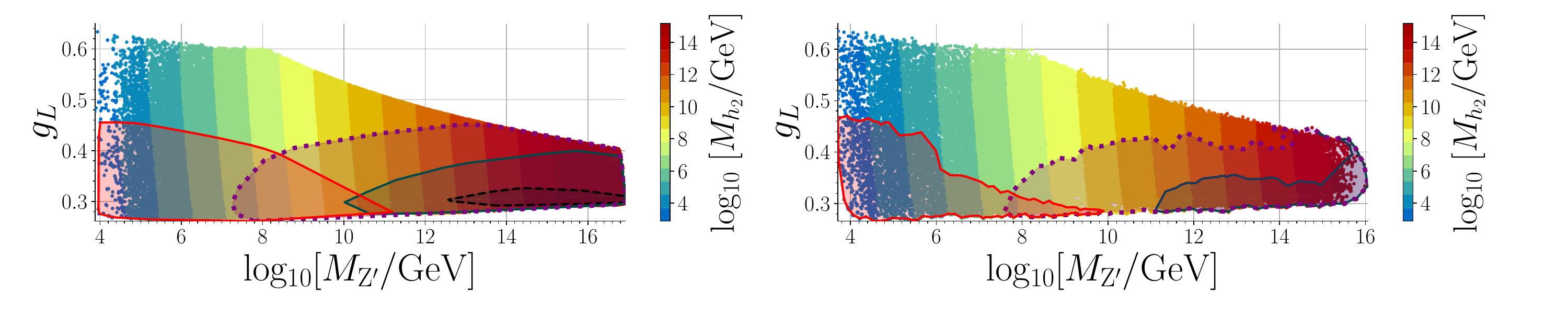}}\\ 
    \subfloat{\includegraphics[width=\textwidth]{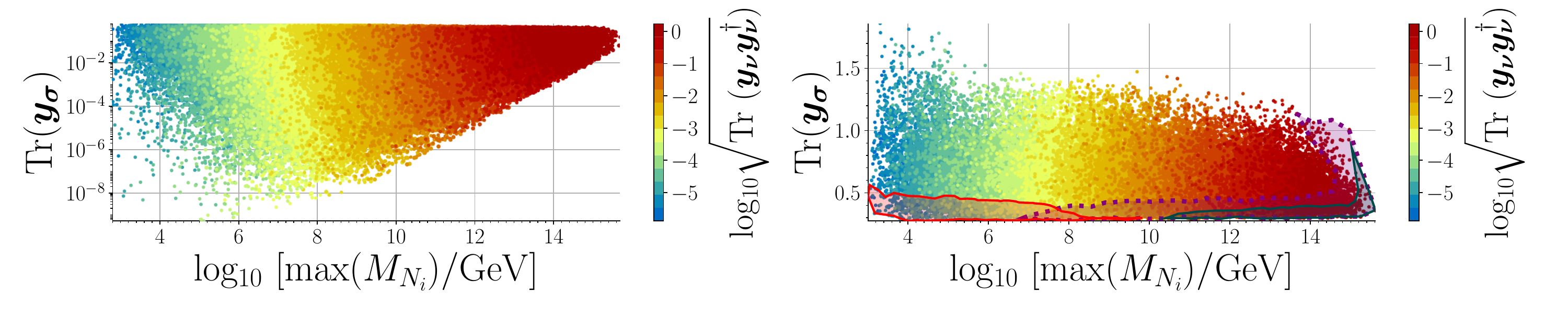}}
    \caption{\footnotesize 
    Scatter plots for generic $\U{B-L}$ model in the
    $(M_\mathrm{Z^\prime},g_L)$ and $({\rm max}(M_{N_i}),\mathrm{Tr}(\bm{y_\sigma}))$ planes for $\mathrm{Tr}(\bm{y_\sigma}) < g_L$ (left panels) and $\mathrm{Tr}(\bm{y_\sigma}) > g_L$ (right panels). The color scales represent the heavy Higgs mass (top row), and the magnitude of the Dirac Yukawa couplings, $\sqrt{\mathrm{Tr} (\bm{y}_\nu \bm{y}_\nu^\dagger)}$ (bottom row). The closed contours outline the regions with $\mathrm{SNR_{min}} > 10$ for LVK (dashed black), LIGO-O5 (solid green) LISA (solid red), and ET (dotted dark purple).}
	\label{fig:GW_SNR_LVK_B-L}
\end{figure*}

\subsubsection{Sources of gravitational waves}
\label{sec:SWBC}

In the context of strongly supercooled phase transitions, we neglect the contribution from turbulence and focus on sound waves and bubble collisions as the main sources of GWs. As \cref{eq:efficiency_SW} shows, these phenomena are interrelated such that, one dominates over the other in most cases, although they may contribute comparably. From \cref{fig:hGW_spectrum_BCSW}, it is evident that the contribution from sound waves dominates in most of the parameter space. The region where the sound wave contribution becomes negligible corresponds to scenarios in which percolation is not assured at $T_p$, but is possible at a lower temperature (defined as usual by the black dashed curve). 
\begin{figure*}[t]
	\centering
	\subfloat{\includegraphics[width=\textwidth]{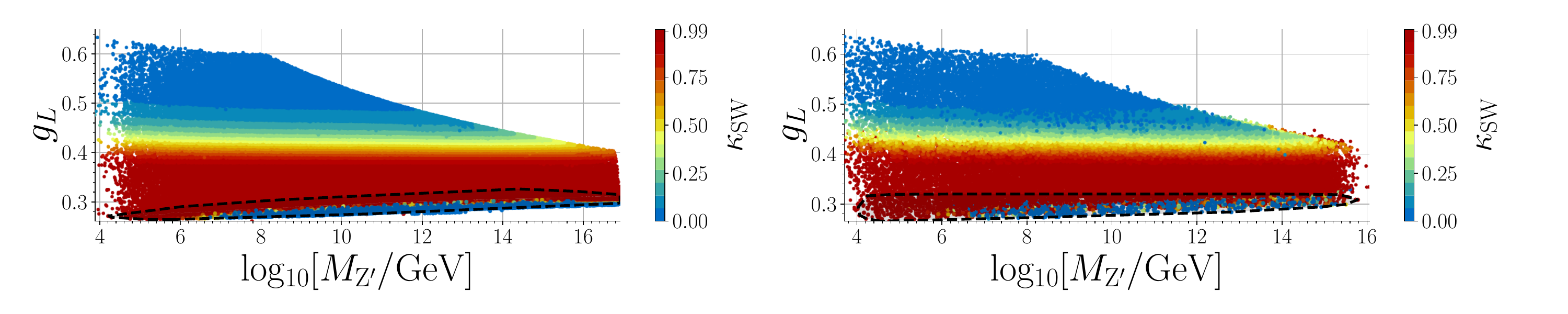}} \\
    \subfloat{\includegraphics[width=\textwidth]{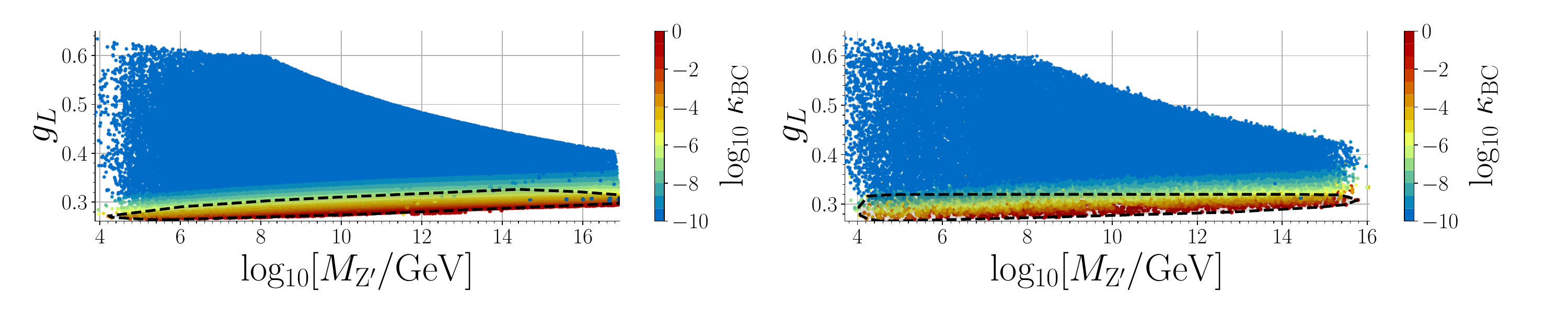}} \\
    \caption{\footnotesize 
    Similar to \cref{fig:GW_SNR_LVK}, but the color scales indicate the efficiency factors for sound waves (top row) and bubble collisions (bottom row). }
	\label{fig:hGW_spectrum_BCSW}
\end{figure*}
\begin{figure*}[t]
	\centering
	\subfloat{\includegraphics[width=\textwidth]{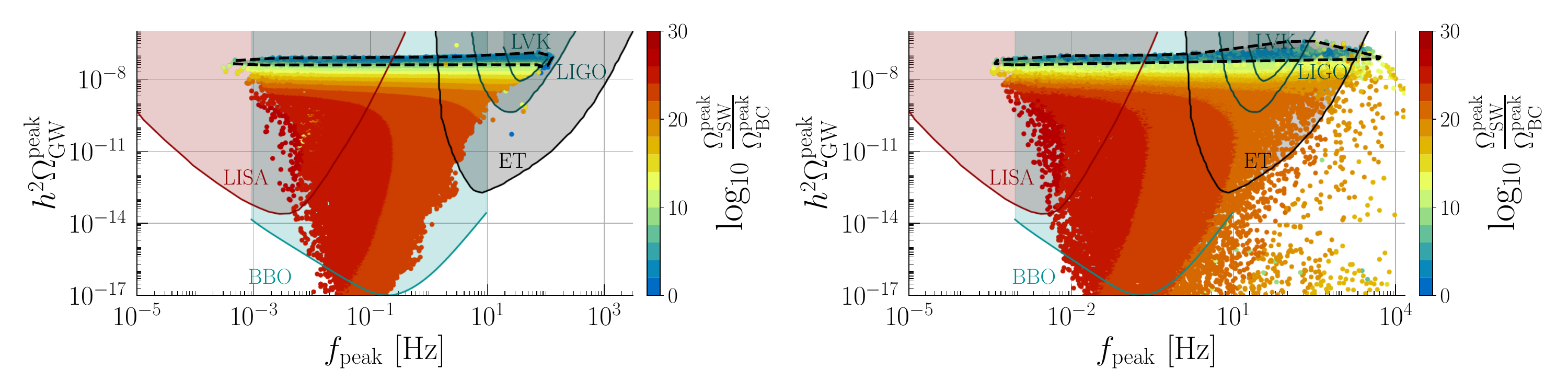}}
    \caption{\footnotesize 
    Similar to \cref{fig:hGW_fpeak_proj_LowHigh_HEP}, but the
    color scales indicate the ratio of the energy density in GWs contributed by sound waves and bubble collisions.  }
	\label{fig:SWoverBC}
\end{figure*}
Figure~\ref{fig:SWoverBC} shows that bubble collisions become the dominant source as $h^2 \Omega_\mathrm{GW}^\mathrm{peak}$ approaches $10^{-7}$. This corresponds to the highest values of $\alpha \gtrsim 10^{15}$, for which supercooling is maximal.

\subsubsection{Comparison with the literature}

To conclude this section, we compare our results with those of Ref.~\cite{Ellis:2020nnr}, which also studied GWs in the classical scale-invariant version of the B$-$L model. We find qualitative agreement in the thermodynamic parameters, but we obtain roughly an order of magnitude lower percolation temperatures. Furthermore, the stronger FOPTs that we obtain do not satisfy the percolation condition in \cref{eq:perc_condition}, whereas Ref.~\cite{Ellis:2020nnr} finds percolation to always occur at $T_p$. We attribute these differences to how the potential is minimized and how the scalar masses are calculated. Specifically, Ref.~\cite{Ellis:2020nnr} considers the RG-improved tree-level potential, $V = \frac{1}{4} \lambda_\sigma(t) \phi_\sigma^4$, whereas we include the Coleman-Weinberg contribution, $i.e.$, $V = \frac{1}{4} \lambda_\sigma(t) \phi_\sigma^4 + V_{\mathrm{CW}}(t,\phi_\sigma)$, and minimize it with all parameters defined at $\mu = M_\mathrm{Z^0}$, as was done in Ref.~\cite{Kierkla:2022odc}. We calculate the mass spectrum at one-loop, including the self-energies at $p^2 \neq 0$ and the second derivatives of the CW potential at $p^2 = 0$; see \cref{eq:mass_matrix_1loop}. Figure~\ref{fig:plot_CW} shows that the one-loop calculation 
not only shifts the minimum (left panel), but also significantly impacts the height of the potential barrier (right panel). While Ref.~\cite{Ellis:2020nnr} does not include Daisy corrections, we find numerically that their impact on the thermal corrections is marginal. Note that we accurately reproduce the results of Ref.~\cite{Ellis:2020nnr} by using its methodology. 
\begin{figure*}[t]
	\centering
	\subfloat{\includegraphics[width=\textwidth]{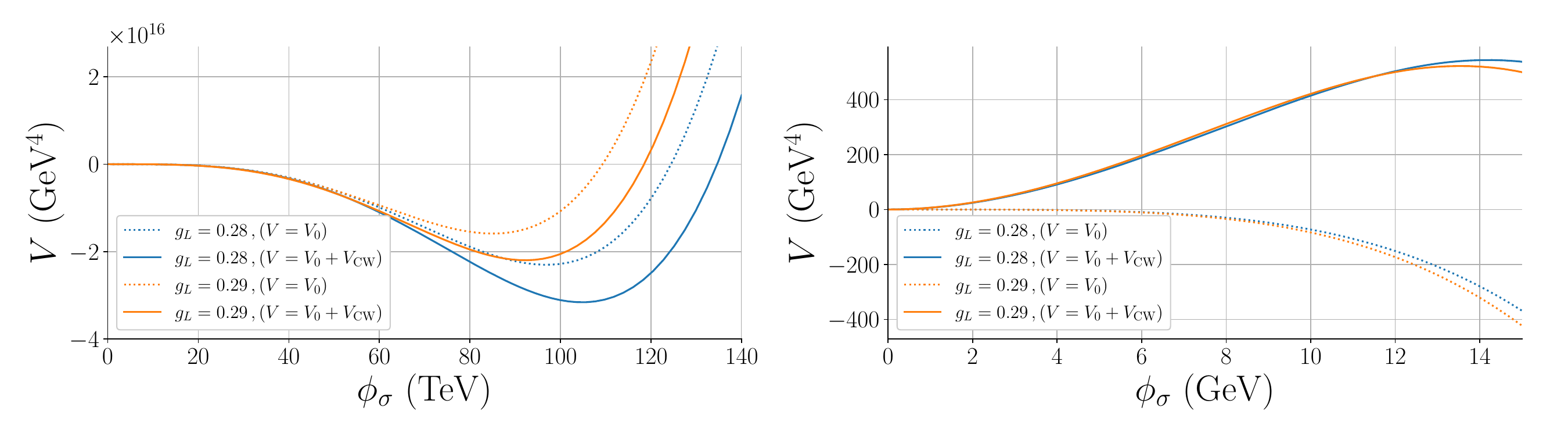}}
    \caption{\footnotesize 
    Comparison of the tree-level ($V = V_0$, dashed curves) and 
    one-loop corrected potential ($V = V_0 + V_\mathrm{CW}$ in \cref{eq:full_4DEFT}, solid curves) for BP(a) at $T=15~\mathrm{GeV}$. The left panel illustrates the behavior near the true minimum, and the right panel focuses on the potential barrier. }
	\label{fig:plot_CW}
\end{figure*}

A major difference is that we explore large masses, ranging from the TeV scale up to the Planck scale. This contrasts with Ref.~\cite{Ellis:2020nnr}, which does not consider masses above $10^8~\mathrm{GeV}$, but studies FOPTs below the QCD scale, which we do not. We have also analyzed the case of large right-handed neutrino Yukawa couplings, while Ref.~\cite{Ellis:2020nnr} neglects this contribution. Importantly, we have established a connection between neutrino physics and SGWB signals, a subject not discussed in Ref.~\cite{Ellis:2020nnr}. 

On a different note, the study of a minimal 
$\U{}$ conformal dark Higgs model~\cite{Gouttenoire:2023pxh} found that a 
matter-dominated period immediately after the phase transition reduces the peak frequency of the SGWB to the LISA sensitivity range, similar to our \cref{fig:hGW_fpeak_fake}. 
 In Ref.~\cite{Gouttenoire:2023pxh}, this happens because the heavy CP-even Higgs boson acts as a thermal inflaton that decays only to SM particles at a significantly suppressed rate via a small portal coupling. Notably, the inclusion of heavy neutrinos in Majoron models fundamentally alters this picture, allowing for GW signals at LIGO and ET.

\subsection{Scenarios with generic charge assignments}
\label{sec:generic_charges}

The $\U{B-L}$ model is a particular example of a broadly defined $\U{}^\prime$ gauge theory with arbitrary $x_\mathcal{H}$ and $x_\sigma$ charges. In this section, we examine generic charge assignments in the ranges in Table~\ref{tab:num_ranges_scan}.

In \cref{fig:spectrum_xHxS}, the color scale represents $g_L x_\sigma$ in the first and third rows, and $g_L x_\mathcal{H}$ in the second and fourth rows.
\begin{figure*}[t]
	\centering
    \subfloat{\includegraphics[width=0.9\textwidth]{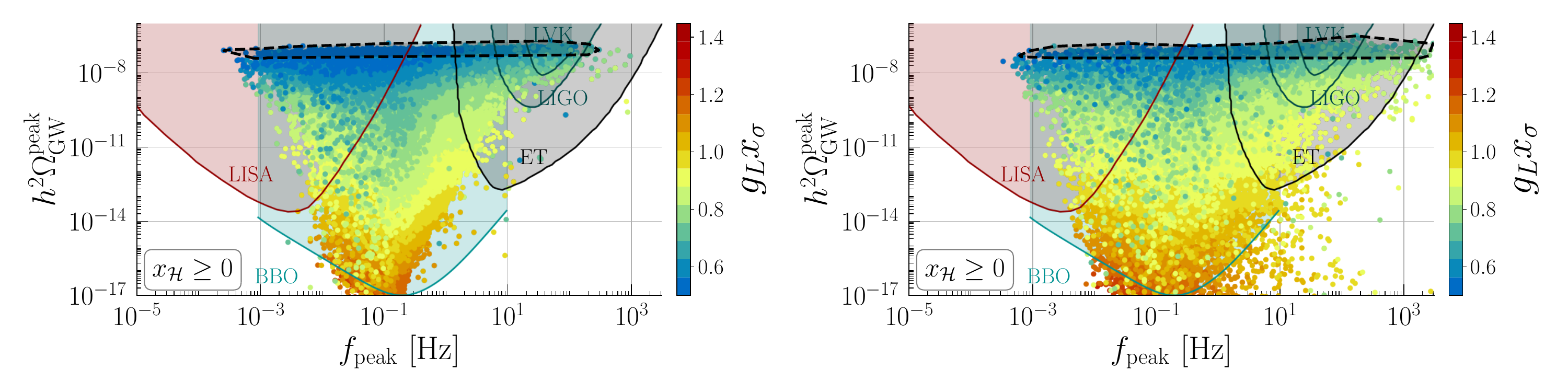}} \\
	\subfloat{\includegraphics[width=0.9\textwidth]{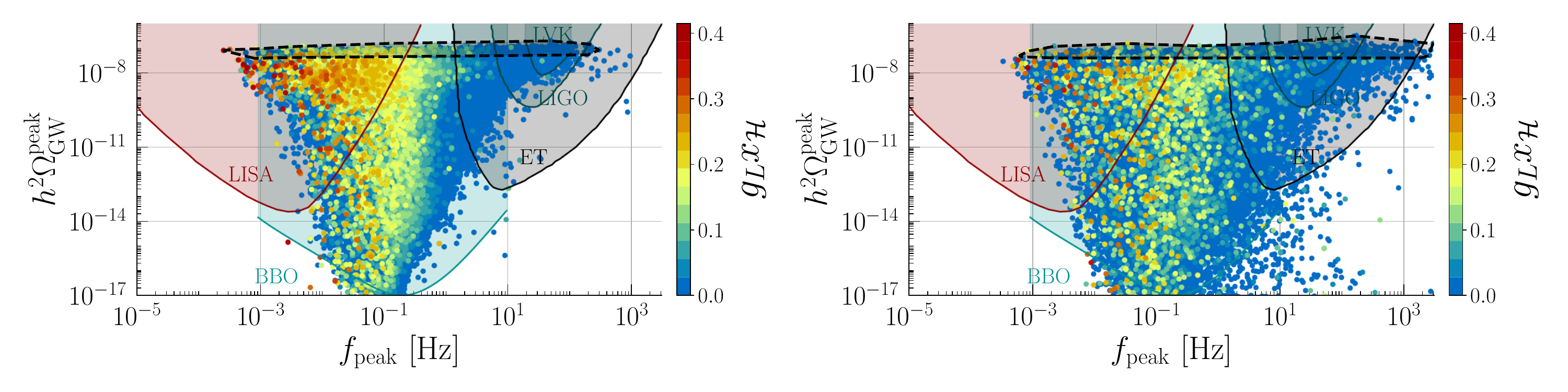}} \\
    \subfloat{\includegraphics[width=0.9\textwidth]{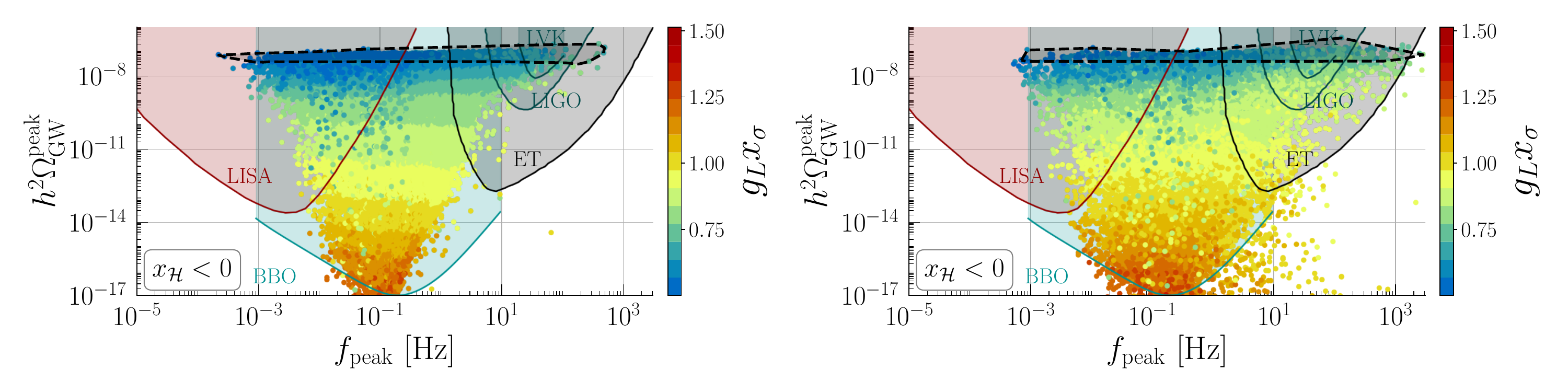}} \\
    \subfloat{\includegraphics[width=0.9\textwidth]{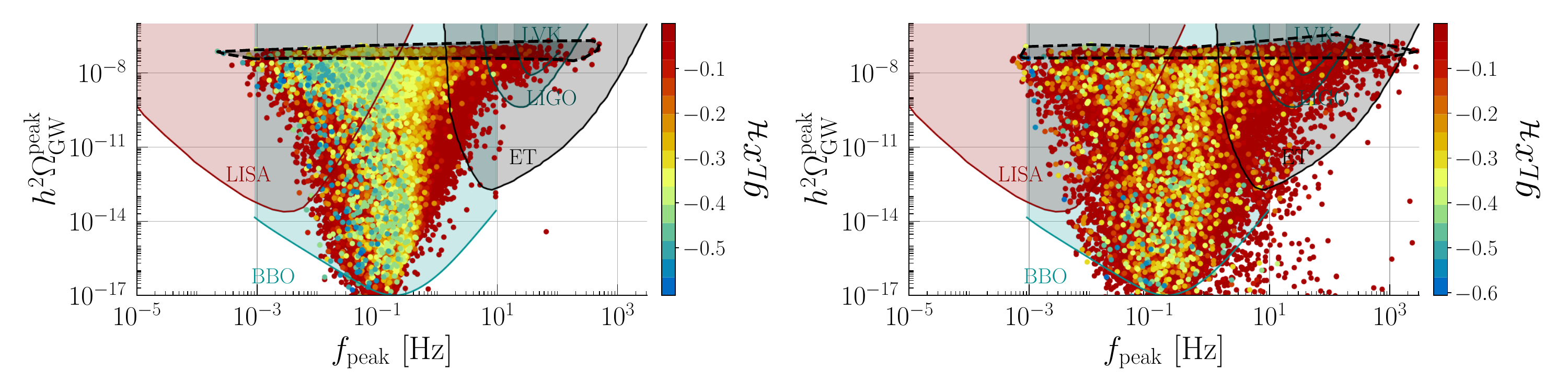}}
    \caption{\footnotesize 
    Similar to \cref{fig:hGW_fpeak_proj_LowHigh_HEP}, but for generic $\U{}^\prime$ models. The color scales represent the Majoron charge times $g_L$ in the first and third rows, and the Higgs doublet charge times $g_L$ in the second and fourth rows. In the top two rows $x_\mathcal{H} \geq 0$, and in the bottom two rows $x_\mathcal{H} < 0$. }
	\label{fig:spectrum_xHxS}
\end{figure*}
To fully visualize the parameter space, we divide our data into two sets. The top two rows have points with $x_\mathcal{H} \geq 0$, and the bottom two rows have points with $x_\mathcal{H} < 0$. From the first row, we observe that the qualitative behavior of $g_L x_\sigma$ is similar to that of $g_L$ in the $\U{B-L}$ case, in that a smaller gauge interaction strength increases the peak amplitude of the SGWB. This is expected as physical observables depend on $g_L x_\sigma$ and not just $g_L$, and the charge can be absorbed in the definition of the gauge coupling. In the second and fourth rows, we find that larger values of $|g_L x_\mathcal{H}|$ correspond to lower peak frequencies and higher peak amplitudes. This stems from the fact that with increasing values of the charges, the gauge and scalar quartic couplings run faster, reaching a Landau pole at lower scales. In the high-frequency region, where the true vacuum takes very large values, the emergence of Landau poles appears below the $\U{}^\prime$ breaking scale, which explains why $g_L x_\mathcal{H} \approx 0$ (dark purple and dark red points in the second and fourth rows). 

 Note that the dispersion of color in the $g_L x_\sigma$ plots, compared to the neatly horizontal distribution of $g_L$ for the $\U{B-L}$ model (fourth row of \cref{fig:hGW_fpeak_proj_LowHigh_HEP}), is caused by $g_L x_\mathcal{H}$. Indeed, there is a close correspondence between the color gradient in rows one and two and in rows three and four, where larger $|g_L x_\mathcal{H}|$ implies smaller $g_L x_\sigma$. This relationship arises from the leading effects in the $g_L$ beta function: $\beta^{(1)}(g_L) \approx \tfrac{g_L^3}{3}(82 x_\mathcal{H}^2 + 31 x_\mathcal{H} x_\sigma + 9 x_\sigma^2)$. A larger $\abs{g_L x_\mathcal{H}}$ must be compensated by a smaller $g_L x_\sigma$, and vice versa, to prevent $g_L$ from reaching a Landau pole. A smaller $g_L x_\sigma$ leads to a larger peak SGWB amplitude, because $\Delta V$ is larger, as discussed in connection with \cref{fig:potential_running}. Since the coefficient of the $x_\mathcal{H}^2$ term is nine times larger than that of the $x_\sigma^2$ term, a Landau pole is reached faster even for $x_\sigma \to 0$. 
 We exclude all points for which a coupling becomes non-perturbative at a scale $\mu < v_\sigma$. As in the B$-$L case, a sizeable $\mathrm{Tr}(\bm{y_\sigma})$ increases the spread of points and gives GW spectra with higher peak frequencies.

In \cref{fig:spectrum_MZMh2}, the color scale is for the $\mathrm{Z^\prime}$ boson and $h_2$ scalar masses. LIGO and ET are sensitive GUT scale masses. While the distribution of points is similar to $\U{B-L}$ case, regions of a given color are not as well defined. This is due to the freedom introduced by $x_\mathcal{H}$, which affects both the heavy Higgs and $\mathrm{Z}^\prime$ masses. For a generic $\U{}^\prime$ model, LISA will be sensitive to $h_2$ masses ranging from $1~\mathrm{TeV}$ to $10^{9}~\mathrm{GeV}$, with the corresponding $\mathrm{Z^\prime}$ masses an order of magnitude larger. For a given mass, the peak frequency is lower
for $\mathrm{Tr}(\bm{y_\sigma}) < g_L$ (left panels) 
than for $\mathrm{Tr}(\bm{y_\sigma}) > g_L$ (right panels). 
 This shift is more pronounced for small $g_L x_\sigma$ and large $\abs{g_L x_\mathcal{H}}$, because the $x_\mathcal{H}^2$ term dominates the RG evolution of $g_L$, so that the tree-level potential $V_0$ is minimized at a lower VEV. This effect is marginal in the $\mathrm{B-L}$ scenario since only  $\bm{y_\sigma}$ modifies $\lambda_\sigma$.  Points with $\mathrm{Tr}(\bm{y_\sigma}) \ll g_L x_\sigma$, have a cleaner distribution  because $g_L x_\mathcal{H} \approx 0$, as observed in \cref{fig:spectrum_xHxS}. However, for $\mathrm{Tr}(\bm{y_\sigma}) > g_L x_\sigma$, the frequency coverage is broader for both high- and low-frequency experiments.
\begin{figure*}[t]
	\centering
    \subfloat{\includegraphics[width=\textwidth]{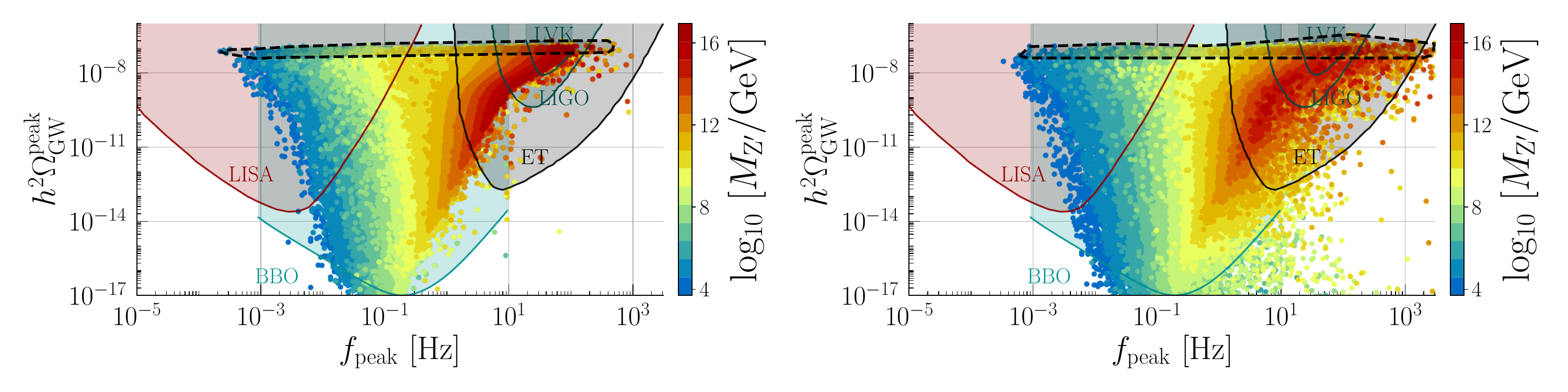}} \\
    \subfloat{\includegraphics[width=\textwidth]{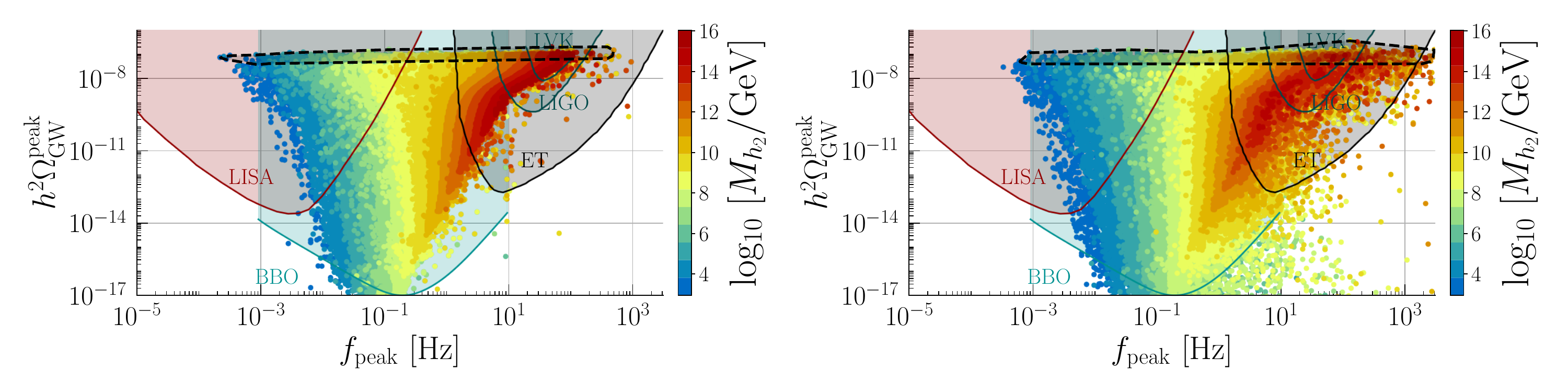}} \\
    \caption{\footnotesize 
    Similar to \cref{fig:spectrum_xHxS}, but the color scales represent the $\mathrm{Z^\prime}$ boson mass (top row), the heavy scalar mass $M_{h_2}$ (bottom row).  }
	\label{fig:spectrum_MZMh2}
\end{figure*}

In \cref{fig:GW_GenCharges_xHxS} we show the $(x_\mathcal{H}, g_L x_\sigma)$ plane for several thermodynamic parameters. Each value of $x_\mathcal{H}$ defines a different $\U{}^\prime$ model.
\begin{figure*}[t]
	\centering
	\subfloat{\includegraphics[width=\textwidth]{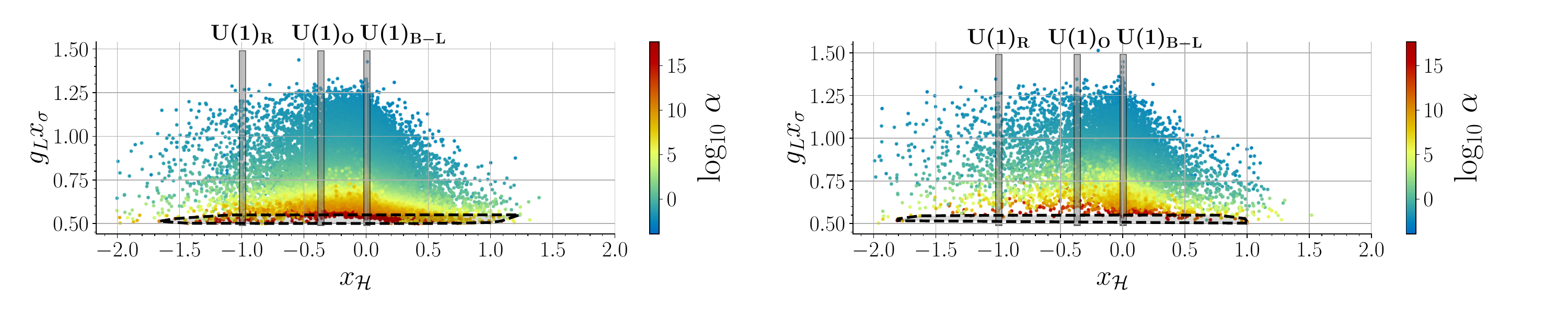}} \\
    \subfloat{\includegraphics[width=\textwidth]{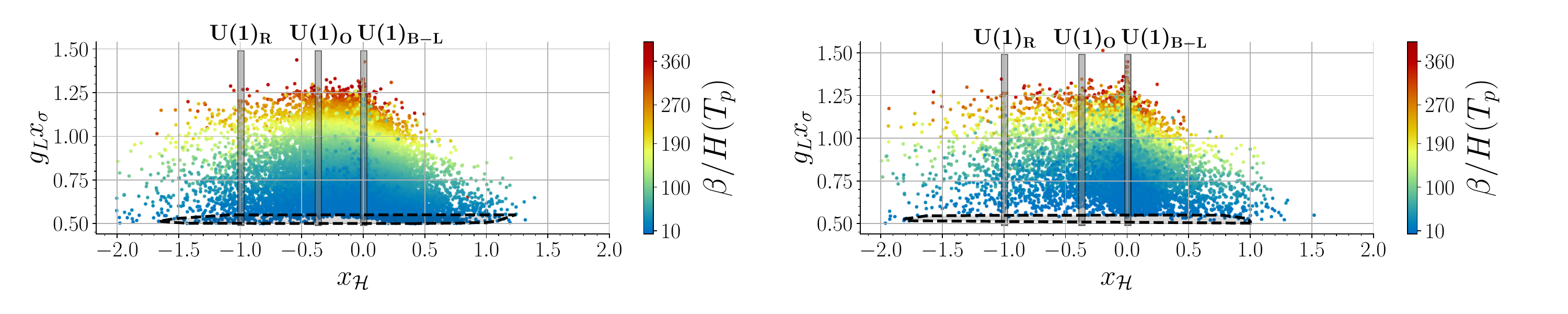}} \\
    \subfloat{\includegraphics[width=\textwidth]{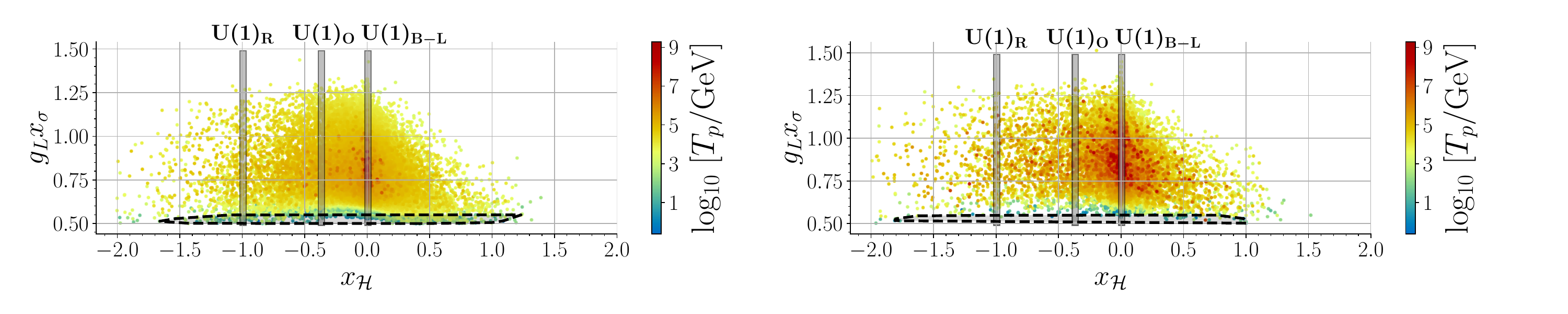}} \\
    \subfloat{\includegraphics[width=\textwidth]{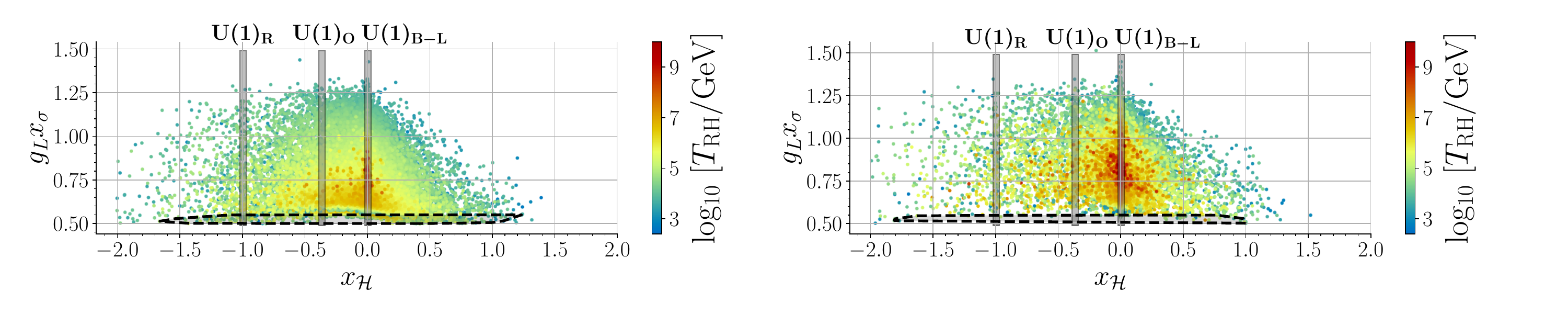}} \\
    \caption{\footnotesize Scatter plots in the $(x_\mathcal{H}, g_L x_\sigma)$ plane for generic $\U{}^\prime$ models. The
    color scales indicate the phase transition strength $\alpha$ (first row), and its inverse time duration $\beta/H(T_p)$ (second row), the percolation temperature $T_p$ (third row), and the reheat temperature $T_{\rm RH}$ (fourth row).
     The $\U{R}$, $\U{B-L}$, and $\U{O}$ models correspond to charges $(x_\mathcal{H}, x_\sigma) = (-1, 2)$, $(0, 2)$, and $(-16/41, 2)$, respectively. The black dashed contour has the same meaning as in previous figures. The left (right) panels show points with $\mathrm{Tr}(\bm{y_\sigma}) < g_L$ ($\mathrm{Tr}(\bm{y_\sigma}) > g_L$). }
	\label{fig:GW_GenCharges_xHxS}
\end{figure*}
We identify three different models: the $\U{B-L}$ model; the $\U{R}$ model with the charge assignment $(x_\mathcal{H},x_\sigma) = (-1,2)$; and the $\U{O}$ model with {\it e.g.,} $(x_\mathcal{H},x_\sigma) = (-16/41,2)$, 
which yields the orthogonality condition between $\U{Y}$ and $\mathrm{U(1)}^\prime$, $41 x_\mathcal{H} + 8 x_\sigma = 0$, so that the kinetic mixing $g_{12}$ does not evolve with energy at one-loop if $g_{12}=0$ at some scale $\mu$ (see Eq.~\eqref{eq:betaf_g12}).
The thermodynamic parameters $\alpha$, $\beta/H(T_p)$, and $T_p$ are almost independent of $x_\mathcal{H}$.
Consequently, the SGWB geometric parameters are also weakly dependent on $x_\mathcal{H}$, as shown in \cref{fig:spectrum_xHxS}, and it is not possible to exclude a specific $\U{}^\prime$ model based on GW data alone. However, for models with large $|x_\mathcal{H}|$
(which suffer a loss of perturbativity below $v_\sigma$), the allowed parameter space shrinks so that the density of points with higher $T_p$ and $T_\mathrm{RH}$ is lower far from $x_\mathcal{H}=0$,  and leads to lower peak frequencies.

In \cref{fig:xH_variation}, we show GW spectra for different $x_\mathcal{H}$ charges with the other parameters fixed.
\begin{figure*}[t]
	\centering
	\subfloat{\includegraphics[width=0.5\textwidth]{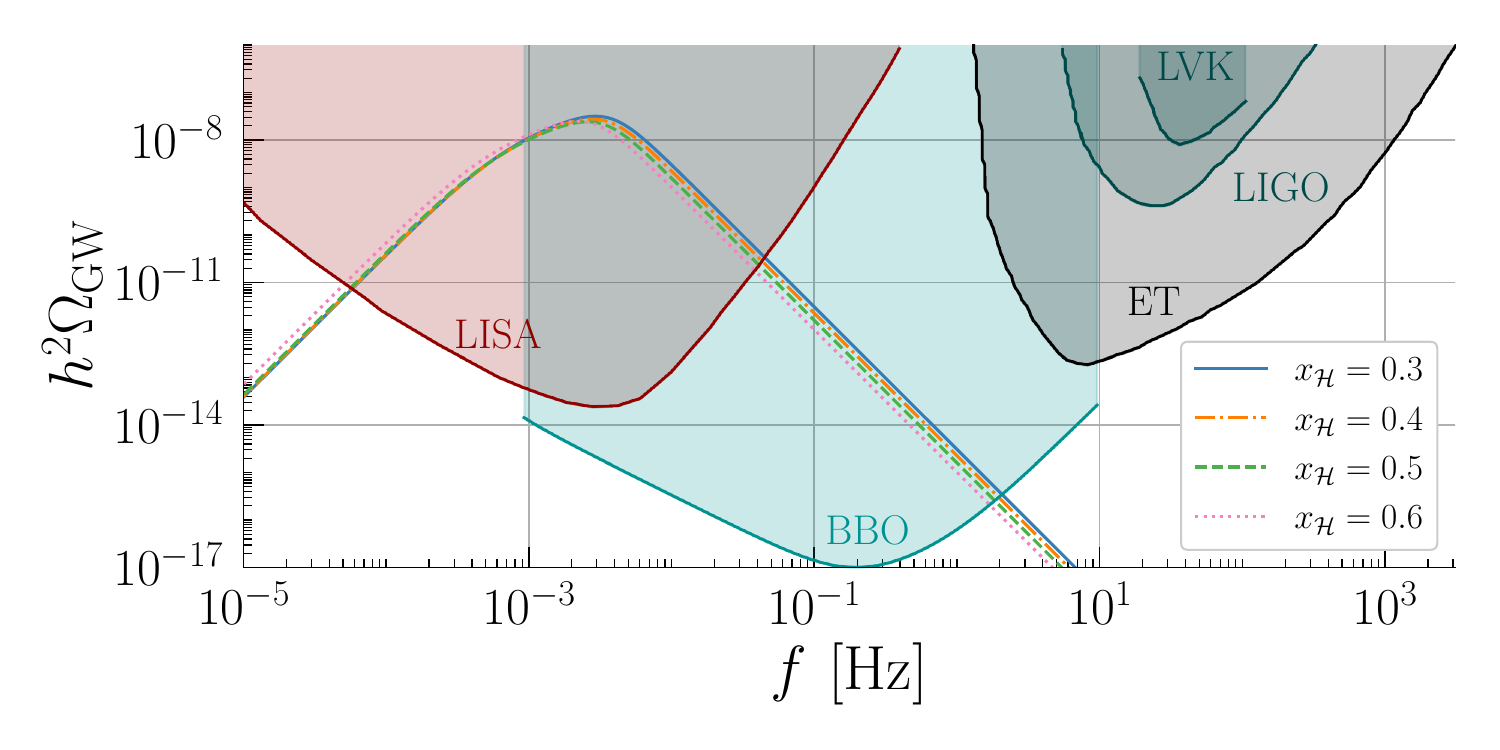}} 
    \subfloat{\includegraphics[width=0.5\textwidth]{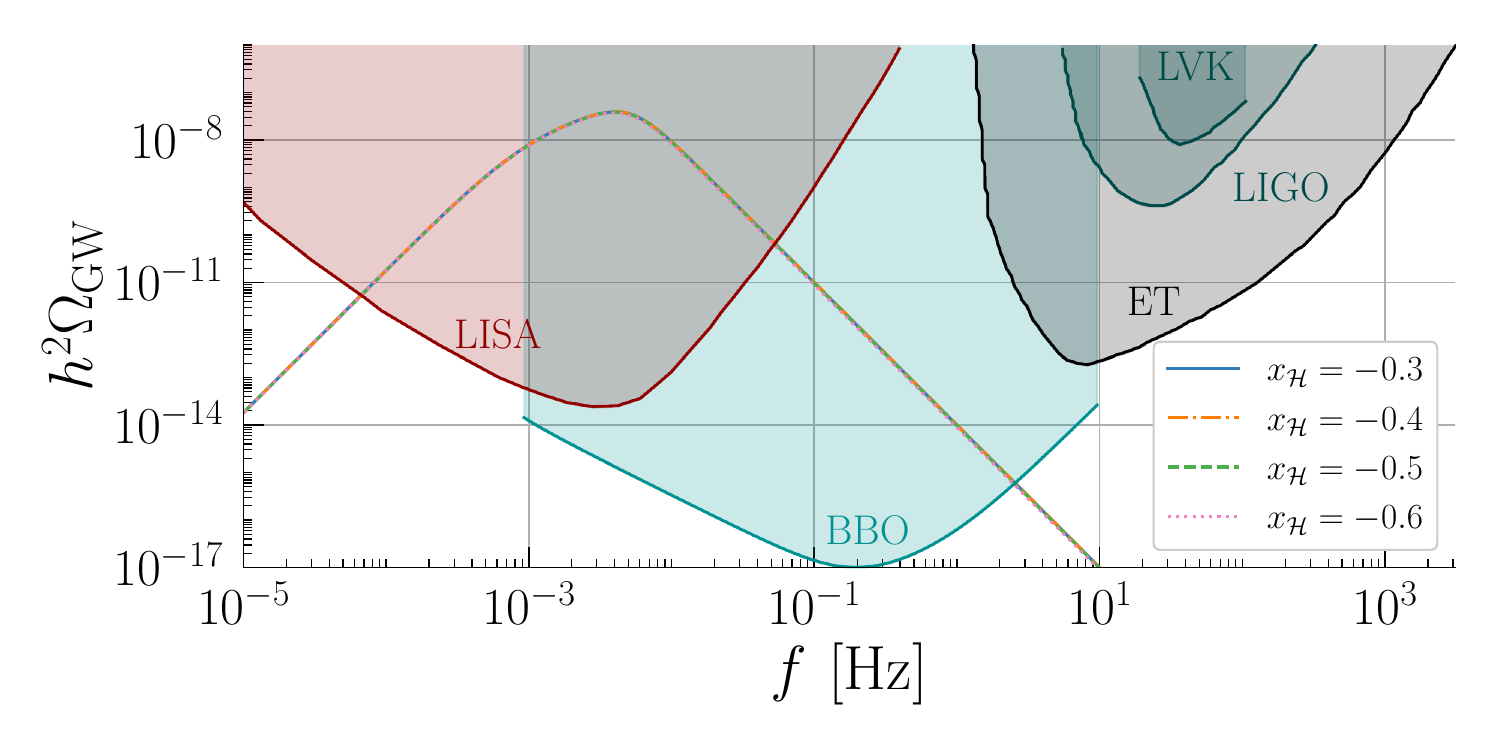}}
    \caption{\footnotesize 
    SGWB spectrum for various values of the Higgs charge in generic $\U{}^\prime$ models. The other parameters are fixed to those of BP(a). }
	\label{fig:xH_variation}
\end{figure*}
We observe that as $x_\mathcal{H}$ decreases, the spectrum shifts towards lower frequencies while maintaining an approximately constant peak amplitude. This shift can be attributed to changes in the percolation temperature due to the modified running of the $g_L$ $\beta$-function for different $x_\mathcal{H}$. However, the shift in the spectrum is small compared to the theoretical uncertainties in \cref{fig:GW_uncer}.

In \cref{fig:GW_SNR_LVK_gen}, we present scatter plots in the $(M_{\mathrm{Z'}}, g_L x_\sigma)$ plane, with the color scale representing the heavy Higgs mass.
\begin{figure*}[t]
	\centering
    \subfloat{\includegraphics[width=\textwidth]{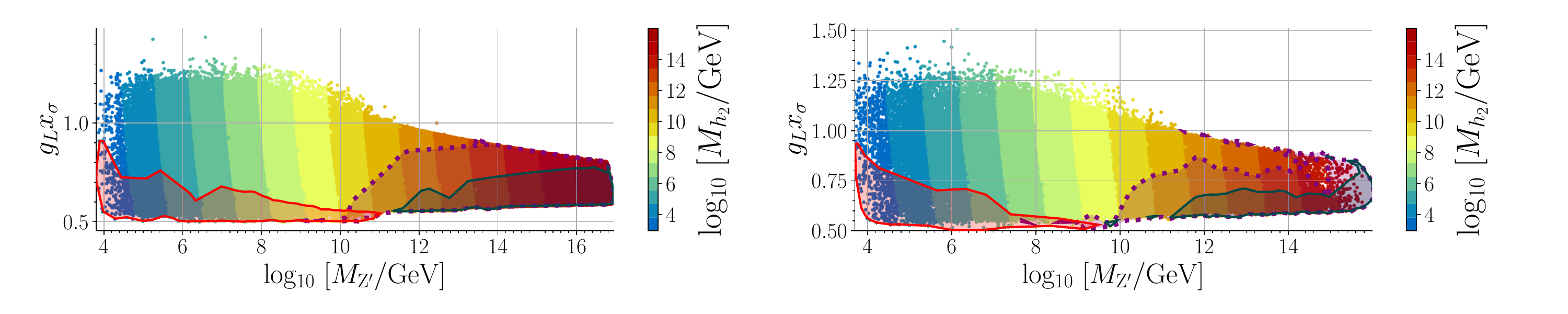}} \\
    \subfloat{\includegraphics[width=\textwidth]{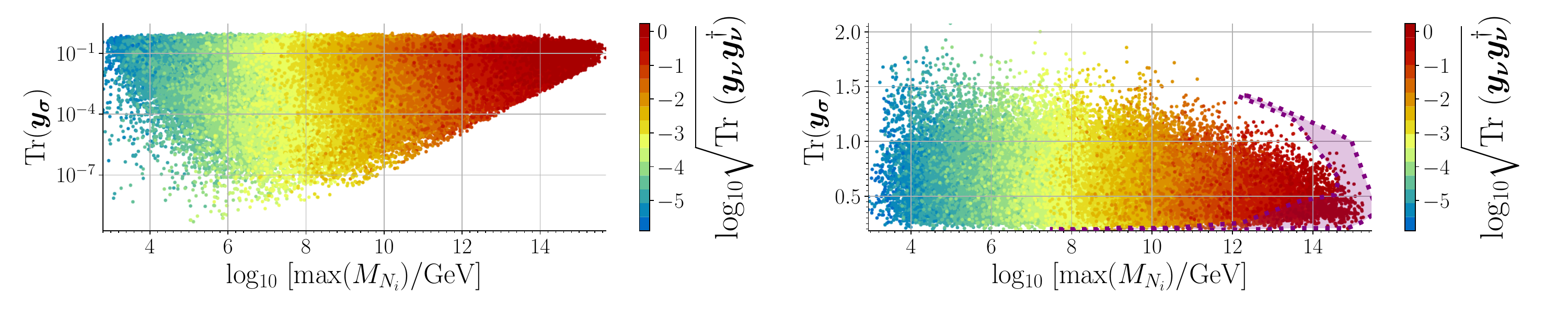}}\\
    \caption{\footnotesize 
    Similar to \cref{fig:GW_SNR_LVK_B-L}, but for generic $\U{}^\prime$ models. 
   }
	\label{fig:GW_SNR_LVK_gen}
\end{figure*}
At high frequencies, LVK data do not exclude any region of parameter space due to the dispersion caused by $x_\mathcal{H}$. The excluded regions for the $\mathrm{Z}^\prime$ and heavy Higgs bosons are similar, but slightly weaker than for the $\U{B-L}$ model. The region within the solid green (dotted dark purple) boundary will be tested by  LIGO-O5 (ET) data with $\mathrm{SNR} > 10$. This figure confirms that Earth-based interferometers will be sensitive to heavy boson masses from $10^{10}~\mathrm{GeV}$ up to the GUT scale, and $g_L x_\sigma$ between 0.5 and 0.9. At lower frequencies, LISA will also cover a wide range of masses (enclosed by the solid red line). The overlapping sensitivity of ET and LIGO is evident in the top-left panel demonstrating an opportunity to test this class of models in more than one experiment. For $\mathrm{Tr}(\bm{y_\sigma}) > g_L$, the spread in points of a given color shrinks the area with $\mathrm{SNR} > 10$.

\section{Summary}\label{sec:conclusions}

We explored the potential to indirectly test a class of classically scale-invariant $\U{}^\prime$ models at GW experiments. We discussed how the parameter space of these models can be studied at energy scales far beyond the reach of collider experiments. Strongly supercooled FOPTs produce a SGWB with a large SNR across a wide range of frequencies, from 0.1~mHz, within reach of LISA, up to a ~kHz, within reach of LIGO and ET.

We examined the $\U{B-L}$ model with gauge charges, $x_\mathcal{H} = 0$ and $x_\sigma = 2$. We find that the peak amplitude of the SGWB is primarily influenced by the gauge coupling $g_L$, which must be in the range $0.26 \lesssim g_L \lesssim 0.63$ for FOPTs to occur; see the fourth row of \cref{fig:hGW_fpeak_proj_LowHigh_HEP}, which also shows that the peak amplitude reaches its highest values, $h^2 \Omega_\mathrm{GW}^\mathrm{peak} \approx 10^{-7}$, for lower values of the $g_L$ range across the entire frequency spectrum. The $\U{B-L}$ breaking scale, which determines the masses of the $\mathrm{Z'}$ and $h_2$ bosons, governs the peak frequency, as can be seen from the second and third rows of \cref{fig:hGW_fpeak_proj_LowHigh_HEP}. We confirmed that a potential barrier between the true and false vacuum develops at finite temperatures for a non-zero gauge coupling; see \cref{fig:potential_vs_temperature} and \cref{eq:potential_HighT}. Generally, the sound wave contribution is the dominant source of GWs in most of the parameter space, although for $\alpha \gtrsim 10^{13}$,  bubble collisions become dominant. For such high values of $\alpha$, percolation is not guaranteed, but may occur at temperatures below  $T_p$.

The heavy neutrinos that participate in the type-I seesaw mechanism, play an important role in the production of GW
spectra that peak at high frequencies.
Large $\bm{y_\sigma}$ can compete with the gauge coupling in the development of a global minimum of the potential, thereby increasing the VEV and the potential energy difference between the true and false vacuum, and pushing the peak frequency of the GW spectrum into the kHz range; see the right panels of \cref{fig:hGW_fpeak_proj_LowHigh_HEP,fig:hGW_fpeak_proj_LowHigh_Th,fig:hGW_fpeak_Yukawas,fig:SWoverBC}. We find that if the $\U{}^\prime$ and SM sectors are in thermal equilibrium, the primary influence of the neutrino sector on the FOPT arises from the $h_2 \to \bar{N_i} N_i$ decay rate, which exceeds the Hubble rate at percolation in most of the parameter space, which makes the Universe promptly enter the radiation-dominated era after percolation; see \cref{fig:GW_GammaH}. With decoupled heavy neutrinos, $h_2$ decays only to SM particles, causing $h_2$ to oscillate around the true minimum for longer than the Hubble time. Consequently, the Universe  enters an early matter-dominated period after percolation, leading to a substantial suppression of the SGWB, with peak frequencies not exceeding 0.1~Hz, well below the sensitivity of Earth-based interferometers like ET and LIGO; see \cref{fig:hGW_fpeak_fake}. Thus, in the context of these models, the observation of high-frequency GWs can be interpreted as a signature of heavy neutrinos. We also find a clear correlation between the peak frequency of the SGWB and the magnitude of $\bm{y_\nu}$; see \cref{fig:hGW_fpeak_Yukawas}. Specifically, ET and LIGO will probe $\bm{y_\nu}$ between $10^{-3}$ and unity, and LISA will probe $\bm{y_\nu}$ between $10^{-6}$ to $10^{-3}$. If $\bm{y_\sigma} \sim \mathcal{O}(1)$, the heavy neutrino mass scale in the type-I seesaw mechanism can also be probed with high frequency GWs. 

We demonstrated the importance of computing the RG-improved potential at the one-loop level because it shifts the minimum and significantly impacts the height of the potential barrier compared to a tree-level calculation~\cite{Ellis:2020nnr}; see \cref{fig:plot_CW}. The larger potential energy difference obtained at one-loop increases $\alpha$ and consequently the degree of supercooling. This also explains why our percolation temperatures are generally an order of magnitude lower than in Ref.~\cite{Ellis:2020nnr}.

For generic charges, our findings align with the results of the B$-$L model. However, higher values of the Higgs doublet charge $x_\mathcal{H}$ are less favorable due to the additional contributions they introduce in the $\beta$ functions, which can lead to Landau poles at lower energies. Consequently, a signal at high frequencies will favor $\U{B-L}$; see \cref{fig:spectrum_xHxS}. Due to the weak dependence of the FOPT on the Higgs doublet charge, the SGWB is not sensitive to $x_\mathcal{H}$. A comparison of  \cref{fig:xH_variation,fig:GW_uncer} shows that the effect of $x_\mathcal{H}$ on the SGWB is minimal, with the resulting frequency shift being much smaller than theoretical uncertainties. Also note that unlike non-Abelian models, in $\U{}^\prime$ models the RG evolution of $g_L$ is asymptotically safe at low energies, so that the percolation temperature can be as high as $10^{9}~\mathrm{GeV}$; see the third row of \cref{fig:GW_GenCharges_xHxS}. 

Our quantitative findings with regards to data from LIGO, LISA and ET are as follows: 
\begin{itemize}
    \item If a SGWB signal is not detected in the entire frequency range of these interferometers, generic conformal $\U{}^\prime$ models (including the $\U{B-L}$ model with $x_\sigma = 2$) will be disfavored for $0.5 \lesssim g_L x_\sigma \lesssim 0.6$ and $\mathrm{Tr}\left(\bm{y_\sigma}\right) \gtrsim 0.1$; see \cref{fig:GW_SNR_LVK_gen}. 
    \item Strong supercooling, defined by $g_L x_\sigma \lesssim 0.8$, will be disfavored for $M_\mathrm{Z^\prime} \approx 10 M_{h_2} \gtrsim 10^{11}~\mathrm{GeV}$, if LIGO and ET do not find a SGWB; see the top panels of \cref{fig:GW_SNR_LVK_gen}. A nonobservation will also exclude a seesaw scale with $M_{N_i} \gtrsim 10^{14}~\mathrm{GeV}$, and 
    $\mathrm{Tr}\left(\bm{y_\sigma}\right) \gtrsim 0.1$ and $\bm{y_\nu} \sim \mathcal{O}(1)$; see the bottom-right panel of \cref{fig:GW_SNR_LVK_gen}.
     \item At low frequencies, LISA will test strong supercooling for $M_\mathrm{Z^\prime} \sim 10 M_{h_2} \sim \mathcal{O}(10)~\mathrm{TeV}$, and a $\mathrm{TeV}$-scale type-I seesaw with $\bm{y_\nu} \sim \mathcal{O}(10^{-6})$; see \cref{fig:GW_SNR_LVK_gen}. At high frequencies, the model can be fully excluded for $g_L x_\sigma \lesssim 0.75$ and $M_\mathrm{Z^\prime} \gtrsim 10^{11}~\mathrm{GeV}$; see the top-right panel of \cref{fig:GW_SNR_LVK_gen}.
     \item Strong supercooling in the $\U{B-L}$ model, defined by $g_L \lesssim 0.4$, will be tested at LIGO and ET for $\mathrm{Z^\prime}$ masses above $10^{9}$~GeV, and at LISA for masses below 100~TeV; see the top panels of \cref{fig:GW_SNR_LVK_B-L}. The model can be entirely excluded for $M_\mathrm{Z^\prime} \approx 10 M_{h_2} \gtrsim 10^{14}~\mathrm{GeV}$ (top panels of \cref{fig:GW_SNR_LVK_B-L}) or for $\mathrm{Tr}(\bm{y_\sigma}) < \mathcal{O}(0.1)$ if no GW signal is found (left panels of \cref{fig:GW_SNR_LVK_B-L}). 
    \item Current LVK data exclude the $\U{B-L}$ model for $M_\mathrm{Z^\prime} \approx 10 M_{h_2} > 10^{13}~\mathrm{GeV}$, with $g_L \approx 0.3$ 
    and $\mathrm{Tr}(\bm{y_\sigma}) < \mathcal{O}(0.1)$; see the top-left panel of \cref{fig:GW_SNR_LVK_B-L}.
\end{itemize}

Current and upcoming GW experiments have the ability to test a class of classically scale-invariant $\U{}^\prime$ models of neutrino mass. Because these models have supercooled FOPTs, the GW signal is easily detectable, which may lead to a groundbreaking connection between neutrino physics and GW physics.

\section*{Acknowledgments}

We thank Andreas Ekstedt, Bogumiła Świeżewska and Maciej Kierkla for discussions. %
A.P.M.~expresses gratitude to the CERN TH Department for supporting scientific visits, which have contributed to the development of the work presented in this article and facilitated collaboration in this project.
J.G. and A.P.M.~are supported by the Center for Research and Development in Mathematics and Applications (CIDMA) through the Portuguese Foundation for Science and Technology (FCT - Funda\c{c}\~{a}o para a Ci\^{e}ncia e a Tecnologia), references UIDB/04106/2020 (\url{https://doi.org/10.54499/UIDB/04106/2020}) and UIDP/04106/2020 (\url{https://doi.org/10.54499/UIDP/04106/2020}). A.P.M. and  J.G. are also supported by the projects with references CERN/FIS-PAR/0019/2021 (\url{https://doi.org/10.54499/CERN/FIS-PAR/0019/2021}), CERN/FIS-PAR/0021 /2021 (\url{https://doi.org/10.54499/CERN/FIS-PAR/0021/2021}) and CERN/FIS-PAR/0025/2021 (\url{https://doi.org/10.54499/CERN/FIS-PAR/0025/2021}).
J.G. is also directly funded by FCT through the doctoral program grant with the reference 2021.04527.BD (\url{https://doi.org/10.54499/2021.04527.BD}).
A.P.M.~is also supported by national funds (OE), through FCT, I.P., in the scope of the framework contract foreseen in the numbers 4, 5 and 6 of the article 23, of the Decree-Law 57/2016, of August 29, changed by Law 57/2017, of July 19 (\url{https://doi.org/10.54499/DL57/2016/CP1482/CT0016}).
D.M. is supported in part by the U.S. Department of Energy under Grant No.DE-SC0010504.
R.P.~and J.G.~are supported in part by the Swedish Research Council grant, contract number 2016-05996. R.P.~also acknowledges support by the COST Action CA22130 (COMETA).

\appendix
\section{Renormalization group equations}\label{app:rges}

The beta functions,
\begin{equation*}
\beta\left(X\right) \equiv \mu \frac{d X}{d \mu}\equiv \frac{1}{16 \pi^2}\beta^{(1)}(X)\,,
\end{equation*}
for the model's couplings and generic $\mathrm{U(1)^\prime}$ charge assignments are given by
\begin{align}\label{eq:beta_functions}
    \begin{split}
    &\beta^{(1)}(g_L) = \frac{g_L}{30} \Big[123 g_{12}^2 + 4\sqrt{15} g_{12} g_L (41x_\mathcal{H} + 8x_\sigma) + 10 g_L^2 (82x_\mathcal{H}^2 + 31 x_\mathcal{H} x_\sigma + 9x_\sigma^2)\Big]\,,
    \end{split} \\
    \begin{split}
    & \beta^{(1)}(g_1) = \frac{1}{30}\Big[123g_1^3 + g_1 g_{12} (123 g_{12} + 2\sqrt{15} g_L (41x_\mathcal{H} + 8x_\sigma))\Big]\,, 
    \end{split} \\
    \begin{split}
    &\beta^{(1)}(g_2) =- \frac{19}{6} g_2^{3}\,, 
    \end{split} \\
    \begin{split}
    &\beta^{(1)}(g_3) =-7 g_3^{3}\,, 
    \end{split} \\
    \begin{split}\label{eq:betaf_g12}
    &\beta^{(1)}(g_{12}) = \frac{1}{30}\Big[g_1^2\qty{123g_{12} + 2\sqrt{15} g_L (41x_\mathcal{H} + 8x_\sigma)} + \\
    &\hspace{5em} g_{12}(123 g_{12}^2 + 4 \sqrt{15} g_{12} g_L (41x_\mathcal{H} + 8x_\sigma) + 10 g_L^2 \qty{82x_\mathcal{H}^2 + 32x_\mathcal{H} x_\sigma + 9x_\sigma^2})\Big]\,, 
    \end{split} \\
    \begin{split}
    &\beta^{(1)}(\bm{y_\nu}) = -\frac{1}{20}\Big[9g_1^2 + 9g_{12}^2 + 45 g_2^2 + 6\sqrt{15} g_{12} g_L (2x_\mathcal{H} + x_\sigma) + 30 g_L^2 (2x_\mathcal{H}^2 + 2x_\mathcal{H} x_\sigma + x_\sigma^2) - \\ &\hspace{5em} 60 y^2_t - 20 y_t\Tr{\bm{y_\nu} \bm{y_\nu}}\Big] \bm{y_\nu} + \frac{3}{2}(\bm{y_\nu} \bm{y_\nu}^\dagger \bm{y_\nu}) + 2 \bm{y_\nu} \bm{y_\sigma} \bm{y_\sigma}\,, 
    \end{split} \\
    \begin{split}
    &\beta^{(1)}(\bm{y_\sigma}) = (-\frac{3}{2} g_L^2 x_\sigma^2 + 2\Tr{\bm{y_\sigma} \bm{y_\sigma^*}}) \bm{y_\sigma} + \bm{y_\sigma} \bm{y_\nu}^\dagger \bm{y_\nu} + 4 \bm{y_\sigma} \bm{y_\sigma}^*\bm{y_\sigma} + \bm{y_\nu}^{\mathrm{T}} \bm{y_\nu}^* \bm{y_\sigma}\,,
    \end{split} \\
    \begin{split}
    &\beta^{(1)}(y_t) = -\frac{1}{60}\Big[51 g_1^2 + 51 g_{12}^2 + 2\sqrt{15}g_{12}g_L \qty(34x_\mathcal{H} + 5x_\sigma) + \\ &\hspace{5em} 5\qty{27g_2^2 + 96g_3^2 + 2g_L^2\qty(34x_\mathcal{H}^2 + 10x_\mathcal{H} x_\sigma + x_\sigma^2)} - 180y^2_t - 60\Tr{\bm{y_\nu} \bm{y_\nu}^\dagger}\Big]y_t\,,
    \end{split} \\
    \begin{split}
    &\beta^{(1)}(\lambda_h) = \frac{27}{200}g_1^4 + \frac{27}{100}g_1^2 g_{12}^2 + \frac{27}{200} g_{12}^4 + \frac{9}{20}g_1^2 g_2^2 + \frac{9}{20}g_{12}^2g_2^2 + \frac{9}{8}g_2^4 - \frac{9}{5} g_1^2 \lambda_h - \frac{9}{5} g_{12}^2 \lambda_h - \\
    &\hspace{5em} 9g_2^2 \lambda_h + 24 \lambda_h^2 + \lambda_{\sigma h}^2 + \frac{9}{5}\sqrt{\frac{3}{5}}(g_1^2 g_{12} g_L x_\mathcal{H}) + \frac{9}{5}\sqrt{\frac{3}{5}}(g_{12}^3 g_L x_\mathcal{H}) + \\
    &\hspace{5em} 3\sqrt{\frac{3}{5}} g_{12} g_2^2 g_L x_\mathcal{H} - 12\sqrt{\frac{3}{5}} g_{12} g_L \lambda_h x_\mathcal{H} + \frac{9}{5}(9 g_1^2 g_L^2 x_\mathcal{H}^2) + \frac{27}{5}(g_{12}^2g_L^2 x_\mathcal{H}^2) + \\
    &\hspace{5em} 3 g_2^2 g_L^2 x_\mathcal{H}^2 - 12 g_L^2 \lambda_H x_\mathcal{H}^2 + 12\sqrt{\frac{3}{5}} g_{12} g_L^3 x_\mathcal{H}^3 + 6 g_L^4 x_\mathcal{H}^4 + 12 \lambda_h y^2_t + 4 \lambda_h\Tr{\bm{y_\nu} \bm{y_\nu}^\dagger} - \\ 
    &\hspace{5em}6 y_t^4 - 2\Tr{(\bm{y_\nu} \bm{y^\dagger_\nu})^2}\,,
    \end{split}  \\
    \begin{split}\label{eq:lsigma_betafunction}
    &\beta^{(1)}(\lambda_\sigma) = 2\Big[10 \lambda_\sigma^2 + \lambda_{\sigma h}^2 - 6 g_L^2 \lambda_\sigma x_\sigma^2 + 3 g_L^4 x_\sigma^4 + 4\lambda_\sigma \Tr{\bm{y_\sigma} \bm{y_\sigma}^*} - 8\Tr{(\bm{y_\sigma} \bm{y_\sigma^*})^2}\Big]\,,
    \end{split} \\
    \begin{split}
    &\beta^{(1)}(\lambda_{\sigma h}) = -\frac{9}{10}(g_1^2 \lambda_{\sigma h}) - \frac{9}{10}(g_{12}^2 \lambda_{\sigma h}) - \frac{9}{2}(9 g_2^2 \lambda_{\sigma h}) + 12 \lambda_h \lambda_{\sigma h} + 8 \lambda_\sigma \lambda_{\sigma h} + 4\lambda_{\sigma h}^2 - \\
    &\hspace{5em}6 \sqrt{\frac{3}{5}} g_{12} g_L \lambda_{\sigma h} x_\mathcal{H} - 6 g_L^2 \lambda_{\sigma h} x_\mathcal{H}^2 + \frac{9}{5}(g_{12}^2 g_L^2 x_\sigma^2) - 6 g_L^2 \lambda_{\sigma h} x_\sigma^2 + 12\sqrt{\frac{3}{5}} g_{12} g_L^3 x_\mathcal{H} x_\sigma^2 + \\
    &\hspace{5em} 12 g_L^4 x_\mathcal{H}^2 x_\sigma^2 + 6\lambda_{\sigma h} y_t^2 + 2\lambda_{\sigma h} \Tr{\bm{y_\nu} \bm{y_\nu}^\dagger} + 4\lambda_{\sigma h} \Tr{\bm{y_\sigma} \bm{y_\sigma}^*} - 16 \Tr{\bm{y_\nu} \bm{y_\sigma}^* \bm{y_\sigma} \bm{y_\nu}^\dagger} \,.
    \end{split}
\end{align}

The evolution of the VEVs in the Landau gauge is given by
\begin{align}
    \begin{split}
        & \beta^{(1)}(v) = \frac{3}{20} v\qty(3g_1^2 + 3g_{12}^2 + 15g_2^2 + 4\sqrt{15} x_\mathcal{H} g_{12} g_L + 20 x_\mathcal{H}^2 g_L^2) - 3v y_t^2 - v \Tr{\bm{y_\nu} \bm{y_\nu}^\dagger} \,,
    \end{split} \\
    \begin{split}
        & \beta^{(1)}(v_\sigma) = -2 v_\sigma \Tr{\bm{y_\sigma} \bm{y_\sigma}^*}\,.
    \end{split}
\end{align}

\section{Anomalous dimensions}\label{app:anom_dim}
The anomalous dimensions,
\begin{equation}\label{eq:gamma_def}
\gamma\left(X, Y\right) \equiv \frac{1}{16\pi^2}\gamma^{(1)}\left(X, Y\right)\,,
\end{equation}
in the Landau gauge and for generic $\mathrm{U(1)}^\prime$ charges are
\begin{align}
    \begin{split}
        \gamma(\omega_1, \omega_1) &= -\frac{3}{20}\qty( 3g^2_1 + 3g^2_{12} + 15g_2^2 + 4\sqrt{15}x_\mathcal{H} g_{12}g_L + 20 x_\mathcal{H}^2 g_L^2) + 3y_t^2 + \Tr{\bm{y_\nu} \bm{y_\nu}^\dagger}\,,
    \end{split} \\
    \begin{split}
        \gamma(\omega_2, \omega_2) &= -\frac{3}{20}\qty( 3g^2_1 + 3g^2_{12} + 15g_2^2 + 4\sqrt{15}x_\mathcal{H} g_{12}g_L + 20 x_\mathcal{H}^2 g_L^2) + 3y_t^2 + \Tr{\bm{y_\nu} \bm{y_\nu}^\dagger}\,,
    \end{split} \\
    \begin{split}
        \gamma(h_r, h_r) &= -\frac{3}{20}\qty( 3g^2_1 + 3g^2_{12} + 15g_2^2 + 4\sqrt{15}x_\mathcal{H} g_{12}g_L + 20 x_\mathcal{H}^2 g_L^2) + 3y_t^2 + \Tr{\bm{y_\nu} \bm{y_\nu}^\dagger}\,,
    \end{split} \\
    \begin{split}
        \gamma(\eta, \eta) &= -\frac{3}{20}\qty( 3g^2_1 + 3g^2_{12} + 15g_2^2 + 4\sqrt{15}x_\mathcal{H} g_{12}g_L + 20 x_\mathcal{H}^2 g_L^2) + 3y_t^2 + \Tr{\bm{y_\nu} \bm{y_\nu}^\dagger}\,,
    \end{split} \\
    \begin{split}\label{eq:anom_dim}
        \gamma(h_r^\prime, h_r^\prime) &= 3x_\sigma^2g_L^2 - 2\Tr{\bm{y_\sigma} \bm{y_\sigma}^*}\,,
    \end{split} \\
    \begin{split}
        \gamma(J, J) &= 3x_\sigma^2g_L^2 - 2\Tr{\bm{y_\sigma} \bm{y_\sigma}^*}\,.
    \end{split}
\end{align}

\section{One-loop self-energy for physical scalar particles}\label{app:self_energy}

We provide a summary of all self-energy contributions to the one-loop masses of the scalar fields, which includes diagrams involving physical scalar fields, Goldstone bosons, $W^\pm$ and $\mathrm{Z^0}$ bosons, the top quark, and right-handed neutrinos. These contributions are expressed in terms of Passarino-Veltman loop functions\footnote{We utilize \texttt{Package-X}~\cite{Patel:2016fam} for the computation of all one-loop integrals. It can be downloaded from \url{https://gitlab.com/mule-tools/package-x}.}
\begin{equation}\label{eq:PV_functions}
   \begin{aligned}
        &B_0(s, M_1, M_2) = \mu^{2\epsilon} e^{\gamma_E \epsilon} \frac{1}{2} \Gamma(\epsilon) \lim_{\varepsilon \rightarrow 0^+} \int_0^1 dx (sx^2 + (-s + M^2_2 + M^2_1)x + M^2_1 - i\varepsilon)^\epsilon\,, \\
        &A_0(M) = \mu^{2\epsilon} e^{\gamma_E \epsilon} \Big[ -\frac{1}{2} \Gamma(-1+\epsilon) M^2 \Big] \Big(\frac{1}{M^2}\Big)^{-1+\epsilon}\,,
   \end{aligned}
\end{equation}
where $\Gamma(x)$ is the gamma function. We present all self-energy contributions for the Higgs boson in the Landau gauge. The same diagrams contribute to the mass of the heavy Higgs with appropriate couplings and masses. We denote all scalar fields by $\Phi = h_1, h_2, G_1^0, G_2^0, G^\pm$. Note that all couplings should be interpreted as physical couplings since they are determined after symmetry breaking in the mass basis.
\vskip1.mm
\begin{align}\label{eq:self_energies}
    &\smash{\parbox{85pt}{\includegraphics{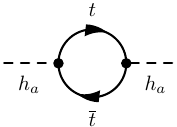}}} = && -\frac{i}{16 \pi ^2} \Big[9 Y_{t,a}^2 \Big(\left(4 M_t{}^2-p^2\right) B_0(p^2,M_t,M_t)+2 A_0(M_t)\Big)]\,, \\[4.0em]
    &\smash{\parbox{85pt}{\includegraphics{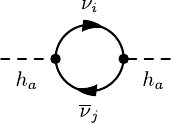}}} = && -\frac{i}{16 \pi ^2} \Big[\left(Y_{\nu }.Y^\mathrm{T}_{\nu})_{ij, a}\right. \left((M_{\nu _i}+M_{\nu _j}){}^2-p^2\right) B_0(p^2,M_{\nu _i},M_{\nu _j})+ \\\nonumber&{} && \left(Y_{\nu }.Y^\mathrm{T}_{\nu})_{ij, a}\right. A_0(M_{\nu _i})+\left(Y_{\nu }.Y^\mathrm{T}_{\nu})_{ij, a}\right. A_0(M_{\nu_j})\Big]\,, \\[2.5em] 
    &\smash{\parbox{85pt}{\includegraphics{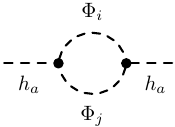}}} = && \frac{i}{32 \pi ^2} \Big[ \left(\lambda^2_{h_aG_1^0G_1^0}+\lambda^2_{h_aG_2^0G_2^0}+\lambda^2_{h_aG^{\pm }G^{\mp}}\right) B_0(p^2, 0 , 0) + \\\nonumber &{} && \lambda^2_{h_ah_1h_1} B_0(p^2, M_{h_1}, M_{h_1}) + \lambda^2_{h_ah_1h_2} B_0(p^2, M_{h_2},M_{h_1}) + \\\nonumber &{} && \lambda^2_{h_ah_2h_2} B_0(p^2,M_{h_2},M_{h_2})\Big]\,, \\[2.0em]
    &\smash{\parbox{85pt}{\includegraphics{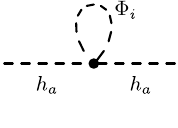}}} = && \frac{i}{32 \pi ^2} \Big[2 \lambda _{h_ah_ah_1h_1} A_0(M_{h_1}) + 2 \lambda _{h_ah_ah_2h_2} A_0(M_{h_2})\Big]\,, \\[2.0em]
    &\smash{\parbox{85pt}{\includegraphics{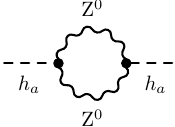}}} && = -i\frac{g^2_{h_a \mathrm{Z^0 Z^0}}}{64\pi^2 M^2_{\mathrm{Z^0}}}\Big[ - 2 M_{\mathrm{Z^0}}^2 A_0(M_{\mathrm{Z^0}}) + 2(M^2_{\mathrm{Z^0}} - p^2)^2 B_0(p^2, M_{\mathrm{Z^0}},0) - \\\nonumber &{} &&(12M^4_{\mathrm{Z^0}} - 4M^2_{\mathrm{Z^0}}p^2 + p^4) B_0(p^2, M_{\mathrm{Z^0}}, M_{\mathrm{Z^0}}) - p^4 B_0(p^2, 0, 0) 
    \Big]\,, \\[2.0em]
    &\smash{\parbox{85pt}{\includegraphics{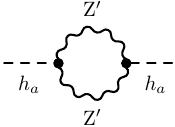}}} = && -i\frac{g^2_{h_a \mathrm{Z^\prime Z^\prime}}}{64\pi^2 M^2_{\mathrm{Z^\prime}}}\Big[ - 2 M_{\mathrm{Z^\prime}}^2 A_0(M_{\mathrm{Z^\prime}}) + 2(M^2_{\mathrm{Z^\prime}} - p^2)^2 B_0(p^2, M_{\mathrm{Z^\prime}},0) - \\ \nonumber &{} &&(12M^4_{\mathrm{Z^\prime}} - 4M^2_{\mathrm{Z^\prime}}p^2 + p^4) B_0(p^2, M_{\mathrm{Z^\prime}}, M_{\mathrm{Z^\prime}}) - p^4 B_0(p^2, 0, 0) 
    \Big]\,, \\[2.0em]
    &\smash{\parbox{85pt}{\includegraphics{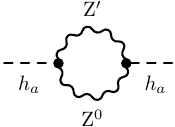}}} = && i\frac{g^2_{h_a \mathrm{Z^\prime Z^0}}}{64\pi^2 M_{\mathrm{Z^\prime}} M_{\mathrm{Z^0}}}\Big[ M_\mathrm{Z^0}^4 B_0(p^2, M_{\mathrm{Z^0}}, M_{\mathrm{Z^\prime}}) + \Big(-2p^2 (M^2_\mathrm{Z^0} + M^2_\mathrm{Z^\prime}) + p^4 \\\nonumber & {} && + 10 M^2_\mathrm{Z^0} M^2_\mathrm{Z^\prime} + M^4_\mathrm{Z^\prime} \Big) B_0(p^2, M_{\mathrm{Z^0}}, M_{\mathrm{Z^\prime}}) - (M^2_{\mathrm{Z^0}} - p^2)^2 B_0(p^2, M_{\mathrm{Z^0}}, 0) \\\nonumber & {} && - (M^2_{\mathrm{Z^\prime}} - p^2)^2 B_0(p^2, M_{\mathrm{Z^\prime}}, 0) + p^4 B_0(p^2, 0, 0) +  M^2_{\mathrm{Z^\prime}} A_0(M_{\mathrm{Z^\prime}}) \\\nonumber & {} && + M^2_{\mathrm{Z^0}} A_0(M_{\mathrm{Z^0}}), \Big]\,, \\[2.0em]
    &\smash{\parbox{85pt}{\includegraphics{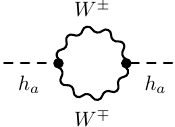}}} = && -i\frac{g^2_{h_a \mathrm{W^+ W^-}}}{32\pi^2 M^2_{\mathrm{W}}}\Big[ - 2 M_{\mathrm{W}}^2 A_0(M_{\mathrm{W}}) + 2(M^2_{\mathrm{W}} - p^2)^2 B_0(p^2, M_{\mathrm{W}},0) - \\\nonumber &{} &&(12M^4_{\mathrm{W}} - 4M^2_{\mathrm{W}}p^2 + p^4) B_0(p^2, M_{\mathrm{W}}, M_{\mathrm{W}}) - p^4 B_0(p^2, 0, 0)
    \Big]\,,\\[3.0em] 
    &\smash{\parbox{85pt}{\includegraphics{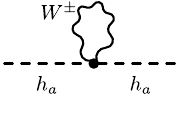}}}
    = && \frac{3ig^2_{h_a \mathrm{W^+ W^-}}}{16\pi^2} A_0 (M_\mathrm{W})\,, \\[4em]
    &\smash{\parbox{85pt}{\includegraphics{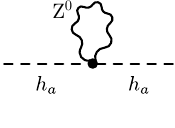}}} = && \frac{3ig^2_{h_a \mathrm{Z^0 Z^0}}}{32\pi^2} A_0 (M_\mathrm{Z^0})\,, \\[4em]
    &\smash{\parbox{85pt}{\includegraphics{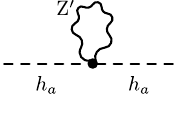}}} = && \frac{3ig^2_{h_a \mathrm{Z^\prime Z^\prime}}}{32\pi^2} A_0 (M_\mathrm{Z^\prime})\,, \\[3.0em] 
    &\smash{\parbox{85pt}{\includegraphics{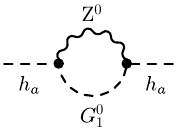}}} = && -\frac{i g_{h_a G_1^0 \mathrm{Z^0}}^2}{16 \pi ^2 M_{\mathrm{Z^0}}^2} \Big[\left(-4 p^2 M^2_{\mathrm{Z^0}} + p^4 - M^4_{\mathrm{Z^0}}\right) B_0\left(p^2; M_{\mathrm{Z^0}} , 0\right) - \\\nonumber &{} && p^4 B_0\left(p^2, 0, 0\right) + \left(M^2_{\mathrm{Z^0}}-p^2\right) A_0(M_{\mathrm{Z^0}})\Big]\,, \\[2.0em]
    &\smash{\parbox{85pt}{\includegraphics{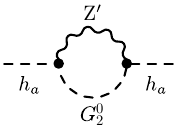}}} = && -\frac{i g_{h_a G_1^0 \mathrm{Z^\prime}}^2}{16 \pi ^2 M_{\mathrm{Z^\prime}}^2} \Big[\left(-4 p^2 M^2_{\mathrm{Z^\prime}} + p^4 - M^4_{\mathrm{Z^\prime}}\right) B_0\left(p^2; M_{\mathrm{Z^\prime}} , 0\right) - \\\nonumber &{} && p^4 B_0\left(p^2, 0, 0\right) + \left(M^2_{\mathrm{Z^\prime}}-p^2\right) A_0(M_{\mathrm{Z^\prime}})\Big]\,, \\[2.0em]
    &\smash{\parbox{85pt}{\includegraphics{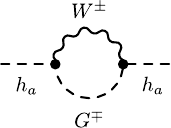}}} = && -\frac{i g_{h_a G^\mp W^\pm}^2}{8 \pi ^2 M_{\mathrm{W}}^2} \Big[\left(-4 p^2 M^2_{\mathrm{W}} + p^4 - M^4_{\mathrm{W}}\right) B_0\left(p^2; M_{\mathrm{W}} , 0\right) - \\\nonumber &{} && p^4 B_0\left(p^2, 0, 0\right) + \left(M^2_{\mathrm{W}}-p^2\right) A_0(M_{\mathrm{W}})\Big]\,.
\end{align}

\section{Numerical procedure for determining physical observables}\label{app:numeric_procedure}

\subsection{One-loop minimization and masses}\label{sec:one_loop_min_masses}

To determine the couplings and physical masses, we minimize the full one-loop potential, including the self-energy corrections to the mass spectrum. While some aspects of this procedure were discussed in the main text, we now delve deeper into the technical details. By neglecting scalar contributions to the one-loop masses, we can derive the following analytical expressions for the quartic couplings by solving \cref{eq:tadpoles_1loop}:
\begingroup
\allowdisplaybreaks
\begin{align}
	&\nonumber \lambda_\sigma = \frac{1}{256 \pi ^2 v_\sigma ^4} \Bigg[3 v^4 \left(\left(g_1^2+g_2^2\right)^2 \mathrm{ln} \left(\frac{v^2 \left(g_1^2+g_2^2\right)}{4 \mu ^2}\right)+2 g_2^4 \mathrm{ln} \left(\frac{g_2^2 v^2}{4 \mu^2}\right)-16 Y_t^4 \mathrm{ln} \left(\frac{Y_t^2 v^2}{2 \mu ^2}\right)\right)- \\
	&\nonumber \hspace*{2em} \left(v^4\left(g_1^4+2 g_1^2 g_2^2+3 g_2^4+(2 g_L x_\mathcal{H}+g_{12})^4-256 \pi ^2 \lambda_h-48 Y_t^4\right)\right)+16 v_\sigma ^4 \left(g_L^4 x_\sigma^4  - {\rm Tr}(\bm{y_\sigma}^4) \right) + \\
	&\nonumber \hspace*{2em} 3 (-2 g_L x_\sigma v_\sigma +2 g_L x_\mathcal{H} v+g_{12} v) \left(4 g_L^2 x_\sigma^2 v_\sigma ^2+v^2 (2 g_L x_\mathcal{H}+g_{12})^2\right) (2 g_L (x_\sigma v_\sigma +x_\mathcal{H} v) + \\
	&\hspace*{2em} g_{12} v) \mathrm{ln} \left(\frac{g_L^2 x_\sigma^2 v_\sigma ^2+\frac{1}{4} v^2 (2 g_L x_\mathcal{H}+g_{12})^2}{\mu ^2}\right)+16 v_\sigma ^4 {\rm Tr}(\bm{y_\sigma}^4) \mathrm{ln} \left(\frac{{\rm Tr}(\bm{y_\sigma}^2) v_\sigma ^2}{2\mu ^2}\right)\Bigg]\,, \\
	&\nonumber \lambda_{\sigma h} = -2 \lambda_h\frac{v^2}{v_\sigma ^2} -\frac{1}{256 \pi ^2}\frac{v^2}{v_\sigma ^2} \Bigg[ 3 \Big(2 \left(g_1^2+g_2^2\right)^2 \left(\mathrm{ln} \left(\frac{v^2 \left(g_1^2+g_2^2\right)}{4 \mu ^2}\right)-\frac{5}{6}\right)+	\left(g_1^2+g_2^2\right)^2+2 g_2^4 + \\
	&\nonumber \hspace*{3em} 4 g_2^4 \left(\mathrm{ln} \left(\frac{g_2^2 v^2}{4 \mu ^2}\right)-\frac{5}{6}\right)+8 (2 g_L x_\mathcal{H}+g_{12})^2 \left(g_L^2 x_\sigma^2 \frac{v_\sigma ^2}{v^2}+\frac{1}{4} (2 g_L x_\mathcal{H}+g_{12})^2\right) \times \\
	&\nonumber \hspace*{3em} \left(\mathrm{ln} \left(\frac{g_L^2 x_\sigma^2 v_\sigma ^2+\frac{1}{4} v^2 (2 g_L x_\mathcal{H}+g_{12})^2}{\mu ^2}\right)-\frac{5}{6}\right)+4 g_L^2 x_\sigma^2 \frac{v_\sigma^2}{v^2} (2 g_L x_\mathcal{H}+g_{12})^2+ (2 g_L x_\mathcal{H}+g_{12})^4 - \\
	&\hspace*{3em} 16 Y_t^4 +16 Y_t^4 \left(-2 \mathrm{ln} \left(\frac{Y_t^2 v^2}{\mu ^2}\right)+3+\mathrm{ln}(4)\right)\Big) \Bigg]\,.
\end{align}
\endgroup
The calculations are performed at the Z$^0$ mass scale, $\mu = 91~\mathrm{GeV}$. The remaining parameters, $\lambda_h$ and $v_\sigma$, are determined using additional equations derived from the one-loop mass spectrum. Unlike the tadpole equations, obtaining analytical expressions for these parameters is not feasible, necessitating the use of numerical methods. In determining the one-loop masses, mixed self-energies are neglected. Instead, the scalar mixing angle is derived from the zero-momentum part of the mass matrix (i.e., from the first and second terms of \cref{eq:mass_matrix_1loop}). A root-finding algorithm is subsequently employed to determine the values of $\lambda_h$ and $v_\sigma$ using \cref{eq:loop_corrected_masses} with the SM-like Higgs boson, $h_1$, and heavy Higgs, $h_2$, masses as free parameters, ensuring that $M_{h_1} > M_{h_2}$. Valid solutions to this set of equations are then subjected to theoretical constraints. These include the perturbativity of the quartic couplings and the absence of Landau poles both in the infrared (down to the QCD scale, approximately $0.17~\mathrm{GeV}$) and in the ultraviolet. In the RG-improved method, the ultraviolet cutoff is determined by ensuring that no Landau poles exist for field values at the true vacuum. Additionally, we require that the Higgs quartic coupling $\lambda_h$ remains positive up to the highest mass scale set by $M_\mathrm{Z^\prime}$.

\subsection{Phase tracing}

In addition to calculating the bounce action, \texttt{CosmoTransitions} includes a phase tracing module that tracks the true and false vacuum in field space and temperature. However, since we work with a single field and the false vacuum is always located at zero due to classical conformal invariance, the phase tracing module is unnecessary. Instead, we employ our own code to compute the location of the false vacuum at different temperatures. This can be done with any publicly available numerical minimization routine, such as \texttt{fmin} from the \texttt{scipy} package.

If one opts to use the phase tracing module from \texttt{CosmoTransitions}, it is important to adjust the \texttt{x\_eps} parameter of the \texttt{generic\_potential} subclass. While \texttt{CosmoTransitions} is generally designed to be scale-invariant, there are some limitations. The \texttt{x\_eps} parameter effectively controls the temperature scale of the problem and is optimized for electroweak scale temperatures by default. Therefore, if the false and true vacua are well separated, lowering the value of \texttt{x\_eps} is necessary. We find that this adjustment becomes important if $\Delta\phi > 2~\mathrm{TeV}$, where $\Delta \phi$ is the difference in field values between the two phases. This parameter can be modified within the \texttt{\_\_init\_\_} method of the \texttt{generic\_potential}. To illustrate this, we consider the following example subclass:
\begin{python}
from cosmoTransitions import generic_potential

class MyPotential(generic_potential.generic_potential):
    def __init__(self, parameter_values):
        self.parameter_values = parameter_values
        self.x_eps = 1e-5  # Adjust this value as needed
        
    def boson_massSq(self, X, T):
        # Define the thermal corrections here
        pass

    def V0(self, X):
        # Define the tree-level potential here
        pass

    def VT(self, T):
        # Define the thermal corrections here
        pass
\end{python}
Note that if \texttt{x\_eps} is too large, then \texttt{CosmoTransitions} may skip over phases for low field/temperature values. Similarly, if \texttt{x\_eps} is too small, then \texttt{CosmoTransitions} may skip over phases located at large field/temperature values. Therefore, for a generic analysis, one must treat \texttt{x\_eps} as a dynamical variable according to the scales involved in the calculation.

\subsection{Calculating the action}

To guarantee that in conformal models the action goes to zero at zero temperature, the default parameters of \texttt{CosmoTransitions} must be modified. When calculating the action with the \texttt{pathDeformation.fullTunneling} method,\footnote{Since we work in a single field direction, the \texttt{tunneling1D.SingleFieldInstanton} method may be used. Both approaches give identical results. For multi-field cases, however, \texttt{pathDeformation.fullTunneling} must be used.} the tolerances and ranges of the integration limits need to be tuned. Consider the following example:
\begin{python}
import pathDeformation as pd

Find_profile_params = {"phitol":1e-10,"xtol":1e-10,
                       "rmin":1e-4,"rmax":1e4,"npoints":500}
Instanton_params    = {"phi_eps":1e-6,"rscale":None}
deformation_params  = {"verbose":False}

S = pd.fullTunneling(np.array([XTrue, XFalse]), V, dV,
                     deformation_deform_params=deformation_params,
                     tunneling_init_params=Instanton_params,
                     V_spline_samples = None,
                     tunneling_findProfile_params=Find_profile_params).action
\end{python}
Here, \texttt{XTrue} and \texttt{XFalse} are the field coordinates for the true and false vacuum, respectively, \texttt{V} is the full scalar potential, and \texttt{dV} its field derivative. We define three dictionaries to store relevant tolerances, and emphasize the necessity for small values of \texttt{phitol} and \texttt{xtol}, which control the step size during the integration of the bounce equation \eqref{eq:euclidean_action}. We find that for the default parameters of $10^{-6}$, as $T \to 0$, the action $S \rightarrow \infty$ in conformal models. Therefore, it is crucial to set these parameters to $10^{-10}$ even though this significantly slows down calculations. Since this adjustment is essential for low-temperature calculations, we relax these tolerances at higher temperatures. Above $T_c/5$, default tolerances suffice. The other parameters have minimal impact and their default settings are acceptable.

 By default, \texttt{CosmoTransitions} creates a spline function of the user-provided potential to speed up computation. While this is adequate for polynomial-like potentials, it is inadequate for conformal models because the potential is nearly flat in the vicinity of the true vacuum. The default number of spline points is insufficient to accurately capture the potential's behavior in such cases. Therefore, setting \texttt{V\_spline\_samples} to \texttt{None} ensures that the full potential is employed without approximation. Alternatively, similar results can be obtained by specifying a high density of sample points ($e.g.$, \texttt{V\_spline\_samples = 50\_000}). However, based on our findings, setting it to \texttt{None} is preferable as it eliminates the need for approximating the potential, thereby enhancing both precision and computing speed.

To test the validity of our code, we have cross-checked our implementation against previous work on conformal models. In particular, we were able to reproduce the results of Refs.~\cite{Kierkla:2022odc,Kierkla:2023von} for an $\SU{2}{}$ conformal extension of the SM, and Ref.~\cite{Ellis:2020nnr} for a $\U{B-L}$ conformal model.

\newpage
\bibliographystyle{JHEP}
\bibliography{Refs}

\end{document}